# Dual-band *in situ* molecular spectroscopy using single-sized Al-disk perfect absorbers

Thang Duy Dao*[1], Kai Chen[1,2] and Tadaaki Nagao*[1,3]

[1]*International Center for Materials Nanoarchitectonics, National Institute for Materials Science (NIMS), 1-1 Namiki, Tsukuba, Ibaraki 305-0044, Japan.*

[2]*Institute of Photonics Technology, Jinan University, Guangzhou, 510632, China*

[3]*Department of Condensed Matter Physics, Graduate School of Science, Hokkaido University, Kita 10, Nishi 8, Kita-ku, Sapporo 060-0810, Japan*

*Corresponding Authors:*

*Thang Duy Dao: Dao.duythang@nims.go.jp; katsiusa@gmail.com*

*Tadaaki Nagao: Nagao.Tadaaki@nims.go.jp*

**Graphical abstracts**

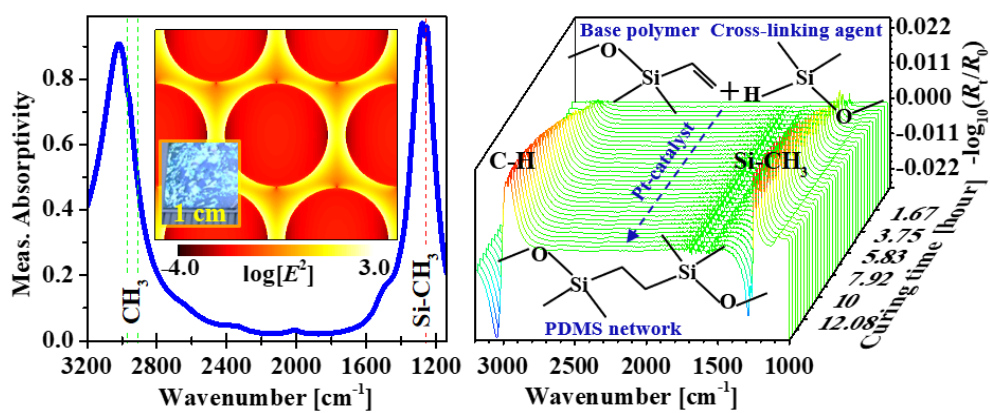

## Abstract

We propose antenna-enhanced infrared vibrational spectroscopy by adopting single-sized Al disks on $Al_2O_3$-Al films fabricated by colloidal-mask lithography. The precisely designed plasmonic resonator with dual-band perfect absorption (DPA) shows strongly-enhanced nearfield intensity and polarization independence, at both resonances, providing a powerful antenna platform for the multi-band vibrational sensing. As a proof of concept, we experimentally apply the plasmonic DPAs in bond-selective dual-band infrared sensing of an ultrathin polydimethhylsiloxine (PDMS) film, simultaneously amplifying two representative vibrational bands (asymmetric C–H stretching of $CH_3$ at 2962 $cm^{-1}$ and $CH_3$ deformation of Si-$CH_3$ at 1263 $cm^{-1}$) by surface-enhanced infrared absorption spectroscopy (SEIRA). The plasmonic DPA was successfully adopted for the *in situ* monitoring of reaction kinetics, by recording the spectral changes in C-H stretching and Si-$CH_3$ deformation modes of a 10-nm PDMS elastomer, which are selectively enhanced by the two antenna resonances, during its gelation process. Our systematic study of the SEIRA spectra has demonstrated mode splitting and a clear avoided-crossing in the dispersion curve as a function of resonance frequency of DPA, manifesting itself as a promising basis for future polaritonic devices utilizing the hybridization between the molecular vibrational states and the enhanced light field.

**Introduction**

Engineering of efficient light absorption and nearfield enhancement at desired wavelengths by metamaterial perfect absorbers (PAs) provides us a wide variety of applications that would be impossible using natural materials. Since the first PA was experimentally realized by Landy *et al.* in 2008 in the gigahertz frequency region, the concept of the PA has enjoyed it widespread applicability from the terahertz and infrared (IR), to the visible spectral regions owing to its unique ability to achieve 100% absorptance, or unity absorptivity[1]. In the past decade, PAs have attracted much attention in the IR spectral range owing to their significant impact on thermal devices and IR spectroscopy applications, such as in thermal emitters[2–7], IR detectors[8–13], gas sensing[14,15] thermophotovoltaic[16–19] and radiative cooling[20–23], and especially in surface-enhanced infrared absorption spectroscopy (SEIRA)[24–32]. Most structural designs for PAs have focused on the controllability and tunability of their resonant frequency, bandwidth, polarization, and working angle as well as their reconfigurability. In this regard, increasing demand exists not only on single-band PAs, but also on dual- and multiple-band PAs due to their great potential in non-destructive chemical analysis and in remote sensing applications such as non-dispersive infrared spectroscopy (NDIR) and SEIRA.[14,29,32]

Among those dual-band PAs designs, many structures have been proposed, including patterned symmetric cross-shapes arrays[5,33,34], asymmetric cross-shape arrays[29], elliptical nanodisk arrays[35], two different nanopatterns[14,32], T-shaped plasmonic arrays[36], and distinct dielectric spacing layers[37]. These structures rely on the first order of the magnetic resonances within two different resonators simultaneously arranged in the top layers. Several works have employed higher orders of the magnetic modes in plasmonic metal PAs for dual- or multiple-band resonance.[38] Yoo *et al.* experimentally demonstrated single-sized disk or donut-type structures utilizing the third-order magnetic resonance for dual-band PAs in the gigahertz region;

however, the polarization properties and the tunability of these two resonances were not fully examined.[39] Recently, Ito *et al.* reported a proximity-coupled-resonators-based dual-band PA, in which the resonators were carefully-arranged in the top layer to control both the fundamental and second-order magnetic modes.[40] However, the second-order resonance was realized only at oblique incident angles. Thus, development of a simpler device composed of single-sized building blocks, which exhibits polarization independence, a wide working angle, as well as high tunability and reconfigurability is still required for practical applications.

In this work, we demonstrate dual-band perfect absorbers (DPAs) with high tunability in the mid-infrared (MIR) region for effective use in detecting a trace amount of molecules as well as their reaction processes. The structural design of the DPAs comprises an array of single-sized symmetric Al disk resonators fabricated by colloidal-mask lithography and a metal-insulator bilayer placed beneath them. The DPAs are designed by high-precision numerical electromagnetic simulations which adopts both the surface plasmon polariton (SPP)-coupled (third-order) magnetic resonance and the fundamental magnetic resonance. Remarkably, the proposed plasmonic DPAs exhibit excellent absorptivity (i.e. 98%), with polarization independence at both resonances. Furthermore, with wide tunability in the mid-IR region, the resonances of DPAs' span the majority of the molecular fingerprint region. In particular, with strong nearfield coupling and confinement in the subwavelength plasmonic cavity, the fabricated DPAs show exceptional success as IR plasmonic antennas for the multispectral SEIRA spectroscopy. We exemplify the wavelength-selective SEIRA detection capability using plasmonic DPAs to simultaneously enhance two vibrational bands (asymmetric C-H stretching in $CH_3$ at 2962 cm$^{-1}$ and $CH_3$ deformation in Si-$CH_3$ at 1263 cm$^{-1}$) of a 10-nm polydimethylsiloxane (PDMS) film. We successfully applied our DPA antennas to the *in situ* monitoring of the gelation process and reaction kinetics of a 10-nm thick PDMS film. Our further analyses revealed Fano coupling between the molecular vibrations and the antenna resonance as evidenced by the remarkable mode splitting and anti-crossing in the polaritonic bands. The obtained knowledge here can serve as a promising basis for the polariton-based mid-infrared devices based on the hybridization between the vibrational states of matter and plasmonic excitations.

## Results and discussion

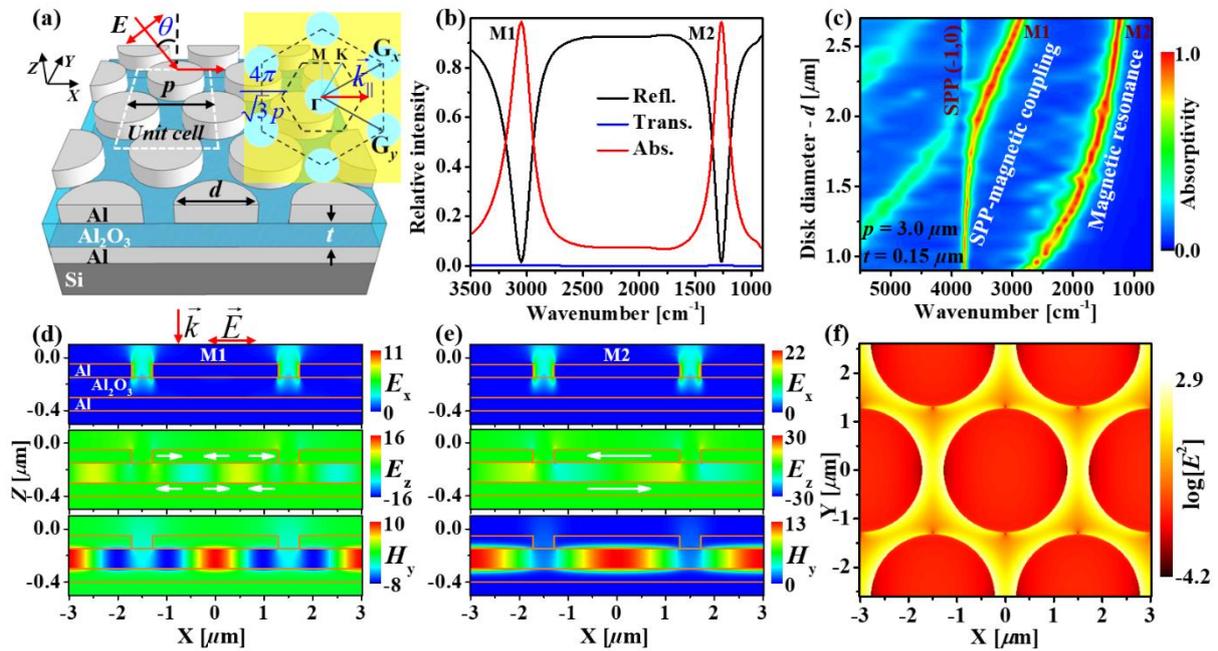

**Figure 1.** (a) A schematic illustration of DPAs with definition of a simulation unit cell (dashed rectangle) and geometrical parameters: periodicity - $p$, Al disk diameter - $d$ and insulating $Al_2O_3$ thickness - $t$. The inset in (a) describes a reciprocal lattice of the top Al disk array with a lattice constant of $\frac{4\pi}{\sqrt{3}p}$. (b) Simulated spectra of a DPA ($p = 3.0$ $\mu$m, $d = 2.55$ $\mu$m, $t = 0.15$ $\mu$m, and 0.1 $\mu$m thick Al disks bottom film) manifest two perfect absorption resonant peaks at 3044 cm$^{-1}$ (M1) and 1280 cm$^{-1}$ (M2). (c) Simulated absorptivity of DPAs with different disk diameters ($p$ and $t$ are fixed at 3.0 and 0.15 $\mu$m, respectively) shows the tunability of DPAs covering the MIR region. (d) – (e) Electric fields ($E_x$, top), ($E_z$, middle) and magnetic field ($H_y$, bottom) distribution in a DPA ($p = 3.0$ $\mu$m, $d = 2.55$ $\mu$m,

and $t = 0.15$ $\mu$m) excited at both M1 (d) and M2 (e) resonances shown in (b). (f) Simulated total electric field distribution at M2 evidences strongly enhanced nearfield intensity at vicinity of Al disks. In all simulations, the incident electric field oscillates along the *X*-axis and propagates along the *-Z*-axis; the amplitudes of the incident fields and are normalized to unity.

**Structure and optical properties of the DPAs**

Figure 1a illustrates a schematic design of the DPA with periodicity *p*, disk diameter *d*, and insulating layer thickness *t* as geometrical parameters. In this work, the Al film is fixed at 0.1 $\mu$m while the thickness of the Al disks is optimized at 0.1 $\mu$m (Figure S1a, Supporting Information). Therefore, the optical properties of the DPA are mainly controlled by changing *p*, *d* and *t*. The inset in Figure 1a depicts the reciprocal lattice space (momentum space) of the upper Al disk hexagonal array, where the lattice constant *k* is transformed from the corresponding real space lattice constant, i.e. $k = \frac{4\pi}{\sqrt{3}p}$. Figure 1b presents the simulated transmittance, reflectance, and absorptivity spectra of a DPA with $p = 3.0$ $\mu$m, $d = 2.55$ $\mu$m, $t = 0.15$ $\mu$m (optimized thickness for 3.0-$\mu$m periodicity DPAs, see Figure S1b in the Supporting Information). As seen in Figure 1b, the DPA exhibits a pair of narrow absorption bands resonating at 3044 cm$^{-1}$ (M1) and 1280 cm$^{-1}$ (M2) with nearly perfect absorptivity (0.98). It has been shown that in a metal-insulator-metal dipole absorber, the induced electric dipole at the lower metal film oscillates with opposite phase compared to that of the electric dipole in the upper metallic resonator, resulting in an enhancement of the magnetic field between the upper and the lower metals layers. This resonance is therefore called the magnetic resonance. In addition to the fundamental magnetic resonance (the smaller wavenumber resonance, namely M2), an additional absorption band is observed in the higher wavenumber range, which is the SPP-coupled magnetic resonance resulting from hybridization between the third-order magnetic resonance and the photonic mode (SPP) of the periodic Al disk array (the higher wavenumber resonance, namely M1).

To further understand the origin of these two resonant bands, we performed an angle-resolved simulation (plotted in wavelength). The result is shown in Figure S2a (Supporting Information). It was found that, for the fundamental magnetic mode, M2, at longer wavelength (lower wavenumber), the resonant peak position and peak intensity is almost unchanged across a wide incident angle range up to 80 degrees (°). By contrast, as a result of SPP-coupling with the magnetic resonance, the shorter-wavelength resonance (higher wavenumber), M1, changes dramatically when the incident angle increases beyond 15°: in addition, it strongly depends on the periodicity of the hexagonal lattice (Figure S2b, Supporting Information). In a plasmonic hexagonal lattice, SPPs at the metal-air interface are excited if their momentum $|\vec{k}_{spp}| = k_0 \sqrt{\frac{\varepsilon_m}{\varepsilon_m + 1}}$ ($\varepsilon_m$ is the complex permittivity of the metal, $k_0 = \frac{2\pi}{\lambda}$) matches the momentum of the incident photon and the hexagonal lattice: $|\vec{k}_{spp}| = |\vec{k}_\parallel + i\vec{G}_x + j\vec{G}_y|$, where $|\vec{k}_\parallel| = k_0 \sin\theta$ is the projection of momentum of the excitation photon with an incident angle of $\theta$ on the metal surface, $\vec{G}_x$ and $\vec{G}_y$ are two primitive lattice vectors ($|\vec{G}_x| = |\vec{G}_y| = \frac{4\pi}{\sqrt{3}p}$), and $i$ and $j$ are integers. The angular-dependent dispersion relation for a hexagonal lattice along the $(\vec{G}_x + \vec{G}_y)$ direction can be written as (see the Supporting Information for the detailed derivation): $\frac{\varepsilon_m}{\varepsilon_m + 1} = \sin^2\theta + \frac{2}{p}(i+j)\lambda\sin\theta + \frac{4}{3p^2}(i^2 + ij + j^2)\lambda^2$. At normal incidence, the resonant wavelength is easily calculated using the expression $\lambda_{spp} = \frac{\sqrt{3}p}{2\sqrt{(i^2 + ij + j^2)}}\sqrt{\frac{\varepsilon_m}{\varepsilon_m + 1}}$.

Here, with $p = 3.0\ \mu m$, the SPP resonance is found at 2.6 $\mu$m (3846 cm$^{-1}$). As shown in Figure

S2 (Supporting Information), M1 is clearly originated from the coupling between the SPP mode (-1,0) (dashed black curves) and the third-order magnetic resonance. By tuning the size of Al disk, both resonant bands are readily tuned in the MIR region to span the most interesting range of molecular vibrations (Figure 1c and Figure S2c in the Supporting Information). Here we used the half-wave dipole antenna model to exemplify the dependence of the resonance of the DPA on the diameter of the Al disk and to estimate the coupling energy strength (i.e. 28 meV) between the SPP and the third-order magnetic resonance (see details in the Supporting Information). The electromagnetic field distribution was also simulated on the resonances to confirm the optical properties of the DPA. As shown in Figures 1d-f, the electric fields ($E_z$) excited at both M1 and M2 display opposite phase oscillations at the upper Al disk and the lower film, resulting in dramatic enhancement of magnetic field (i.e. 10 times). As stated above, M1 originates from the third-order magnetic resonance coupled with the SPP and the electric field distribution for this resonance clearly shows three spots with enhanced resonant magnetic field below each resonator (Figure 1d) while M2, which results from the fundamental magnetic resonance, exhibits only one spots with enhanced resonant magnetic field (Figure 1e). Furthermore, the total electric field intensity distribution (Figure 1f) indicates strong near-field intensity enhancement (i.e. 1000 times), suggesting that the DPA can be a good antenna platform for near-field-enhanced spectroscopy applications such as SEIRA, especially for dual-band SEIRA utilizing both the M1 and M2 resonances.

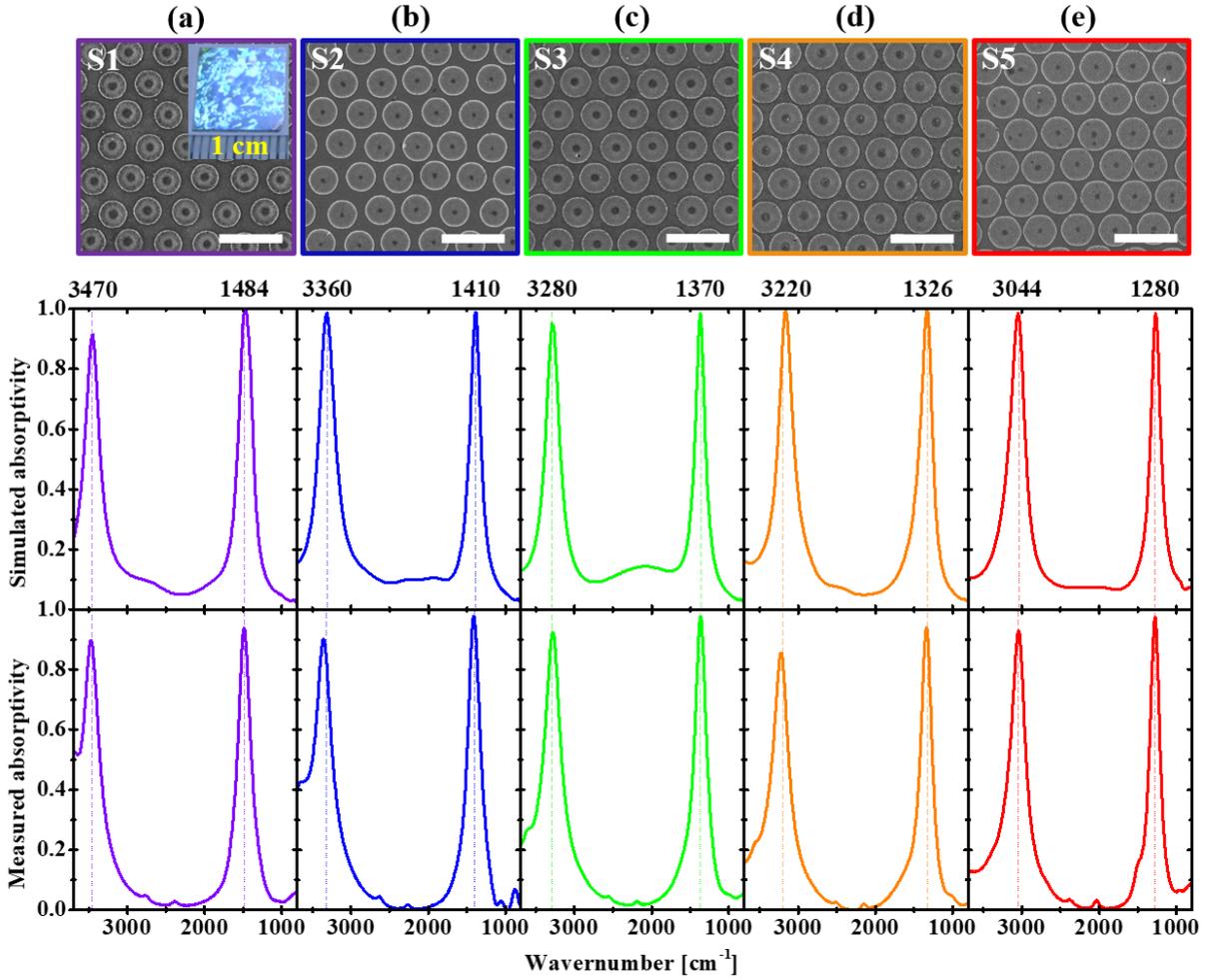

**Figure 2.** From top to bottom: SEM images, simulated and measured absorptivity spectra of 5 different DPAs having the same periodicity and insulator thickness ($p = 3.0$ $\mu$m, $t = 0.15$ $\mu$m) but with different disk diameters. (a) S1 with $d = 2.08$ $\mu$m resonates at 3470 cm$^{-1}$ and 1484 cm$^{-1}$. (b) S2 with $d = 2.21$ $\mu$m resonates at 3360 cm$^{-1}$ and 1410 cm$^{-1}$. (c) S3 with $d = 2.32$ $\mu$m resonates at 3280 cm$^{-1}$ and 1370 cm$^{-1}$. (d) S4 with $d = 2.43$ $\mu$m exhibits two resonances at 3220 cm$^{-1}$ and 1326 cm$^{-1}$. (d) S5 with $d = 2.55$ $\mu$m reveals two resonances at 3044 cm$^{-1}$ and 1280 cm$^{-1}$. The inset in panel (a) displays a photo of a 1×1 cm$^2$ fabricated S1. The scale bars in all the SEM images correspond to 5 $\mu$m.

To realize the DPA structure (i.e. 1×1 cm$^2$ as shown in the inset of Figure 2a), we employed a colloidal-mask lithography process involving two reactive-ion etching (RIE) steps.[6] The details are shown in the Methods section. Here, a set of five DPAs with the same periodicity and insulator thickness ($p = 3.0$ $\mu$m, and $t = 0.15$ $\mu$m) but different diameters are numerically simulated and precisely fabricated (SEM images, Figure 2, upper row of images), namely, S1 ($d$ = 2.08 $\mu$m), S2 ($d$ = 2.21 $\mu$m), S3 ($d$ = 2.32 $\mu$m), S4 ($d$ = 2.43 $\mu$m) and S5 ($d$ = 2.55 $\mu$m). The absorptivity (1 – reflectance) spectra of the fabricated DPAs were measured using a Fourier-transform infrared (FTIR) spectrometer equipped with a reflection compartment accessory. Figure 2, from top to bottom, presents the SEM images, and the simulated and measured absorptivity spectra of the DPAs, respectively. Typical SEM images obtained for each sample indicate the well-defined periodic structures. Using this scalable colloidal-mask lithography technique, the domain size of the Al disk lattice is from a few tens to a few hundreds of unit cells, which agrees well with the infinite periodic lattice model used in the simulation. As seen in Figure 2, the measured results agree well with the simulated spectra. All five fabricated DPAs reveal two narrow resonances (the quality factor – $Q \approx$ 13 for M1 resonant modes, $Q \sim$ 9 for M2 resonant modes) with high absorptivities (0.86 – 0.93 for the M1 modes, 0.94 – 0,98 for the M2 modes). The $Q$ factors of M1 resonant modes are higher than those of the M2 modes in all five fabricated DPAs because the M1 resonance results from the grating-mediated coupling between the SPP and the third-order magnetic resonance. The resonant peak positions of M1 modes and M2 modes of the DPAs are redshifted with increasing the diameter of the Al disk. The wavenumbers of the M1 resonance decreases from 3470 cm$^{-1}$ (S1) to 3360 cm$^{-1}$ (S2), 3280 cm$^{-1}$ (S3), 3220 cm$^{-1}$ (S4) and to 3044 cm$^{-1}$ (S5) with high absorptivities ranging from 0.86 (S1) to 0.93 (S5). Similarly, the wavenumbers of M2 resonance also decreases with increasing disk diameter, from 1484 cm$^{-1}$ (S1) to 1410 cm$^{-1}$ (S2), 1370 cm$^{-1}$ (S3), 1326 cm$^{-1}$ (S4) and to 1280 cm$^{-1}$ (S5) with excellent absorptivities from 0.94 (S1) to 0.98 (S5). Interestingly, with a symmetrical geometry, both resonant peaks of the DPAs are independent of the polarization at normal incidence. Indeed, Figures 3a-b display the simulated and measured polarization-independent absorptivity maps for S5. The measured polarization-independent absorptivity maps of S3 and S4 are also provided in the Supporting Information,

Figure S3. The peak positions and intensities of M1 and M2 stay almost unchanged when the electric field polarization is varied in the range of 0 – 90°, which means that the DPA is highly efficient for practical applications such as thermal management as well as IR spectroscopic devices utilizing unpolarized light sources (random polarization). It is worth noting that the absorptivities of the fabricated DPAs shown in Figure 2 were obtained with an unpolarized IR source and without the use of any polarizer.

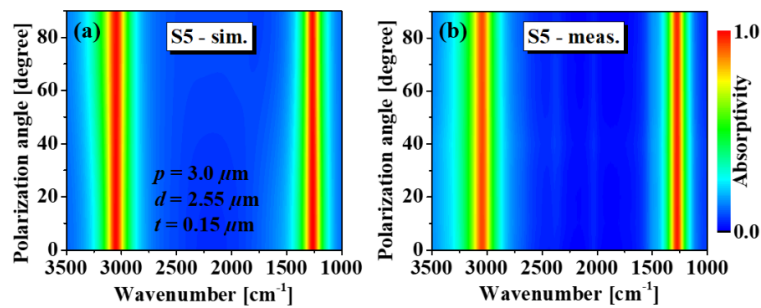

**Figure 3.** Simulated (a) and measured (b) polarization-independent absorptivity of DPA-S5 with the geometrical parameters of $p = 3.0$ $\mu$m, $d = 2.55$ $\mu$m and $t = 0.15$ $\mu$m.

**Dual-band SEIRA study on DPA**

As stated above, the DPAs' resonances cover a wide range of molecular vibrations. Here we exemplify its dual-band SEIRA spectroscopic capability by measuring 10-nm thin polydimethylsiloxane (PDMS) films, simultaneously targeting the $CH_3$ asymmetric C-H stretching at 2962 cm$^{-1}$ and the $CH_3$ deformation in Si-$CH_3$ at 1263 cm$^{-1}$. Prior to the SEIRA study, a spectroscopic ellipsometry measurement on the PDMS film was also performed to verify the optical properties of the PDMS thin film, particularly the vibrations of the PDMS molecules; the details of this measurement are described in the Methods section. The full-range spectroscopic ellipsometry measurement and the fitting model results as well as the retrieved permittivity of the PDMS film, from the UV (50000 cm$^{-1}$) to the MIR (400 cm$^{-1}$) are plotted in Figure S4 (Supporting Information). The measured permittivity data for PDMS is provided in Table S1 (Supporting Information). As seen in Figure S4 (Supporting Information), while the PDMS film is almost transparent in the VIS – MIR region, many absorption features appear in the MIR region due to the vibrations of the PDMS film, including the asymmetric C-H stretching in $CH_3$ featuring at 2962 $cm^{-1}$ and the $CH_3$ deformation in Si-$CH_3$ featured at 1263 $cm^{-1}$. To the best of our knowledge, this is the first systematic spectroscopic ellipsometry measurement for PDMS. The experiment was carried out over a wide spectral range, from the UV to the MIR region (50000 – 400 $cm^{-1}$), providing a good quantitative reference for photonic and materials research. We also used ellipsometry to control the thickness of the PDMS film, which is linearly proportional to the concentration (volume ratio) of the PDMS dissolved in *n*-heptane solvent.

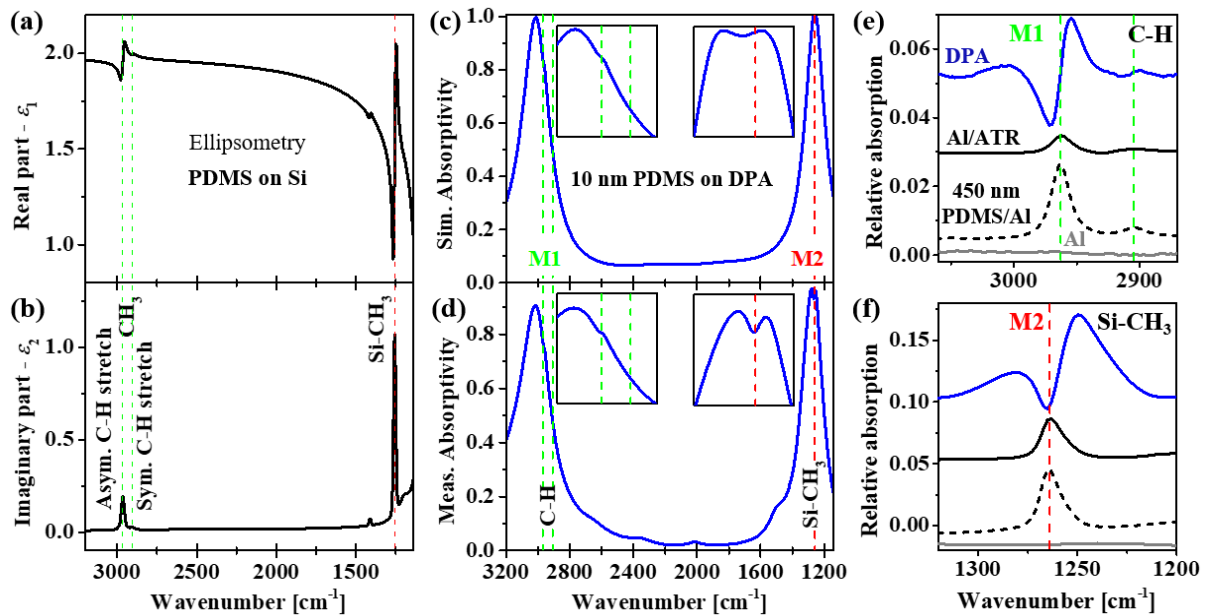

**Figure 4.** (a) Real part and (b) imaginary part of the retrieved complex permittivity of PDMS in the MIR region focusing on two vibrations at 2962 $cm^{-1}$ (asymmetric C-H stretching in $CH_3$) and 1263 $cm^{-1}$ ($CH_3$ deformation in Si-$CH_3$). The full complex permittivity of PDMS is provided in Figure S4 (Supporting Information). (c) Simulated and (d) measured SEIRA spectra of 10 nm PDMS film coated on DPA-S5 reveal simultaneous detection of the two vibrations of PDMS (2962 $cm^{-1}$ and 1263 $cm^{-1}$). Relative absorption SEIRA spectra of the 10 nm PDMS film coated on S5 (blue) and coated on 100 nm

Al film measured by conventional reflection FTIR at normal incidence (grey) and at ATR (black) geometries: (e) at 2962 cm$^{-1}$ (asymmetric C-H stretching in CH$_3$) and (f) at 1263 cm$^{-1}$ (Si-CH$_3$). A shallow peak at 2904 cm$^{-1}$ in (a) – (d) indicates the symmetric C-H stretching in CH$_3$. Dashed black curves in (e) and (f) present relative absorption spectra of a 450 nm PDMS layer coated on 100 nm Al film measured in reflectance mode at normal incidence. Note that the signal from 10 nm PDMS layer coated on DPA-S5 is comparable to the signal from 450 nm PDMS on a planar Al film.

The measured complex permittivity of PDMS focusing on the two different vibrations at 2962 cm$^{-1}$ (CH$_3$) and at 1263 cm$^{-1}$ (Si-CH$_3$) is plotted in Figures 4a-b for our SEIRA study. Figures 4c-d show simulated and measured absorptivity spectra of a 10 nm PDMS film coated on DPA-S5. The two vibrational bands of PDMS at 2962 cm$^{-1}$ and 1263 cm$^{-1}$ are perfectly matched to the M1 and M2 resonant modes, respectively of S5. It is apparent that the two vibrations of PDMS are enhanced and coupled to the two resonant bands of S5, resulting in Fano-like absorption spectra. Since the presence of PDMS caused a redshift of the resonance of the DPA, a polynomial fitting procedure for each resonant band that is truncated at the vibration of PDMS was used as the reference for the subtraction to extract the enhanced spectra (relative absorption) of the PDMS vibrations. The relative absorption spectra at 2962 cm$^{-1}$ (CH$_3$) and 1263 cm$^{-1}$ (Si-CH$_3$) of the PDMS vibrations are plotted in Figures 4e-f, respectively. For comparison, two FTIR spectra of a 10-nm PDMS film coated on a 100-nm sputtered Al film collected using the same reflectance setup at normal incidence as well as separately via a sensitive attenuated total reflectance (ATR) geometry are also displayed. As shown in Figures 4e-f, we do not observe any signals of CH$_3$ and Si-CH$_3$ vibrations from this 10-nm-PDMS-coated Al film via adopting the reflectance geometry at normal incidence, but they are clearly observed by adopting ATR geometry, with the relative absorption intensities of 0.005 at the 2962 cm$^{-1}$ and 0.025 at the 1263 cm$^{-1}$. In DPA-S5, these two vibrational signals are remarkably enhanced and simultaneously observed with the relative intensities of 0.03 (at 2962 cm$^{-1}$) and

0.08 (at 1263 cm$^{-1}$). These signal intensities are 3 – 5 times higher than those of the 10-nm-PDMS-coated Al film obtained using the ATR geometry, indicating that DPA could be a sensitive antenna platform for multi-band SEIRA spectroscopy for molecular fingerprint sensing. For a quantitative estimation of the SEIRA enhancement, the FTIR spectrum of a thick 450-nm PDMS layer coated on a 100-nm Al film substrate was also obtained using the same reflectance setup; the absorption intensities of both the C-H stretching and Si-CH$_3$ vibrations are almost comparable to those of the 10-nm PDMS layer coated on DPA-S5. From the Beer–Lambert law and considering the molecules those who contribute to the SEIRA signal (10-nm PDMS coated on DPA-S5) and bulk PDMS vibration (450-nm PDMS film coated on Al film) intensities, the SEIRA enhancement factor of DPA can be expressed as: $EF = \frac{I_{SEIRA}}{I_{bulk}} \frac{N_{bulk}}{N_{SEIRA}}$ (see Supporting Information for details). Here $I_{SEIRA}$ and $I_{bulk}$ are intensities of the molecular vibration mode from a 10-nm PDMS layer coated on DPA-S5 (SEIRA) and from a 450-nm PDMS film coated on plain Al film, respectively. $N_{bulk}$ and $N_{SEIRA}$ are the number of PDMS molecules contributed to vibrational signals of a 450 nm PDMS film coated on Al film and the SEIRA signal on DPA-S5, respectively. Herein the enhancement factors were found to be as high as 571 at M1 and 642 at M2. These values mean that the DPA will be able to overcome the detection limit of conventional reflectance and ATR measurements together with its excellent mode selectivity.

**Real-time monitoring of PDMS formation**

In addition to identifying the chemical bonds within a molecule by means of its vibrational absorption spectrum, *in situ* IR spectroscopy can be used to monitor real-time reaction process. In this context, we studied the critical behavior of crosslinking polymers[41,42] by extending our *"bond-selective"* SEIRA technique using DPA antennas for the real-time monitoring of the reaction kinetics and gelation process from a 10-nm PDMS elastomer. A 10-nm film of a precursor mixture of the PDMS elastomer film was prepared on DPA-S5 at room temperature. Subsequently, the as-prepared PDMS elastomer film was continuously cured during which time the curing process was monitored by observing changes in the C-H stretching and $CH_3$ deformation signals enhanced by DPA-S5. In this *in situ* SEIRA spectroscopy, the initial reflectance ($R_0$) spectrum of the sample was used as the reference spectrum, and the change in the SEIRA reflectance ($R_t/R_0$) spectra was measured at 25-min intervals over a total time period of 15 h. Figure 5a presents the IR spectral evolution ($-\log_{10}(R_t/R_0)$) as a function of curing time for the sample. As seen in Figures 5a-b, the signal intensities of both the C-H stretching and the $CH_3$ deformation modes remarkably increase as the curing time increases, and then they begin to saturate after approximately 10 h, which indicates that the PDMS elastomer is completely reacted in 10 h. It is worth noting that we did not observe any signal from a 10-nm thick PDMS film coated on a 100-nm Al flat film using the same reflectance geometry.

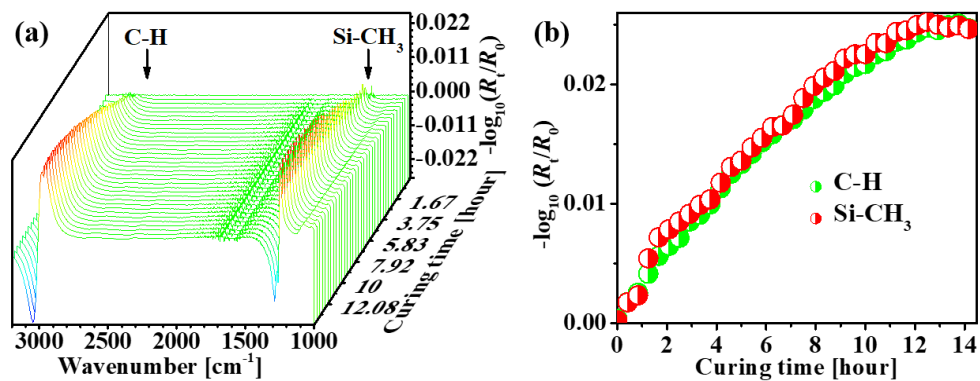

**Figure 5.** (a) Bond-selective *in situ* SEIRA spectroscopy recorded during the curing time of 10-nm PDMS film coated on device S5. (b) The evolution of the SEIRA absorption intensity of both C-H and Si-$CH_3$ vibrations as a function of the curing time.

Here we can consider that the IR absorbance change of the C-H stretching and the CH$_3$ deformation of the PDMS elastomer will be proportional to the percentage of the reacted bonds in PDMS, i.e., the local conformation change associated with the reduction of the crosslinker's S-H bonds and C-Si bond formation during the hydrosilylation reaction (Figure S5a).[41] This local conformational change accompanied with the hydrosilylation reaction inevitably induces local redistribution of the bond charge leading to the alteration in the dynamic dipoles of the C-H stretching and the CH$_3$ deformation modes. The conversions – $\alpha$ can be therefore predicted by the following equation (details are shown in the Supporting Information and Figure S5):

$\alpha_t = \dfrac{\log_{10}(R_t/R_0)}{\log_{10}(R_S/R_0)}$, where $R_S/R_0$ is the saturated SEIRA reflectance. Figure S5a shows the conformational changes of the molecules and Figure S5b plots the degrees of the conversions at both C-H and Si-CH$_3$ vibrations as functions of the reaction times. The measured kinetics of the reaction can be expressed by: $\dfrac{d\alpha}{dt} = K(1-t)^n$, where $t$ is the reaction time, $K$ is the rate constant, $A$ is the preexponential factor.[43] By fitting the degrees of conversions using the above reaction rate equation, we obtained the rate constant – $K$ of about 0.13 with the order – $n$ of the kinetic of about 0.45. It should be noted that both the C-H and the Si-CH$_3$ vibrations follow exactly the same kinetics indicating that the alteration in these two bonds are of the same chemical origin: in this case the cross-linking reaction and the network formation of PDMS elastomer.

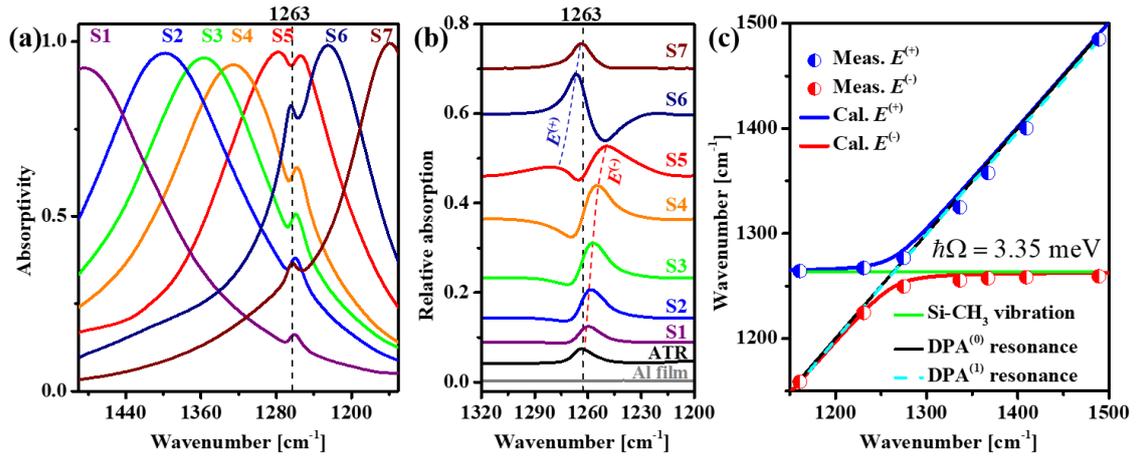

**Figure 6.** (a) SEIRA spectra at around M2 mode of the 10 nm PDMS film coated on a set of DPAs (S1 – S5 with 3.0 μm periodicity and S6 – S7 with 4.4 μm periodicity) with resonances near the Si-CH$_3$ vibration of the PDMS molecule. (b) Relative SEIRA spectra of the Si-CH$_3$ vibration taken from the 10 nm PDMS films coated on DPAs, and on Al film measured by conventional reflectance and ATR geometries. (c) Hybridized modes (half-filled circles) resulted from the coupling between the Si-CH$_3$ vibration and the DPAs resonances plotted against the bare DPAs resonances (black line with slope = 1) and Si-CH$_3$ vibration at 1263 cm$^{-1}$ (green line). The dashed cyan line denotes the DPAs resonances with the presence of the PDMS film. Blue and red lines plot the calculated hybridized modes.

**Coupling between molecular vibrations and DPA's resonances**

Furthermore, it is worth noticing that the coupling between the PDMS vibration and the DPA resonance results not only in the Fano-like resonance but also in a spectral shift of both the DPA resonance (besides a redshift caused by the higher refractive index of PDMS) and the PDMS vibration. The relative spectral-shift (redshift or blueshift) depends on the coupling strength of the PDMS vibration and the DPA resonance with respect to the overlapping between the molecular vibration and the absorber resonance (redshift for the lower energy side, blueshift for the higher energy side). It should be noted that our DPA antenna platform is designed to function at normal incidence as the SPP-coupled (third-order) magnetic resonance M1 works within a small incident angle range (0° – 15°), even though the fundamental magnetic mode M2 is slightly shifted to higher wavenumbers when the angle varies between 0° to 85°. The numerical simulation is shown in Figure S2a (Supporting Information), and the angle-resolved absorptivity measurement is shown in Figures S6a, b (Supporting Information). Here we also performed the angle-resolved SEIRA study of a 10-nm PDMS film coated on the DPA-S5. Figure S6c in the Supporting Information presents the measured angle-resolved SEIRA spectra at the magnetic resonance – M2. As the indent angle increases, both the magnetic resonance M2 and the coupled Si-CH$_3$ vibrations are blue-shifted. To demonstrate the coupling strength between the PDMS vibrations and the DPA resonances, we performed a SEIRA study focusing on the PDMS vibrations (2962 cm$^{-1}$ and 1263 cm$^{-1}$) using the five DPAs (S1 – S5) as well as two additional DPAs with the same periodicity and insulator thickness ($p$ = 4.4 $\mu$m, $t$ = 0.2 $\mu$m): S6 ($d$ = 2.97 $\mu$m) and S7 ($d$ = 3.28 $\mu$m). The details of S6 and S7 are described in the Supporting Information (see Figure S7). The SEIRA spectra of the 10-nm PDMS-film-coated DPAs are plotted in Figure 6a and the relative SEIRA intensities of the Si-CH$_3$ vibration are presented in Figure 6b. It is clearly seen that the coupling strength between the Si-CH$_3$ vibration and the DPA resonance strongly depends on the overlap between the molecular vibration and the plasmon resonance. When the DPA resonance is close to the Si-CH$_3$ vibration, the molecular vibration intensity increases significantly. Compared to the original Si-CH$_3$ vibrational peak position (black line in Figure 6b), the Si-CH$_3$ vibration is redshifted for samples S1 – S5 while it is blue-shifted for samples S6 and S7. Similarly, we also observed the spectral shift of the DPA resonances but in an opposite direction compared to the spectral shift of the molecular vibration (data not shown). In particular, we observed behavior similar to Rabi splitting the resonant peak of DPA-S5, in which the resonance is perfectly matched to the Si-CH$_3$ vibration (see red curve in Figure 6a). SEIRA spectra of the asymmetric C-H stretching vibration were also examined for all the DPAs. The results are shown in Figure S8a in the Supporting Information. The relative absorption intensity of the asymmetric C-H stretching vibration gradually increases from S1 to S5 and from S7 to S6 as the spectral matching between the M1 resonance of DPA and the C-H stretching vibration improves.

The hybridizations of the Si-CH$_3$ vibrations and the DPAs are plotted in Figure 6c (half-filled circles) alongside the bare DPAs resonances (black line, denoted DPA$^{(0)}$) and the original Si-CH$_3$ vibration (at 1263 cm$^{-1}$, green line). As discussed above, the higher refractive index of the PDMS leads to a redshift of the DPA resonance (optical shift) when the DPA is covered by a 10-nm film (denoted DPA$^{(1)}$), and the peak shift of the observed DPA resonance results from a combination of the optical shift and a shift due to coupling the molecular vibration–DPA plasmon coupling. To quantify the optical shift energy (denoted $\Delta\omega$) for each of the DPAs covered by a 10-nm PDMS film, we performed simulations using lossless PDMS permittivity (the imaginary permittivity of PDMS was fixed at 0), for all seven DPAs with a conformal coating consisting of a 10-nm PDMS film. Thus, the energy of each DPA resonance after redshifting due to a 10-nm PDMS film ($E_{DPA^{(1)}} = \hbar\omega_{DPA^{(1)}}$) can be calculated via the measured energy of DPA$^{(0)}$ ($E_{DPA^{(0)}} = \hbar\omega_{DPA^{(0)}}$) and optical shifted energy ($\Delta\omega$) retrieved by simulation: $E_{DPA^{(1)}} = E_{DPA^{(0)}} - \Delta\omega$. The dashed cyan line shown in Figure 6c plots the DPAs' resonances with optical shifts (DPA$^{(1)}$). To elucidate the coupling strength between the DPAs and the Si-CH$_3$ vibration of PDMS, we used a Hamiltonian that follows the Jaynes–Cummings model for the light-matter coupling system:

$\hat{H} = \hat{H}_{DPA^{(1)}} + \hat{H}_{Si-CH_3} + \hat{H}_{int} = \hbar\omega_{DPA^{(1)}}\hat{a}^\dagger\hat{a} + \hbar\omega_{Si-CH_3}\hat{\sigma}^\dagger\hat{\sigma} + \hbar g\left(\hat{a}\hat{\sigma}^\dagger + \hat{a}^\dagger\hat{\sigma}\right)$, where $\hat{a}$ ($\hat{a}^\dagger$) and $\hat{\sigma}$ ($\hat{\sigma}^\dagger$) are the annihilation (creation) operators of the field (plasmon) and the optically active molecular vibration.[44–46] The plasmon-vibration interaction $g$ (coupling constant) is given as

$g = \sqrt{\dfrac{\mu^2 \omega_{Si-CH_3}}{2\hbar V \varepsilon_0 \varepsilon}}$, where $\mu$ is the molecular dipole moment, $V$ is the effective mode volume of the plasmon, $\varepsilon_0$ is the vacuum permittivity and $\varepsilon$ is the effective dielectric constant of PDMS near the DPA surface. It is worth noting that, compared with the Fabry-Perot microcavity, plasmonic absorbers can exhibit a high coupling constant due to the extreme enhancement of the near-field in the vicinity of the resonators, defining a small mode volume; this approach therefore provides a powerful platform for studying light-matter coupling.

Since the vibration-plasmon polariton-coupled system has only one vibration (Si-CH$_3$), the expression of the Hamiltonian can be simplified as: $\hat{H} = \begin{pmatrix} E_{DPA^{(1)}} & \Delta E \\ \Delta E & E_{Si-CH_3} \end{pmatrix}$, where $\Delta E = \hbar g$ is the coupling strength. Through the diagonalization of this matrix, the Hamiltonian yields energy eigenvalues of the two hybridized states: $E^{(\pm)} = \frac{1}{2}\left(E_{DPA^{(1)}} + E_{Si-CH_3}\right) \pm \frac{1}{2}\sqrt{\left(E_{DPA^{(1)}} - E_{Si-CH_3}\right)^2 + 4\Delta E^2}$ (plotted as blue and red curves in Figure 5c). The coupling strength can therefore be calculated as $\Delta E = \frac{1}{2}\sqrt{\left(E^{(+)} - E^{(-)}\right)^2 - \left(E_{DPA^{(1)}} - E_{Si-CH_3}\right)^2}$. From the measured hybridized states energy ($E^{(\pm)}$), the optically shifted DPA resonance energy ($E_{DPA^{(1)}}$), and the Si-CH$_3$ vibration energy ($E_{Si-CH_3}$), the coupling strength ($\Delta E$) was found to be 1.78 meV (14.32 cm$^{-1}$). The splitting energy ($\hbar\Omega = 2\Delta E = 2\hbar g$) was also estimated to be as high as 3.55 meV. Furthermore, the hybridization between the C-H stretching vibration and the M1 resonance was also investigated for all DPAs (Figure S8b in the Supporting Information). The coupling strength between the C-H stretching and M1 was found to be 4.57 meV (36.85 cm$^{-1}$). The relative coupling strength energy defined as the coupling strength and vibrational energy ratio ($\Delta E/E$) was 0.012, which is almost comparable to that of the coupling between the Si-CH$_3$ vibration and the M2 resonance. Thus, this simple aluminum DPA structure could be another interesting plasmonic antenna platform, with high field enhancement, for the study of light-matter interactions such as the vibration-plasmon polariton–coupled system demonstrated in the present study. Although, the splitting energy in this vibration-plasmon polariton in the MIR region is much smaller than that of exciton-plasmon polaritons in UV-VIS region,[47–49] it is comparable to the linewidth of the Si-CH$_3$ vibration (1.67 meV) as well as those values for previously reported perfect absorbers made of gold loaded with molecules.[29] However, the splitting energy value is still far below the bandwidth of DPA resonance (i.e. 30 meV), indicating that the coupling between DPA resonances and PDMS vibrations is in the weak coupling regime.

## Conclusion

We have demonstrated simple, single-sized aluminum disk arrays fabricated by colloidal-mask lithography for dual-band, polarization-independent plasmonic perfect absorbers for high-sensitivity molecular sensing and *in situ* reaction monitoring also capable of selective analysis of their reaction kinetics. The fabricated large-area DPAs exhibited two narrow resonant bands with nearly perfect absorptivities and high tunability in the IR region spanning the most important vibrations of molecular spectroscopy. First, we have demonstrated dual-band SEIRA spectroscopy for selectively detecting two vibrational bands (asymmetric C-H stretching in CH$_3$ at 2962 cm$^{-1}$ and CH$_3$ deformation in Si-CH$_3$ at 1263 cm$^{-1}$) of a 10-nm PDMS film. Second, the DPA antenna platform revealed its outstanding applicability via *real-time* SEIRA observation of the curing and gelation kinetics of the ultrathin PDMS film. Finally, the effects of hybridization between the DPAs resonances and the PDMS vibrations were also studied quantitatively as evidenced by the emergence of avoided-crossings for the polaritonic bands. Overall, the proposed plasmonic DPAs fabricated by the scalable colloidal-mask lithography provide a good plasmonic antenna platform not only for IR molecular/reaction sensors, but can also be effectively adopted for gas sensors and thermal imaging, as well as many other light-matter polaritonic applications from the VIS to the THz regions.

## Methods section

**Fabrication of DPAs.** The DPAs were fabricated using colloidal-mask lithography combined with a two-step RIE process. For each DPA, a tri-layered Al/Al$_2$O$_3$/Al film was deposited on a 1×1 cm$^2$ Si (100) substrate using three sputtering steps during which the sample holder was rotated at 18 rpm, and room temperature, with 20-sccm Ar gas, and 100-W DC power for Al and 300-W RF power for Al$_2$O$_3$ (i-Miller CFS-4EP-LL, Shibaura). To prepare the Al disk array on the Al$_2$O$_3$/Al film, a monolayer of polystyrene (PS) spheres (Polybead Polystyrene Microspheres – Polysciences; the standard deviations: 0.07 $\mu$m for 3.0 $\mu$m PS and 0.015 $\mu$m for 4.4 $\mu$m PS) was deposited on top of the tri-layer films as an RIE mask for fabricating Al disk using the following process. First, the Al/Al$_2$O$_3$/Al substrates were aligned in a 10°-tilted plastic container filled up with deionized water. Subsequently, a monolayer of PS was prepared on the air-water interface using a slow flow of PS solution from a 30°-tilted glass slide. A monolayer of PS spheres was formed on each tri-layered Al/Al$_2$O$_3$/Al film by using a micro-pump to drain the water in the plastic container. The as-made monolayer PS masks were then dried naturally at room temperature. To reduce the size of the PS spheres, oxygen plasma etching with an etching rate of 2 nm/s (20-sccm O$_2$ gas, 1-Pa APC pressure, 200-W RF antenna power, 5-W RF bias power, ULVAC CE-300I) was applied. Here, a short 90-s burst of RIE was repeated on each sample to prevent overheating by the ion bombardment. The etching time on each sample was precisely controlled to achieve different PS sizes corresponding to the designed Al-disk sizes. Then, a second RIE step was carried out to etch Al using the PS masks (a mixture of 3-sccm BCl$_3$ and 3-sccm Cl$_2$ gases, 0.15-Pa APC pressure, 50 W RF antenna power, 10-W RF bias power, and an etching rate of 0.82 nm/s). DPAs were finally achieved by removing the PS mask by washing in ethanol under ultrasonication followed by rinsing in toluene and ethanol. We note that the size of the Al disk was principally determined by the etching time of the PS spheres.

**Characterizations.** SEM images of the fabricated DPAs were obtained using a Hitachi SU8200 SEM system under an accelerating voltage of 5 kV. The reflectance spectra were obtained using an FTIR spectrometer (Nicolet iS50R FT-IR Thermo Scientific) equipped with a variable-angle reflectance compartment accessory and a liquid-nitrogen-cooled mercury cadmium telluride (MCT) detector combined with a KBr beam splitter. For ATR measurement of the 10-nm PDMS coated Al film, an ATR accessory (iS50 ATR) equipped with a deuterated triglycine sulfate detector (DTGS) detector and KBr beam splitter was used. A 200-nm Au film was used as the reference for all the measurements. The absorptivity spectra of the fabricated devices were simply calculated as 1 – reflectance (the Al film at the base of the samples blocked any transmitted light in the MIR region, so transmittance = 0). The spectroscopic ellipsometry study of the PDMS film was carried out using two ellipsometers (SENTECH, SE 850 DUV for UV – NIR region and SENDIRA for MIR region). Further details of the measurements and analysis are provided in the Supporting Information.

**Simulations.** The optical spectra (transmittance, reflectance, and absorptivity) of the DPAs were simulated using the rigorous coupled-wave analysis (RCWA) (DiffractMOD, Synopsys' RSoft). For the electric and magnetic field distributions, full-wave simulations based on the finite-difference time-domain (FDTD) method (FullWAVE, Synopsys' RSoft) were performed. The dielectric functions of Al, $Al_2O_3$ and Si were obtained in the literature,[50] and the dielectric function of PDMS was measured via spectroscopic ellipsometry. The modelling of the DPA was performed in a CAD layout (RSoft CAD) with the unit cell as shown in Figure 1a, and with the mesh size of 20 nm, and with 2 nm as the PDMS layer thickness for the SEIRA study. For the FDTD simulations, periodic boundary conditions were applied to the *X* and *Y* directions and perfectly matched layers were applied to both sides in the *Z* direction. For both RCWA and FDTD simulations, the excitation electromagnetic field propagated along the -*Z* axis and the electric field oscillated along in the *X* direction.

## Supporting Information.

Electronic supplementary information (ESI) available: Additional optical properties of the DPAs (further numerical optimization of structural parameters of DPAs; derivation of SPP in the plasmonic hexagonal lattice and its coupling to the third-order magnetic resonance; measured polarization independence of the absorptivity spectra of samples S3 and S4); spectroscopic ellipsometry study of the PDMS film; detailed calculations of the SEIRA enhancement factors; further analysis of curing dynamics of PDMS by SEIRA; angle-resolved SEIRA study; detailed optical properties of samples S6 and S7; further SEIRA analysis at the M1 resonance and the retrieved permittivity table for PDMS.

## Conflict of interest

The authors declare no competing financial interest.

## Acknowledgements

This work is partially supported by JSPS KAKENHI (16F16315, JP16H06364, 16H03820), and CREST "Phase Interface Science for Highly Efficient Energy Utilization" (JPMJCR13C3) from Japan Science and Technology Agency. Thang. D. Dao would like to acknowledge the JSPS fellowship program (P16315).

## References

1   N. I. Landy, S. Sajuyigbe, J. J. Mock, D. R. Smith and W. J. Padilla, *Phys. Rev. Lett.*, 2008, **100**, 207402.
2   M.-W. Tsai, T.-H. Chuang, C.-Y. Meng, Y.-T. Chang and S.-C. Lee, *Appl. Phys. Lett.*, 2006, **89**, 173116.
3   H. T. Miyazaki, K. Ikeda, T. Kasaya, K. Yamamoto, Y. Inoue, K. Fujimura, T. Kanakugi, M. Okada, K. Hatade and S. Kitagawa, *Appl. Phys. Lett.*, 2008, **92**, 141114.
4   Y.-H. Ye, Y.-W. Jiang, M.-W. Tsai, Y.-T. Chang, C.-Y. Chen, D.-C. Tzuang, Y.-T. Wu and S.-C. Lee, *Appl. Phys. Lett.*, 2008, **93**, 033113.
5   X. Liu, T. Tyler, T. Starr, A. F. Starr, N. M. Jokerst and W. J. Padilla, *Phys. Rev. Lett.*, 2011, **107**, 045901.
6   T. D. Dao, K. Chen, S. Ishii, A. Ohi, T. Nabatame, M. Kitajima and T. Nagao, *ACS Photonics*, 2015, **2**, 964–970.
7   T. Yokoyama, T. D. Dao, K. Chen, S. Ishii, R. P. Sugavaneshwar, M. Kitajima and T. Nagao, *Adv. Opt. Mater.*, 2016, **4**, 1987–1992.
8   F. B. P. Niesler, J. K. Gansel, S. Fischbach and M. Wegener, *Appl. Phys. Lett.*, 2012, **100**, 203508.
9   S. Ogawa, K. Okada, N. Fukushima and M. Kimata, *Appl. Phys. Lett.*, 2012, **100**, 021111.
10  F. Zhao, C. Zhang, H. Chang and X. Hu, *Plasmonics*, 2014, **9**, 1397–1400.
11  S. Ogawa, D. Fujisawa, H. Hata, M. Uetsuki, K. Misaki and M. Kimata, *Appl. Phys. Lett.*, 2015, **106**, 041105.

12  T. D. Dao, S. Ishii, T. Yokoyama, T. Sawada, R. P. Sugavaneshwar, K. Chen, Y. Wada, T. Nabatame and T. Nagao, *ACS Photonics*, 2016, **3**, 1271–1278.

13  J. Y. Suen, K. Fan, J. Montoya, C. Bingham, V. Stenger, S. Sriram and W. J. Padilla, *Optica*, 2017, **4**, 276.

14  H. T. Miyazaki, T. Kasaya, M. Iwanaga, B. Choi, Y. Sugimoto and K. Sakoda, *Appl. Phys. Lett.*, 2014, **105**, 121107.

15  A. Lochbaum, Y. Fedoryshyn, A. Dorodnyy, U. Koch, C. Hafner and J. Leuthold, *ACS Photonics*, 2017, **4**, 1371–1380.

16  E. Rephaeli and S. Fan, *Opt Express*, 2009, **17**, 15145–15159.

17  L. P. Wang and Z. M. Zhang, *Appl. Phys. Lett.*, 2012, **100**, 063902.

18  B. Zhao, L. Wang, Y. Shuai and Z. M. Zhang, *Int. J. Heat Mass Transf.*, 2013, **67**, 637–645.

19  J.-Y. Chang, Y. Yang and L. Wang, *Int. J. Heat Mass Transf.*, 2015, **87**, 237–247.

20  E. Rephaeli, A. Raman and S. Fan, *Nano Lett.*, 2013, **13**, 1457–1461.

21  A. P. Raman, M. A. Anoma, L. Zhu, E. Rephaeli and S. Fan, *Nature*, 2014, **515**, 540–544.

22  L. Zhu, A. Raman, K. X. Wang, M. A. Anoma and S. Fan, *Optica*, 2014, **1**, 32.

23  T. Liu and J. Takahara, *Opt. Express*, 2017, **25**, A612.

24  A. Hartstein, J. R. Kirtley and J. C. Tsang, *Phys. Rev. Lett.*, 1980, **45**, 201–204.

25  D. Enders, T. Nagao, T. Nakayama and M. Aono, *Langmuir*, 2007, **23**, 6119–6125.

26  F. Neubrech, A. Pucci, T. W. Cornelius, S. Karim, A. García-Etxarri and J. Aizpurua, *Phys. Rev. Lett.*, 2008, **101**, 157403.

27  D. Enders, T. Nagao, A. Pucci, T. Nakayama and M. Aono, *Phys. Chem. Chem. Phys.*, 2011, **13**, 4935–4941.

28  Tadaaki Nagao and Gui Han and ChungVu Hoang and Jung-Sub Wi and Annemarie Pucci and Daniel Weber and Frank Neubrech and Vyacheslav M Silkin and Dominik Enders and Osamu Saito and Masud Rana, *Sci. Technol. Adv. Mater.*, 2010, **11**, 054506.

29  K. Chen, R. Adato and H. Altug, *ACS Nano*, 2012, **6**, 7998–8006.

30  C. V. Hoang, M. Oyama, O. Saito, M. Aono and T. Nagao, *Sci. Rep.*, 2013, **3**, 1175.

31  K. Chen, T. D. Dao, S. Ishii, M. Aono and T. Nagao, *Adv. Funct. Mater.*, 2015, **25**, 6637–6643.

32  A. E. Cetin, S. Korkmaz, H. Durmaz, E. Aslan, S. Kaya, R. Paiella and M. Turkmen, *Adv. Opt. Mater.*, 2016, **4**, 1274–1280.

33  Z. H. Jiang, S. Yun, F. Toor, D. H. Werner and T. S. Mayer, *ACS Nano*, 2011, **5**, 4641–4647.

34  H. Cheng, S. Chen, H. Yang, J. Li, X. An, C. Gu and J. Tian, *J. Opt.*, 2012, **14**, 085102.

35  B. Zhang, Y. Zhao, Q. Hao, B. Kiraly, I.-C. Khoo, S. Chen and T. J. Huang, *Opt. Express*, 2011, **19**, 15221–15228.

36  R. Feng, W. Ding, L. Liu, L. Chen, J. Qiu and G. Chen, *Opt. Express*, 2014, **22**, A335.

37  N. Zhang, P. Zhou, D. Cheng, X. Weng, J. Xie and L. Deng, *Opt. Lett.*, 2013, **38**, 1125–1127.

38  M. Pu, C. Hu, M. Wang, C. Huang, Z. Zhao, C. Wang, Q. Feng and X. Luo, *Opt Express*, 2011, **19**, 17413–17420.

39  Y. J. Yoo, Y. J. Kim, P. Van Tuong, J. Y. Rhee, K. W. Kim, W. H. Jang, Y. H. Kim, H. Cheong and Y. Lee, *Opt. Express*, 2013, **21**, 32484.

40  K. Ito, H. Toshiyoshi and H. Iizuka, *J. Appl. Phys.*, 2016, **119**, 063101.

41  S. K. Venkataraman, L. Coyne, F. Chambon, M. Gottlieb and H. H. Winter, *Polymer*, 1989, **30**, 2222–2226.
42  J. E. Dietz, B. J. Elliott and N. A. Peppas, *Macromolecules*, 1995, **28**, 5163–5166.
43  A. K. Burnham and L. N. Dinh, *J. Therm. Anal. Calorim.*, 2007, **89**, 479–490.
44  E. T. Jeynes and F. W. Cummings, *Proc. IEEE*, 1963, **51**, 89–109.
45  J. del Pino, J. Feist and F. J. Garcia-Vidal, *New J. Phys.*, 2015, **17**, 053040.
46  W. Ahn, I. Vurgaftman, A. D. Dunkelberger, J. C. Owrutsky and B. S. Simpkins, *ACS Photonics*, 2018, **5**, 158–166.
47  G. Khitrova, H. M. Gibbs, M. Kira, S. W. Koch and A. Scherer, *Nat. Phys.*, 2006, **2**, 81.
48  S. Savasta, R. Saija, A. Ridolfo, O. Di Stefano, P. Denti and F. Borghese, *ACS Nano*, 2010, **4**, 6369–6376.
49  D. G. Baranov, M. Wersäll, J. Cuadra, T. J. Antosiewicz and T. Shegai, *ACS Photonics*, 2018, **5**, 24–42.
50  E. D. Palik, *Handbook of Optical Constants of Solids*, Academic Press: New York, 3rd edn., 1998.

Supporting Information

# Dual-band in situ molecular spectroscopy using single-sized Al-disk perfect absorbers

Thang Duy Dao*[1], Kai Chen[1,2] and Tadaaki Nagao*[1,3]

[1]*International Center for Materials Nanoarchitectonics, National Institute for Materials Science (NIMS), 1-1 Namiki, Tsukuba, Ibaraki 305-0044, Japan.*

[2]*Institute of Photonics Technology, Jinan University, Guangzhou, 510632, China*

[3]*Department of Condensed Matter Physics, Graduate School of Science, Hokkaido University, Kita 10, Nishi 8, Kita-ku, Sapporo 060-0810, Japan*

*Corresponding Authors:*

*Thang Duy Dao: Dao.duythang@nims.go.jp; katsiusa@gmail.com*

*Tadaaki Nagao: Nagao.Tadaaki@nims.go.jp*

**Optimization of the Al disk and insulator thicknesses.** Figure S1a presents the simulated absorptivity spectra of DPAs having the same $p = 3.0$ μm, $d = 2.55$ μm, $t = 0.15$ μm but different Al disk thicknesses. Both resonances of the DPAs slightly red-shift when the Al disk thickness increases. Based on this simulation, we chose Al disk thickness at 0.1 μm for all DPAs. Figure S1b reveals the simulated dependence of DPA's absorptivity on the insulator thickness with $p = 3.0$ μm, $d = 2.55$ μm, and the Al disk thickness of 0.1 μm. Based on several simulations on different disk diameters, the optimal insulator thickness for 3.0 μm periodicity DPAs was found to be 0.15 μm.

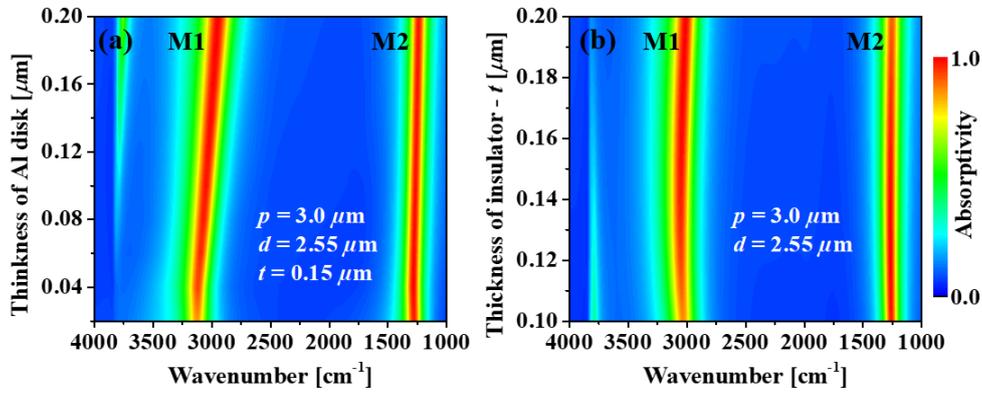

**Figure S1.** (a) Simulated dependence of DPA's absorptivity on the Al disk thickness. (b) Simulated dependence of the DPA's absorptivity on the insulator thickness. In both simulations, the periodicity and the disk diameter were $p = 3.0$ μm, $d = 2.55$ μm; and $t = 0.15$ μm for (a); Al disk thickness chosen at 0.1 μm for (b).

**Surface plasmon polaritons (SPPs) in the plasmonic hexagonal lattice.** In the hexagonal metal lattice, SPPs at the metal-air interface can couple to the incident photon and photonic lattice when SPPs' momentum – $\left|\vec{k}_{spp}\right| = k_0 \sqrt{\dfrac{\varepsilon_m}{\varepsilon_m + 1}}$ matches the surface-parallel component of the momentum of the incident photon – $\left|\vec{k}_\parallel\right| = k_0 \sin\theta$ and photonic lattice – $\left(i\vec{G}_x + j\vec{G}_y\right)$ (momentum conservation):

$$\left|\vec{k}_{spp}\right| = \left|\vec{k}_{\parallel} + i\vec{G}_x + j\vec{G}_y\right| \tag{1}$$

With an azimuthal angle of 0°, as $\vec{k}_{\parallel}$ is oriented along the $(\vec{G}_x + \vec{G}_y)$ direction (it is noted that the angle between two primitive vector is 60° see Figure 1a in the manuscript), $\vec{k}_{\parallel}$ of the incident photon momentum can be written as:

$$\vec{k}_{\parallel} = \frac{\left|\vec{k}_{\parallel}\right|}{2\cos 30°} \frac{\vec{G}_x}{\left|\vec{G}_x\right|} + \frac{\left|\vec{k}_{\parallel}\right|}{2\cos 30°} \frac{\vec{G}_y}{\left|\vec{G}_y\right|} \tag{2}$$

The momentum conservation is now given by:

$$\left|\vec{k}_{spp}\right| = \left|\left(\frac{k_0 \sin\theta}{2\cos 30°} \frac{\vec{G}_x}{\left|\vec{G}_x\right|} + i\vec{G}_x\right) + \left(\frac{k_0 \sin\theta}{2\cos 30°} \frac{\vec{G}_y}{\left|\vec{G}_y\right|} + j\vec{G}_y\right)\right| \tag{3}$$

$$\left|\vec{k}_{spp}\right|^2 = \left(k_0\sqrt{\frac{\varepsilon_m}{\varepsilon_m+1}}\right)^2 = \left(\frac{k_0\sin\theta}{\sqrt{3}} \frac{\vec{G}_x}{\left|\vec{G}_x\right|} + i\vec{G}_x\right)^2 + \left(\frac{k_0\sin\theta}{\sqrt{3}} \frac{\vec{G}_y}{\left|\vec{G}_y\right|} + j\vec{G}_y\right)^2 +$$
$$+ 2\cos 60° \left(\frac{k_0\sin\theta}{\sqrt{3}} \frac{\vec{G}_x}{\left|\vec{G}_x\right|} + i\vec{G}_x\right)\left(\frac{k_0\sin\theta}{\sqrt{3}} \frac{\vec{G}_y}{\left|\vec{G}_y\right|} + j\vec{G}_y\right) \tag{4}$$

Here $\left|\vec{G}_x\right| = \left|\vec{G}_y\right| = \frac{4\pi}{\sqrt{3}p}$, $k_0 = \frac{2\pi}{\lambda}$, the angular dependent dispersion relation for an plasmonic hexagonal lattice along the $\vec{G}_x + \vec{G}_y$ is finally expressed as a function of incident angle and integers $i, j$:

$$\frac{\varepsilon_m}{\varepsilon_m + 1} = \sin^2\theta + \frac{2}{p}(i+j)\lambda\sin\theta + \frac{4}{3p^2}(i^2 + ij + j^2)\lambda^2 \tag{5}$$

**Coupling between SPP (-1,0) and the third-order magnetic resonance.** The resonance (in wavelength) of DPAs can be treated as a function of the geometrical parameters using the half-wave dipole antenna model $\lambda_m = \frac{2}{m}\sqrt{\varepsilon_{eff}}d + C$, where $\lambda_m$ is the resonance at the *m*-order, $\varepsilon_{eff}$ is the effective dielectric function of the antenna insulator, d is the disk diameter, and *C* is a constant. In the periodic circular antenna array, $\varepsilon_{eff}$ can be treated as a function of periodicity – *p*, insulator thickness – *t*, and the dielectric functions of the insulator layer – $\varepsilon_d$ and the metal film $\varepsilon_m$ ($1 < \varepsilon_{eff} < \varepsilon_d$). Based on the simulated dependence of DPA resonance on the Al disk diameter, the dependence of the third-order magnetic resonance on the disk diameter is found to be $\lambda_3 = 0.9876d + 0.7333$ ($\varepsilon_{eff} = 1.481$) (dashed white line in Figure S2d).

The simple Hamiltonian described the coupling between the third-order magnetic resonance energy ($\hbar\omega_3 = \frac{hc}{\lambda_3}$) and the SPP (-1,0) resonance energy ($\hbar\omega_{spp} = \frac{hc}{\lambda_{spp}}$) can be written as:

$\hbar\begin{pmatrix} \omega_3 & g_C \\ g_C & \omega_{spp} \end{pmatrix}$, where $g_C$ is the coupling constant. The eigenvalues of this Hamiltonian are:

$E_C^{(\pm)} = \frac{1}{2}\hbar(\omega_3 + \omega_{spp}) \pm \frac{1}{2}\hbar\sqrt{(\omega_3 - \omega_{spp})^2 + 4g_C^2}$. With the coupling strength energy – $\hbar g_C$ of 28 meV estimated from the simulation of the diameter-dependent resonance (Figure S2c), we plotted the polaritonic hybridized energies $E_C^{(+)}$ (brown solid curve) and $E_C^{(-)}$ (violet solid curve) shown in Figure S2d. It is clear to see the mode coupling and energy splitting behavior (Rabi energy splitting of 56 meV) of M1 resulted from the coupling between the third-order magnetic resonance and the SPP (-1,0) mode.

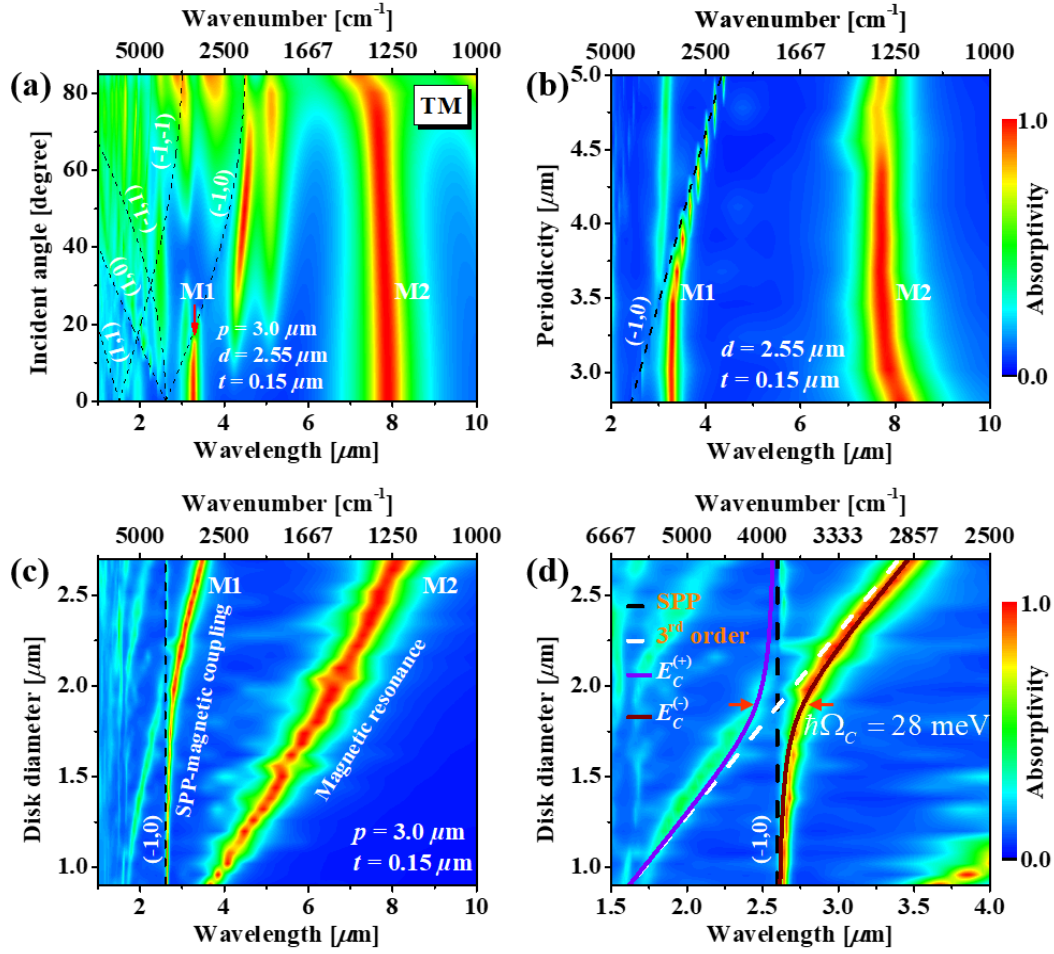

**Figure S2.** (a) Simulated angle-dependent DPA's absorptivity. (b) Simulated dependence of DPA's absorptivity on the periodicity. The same geometrical parameters of DPA were chosen for (a) and (b) with $p = 3.0\ \mu m$, $d = 2.55\ \mu m$ and $t = 0.15\ \mu m$. (c) The simulated dependence of DPA absorptivity on the disk diameter (replotted Figure 1c in the wavelength scale) (with $d = 2.55\ \mu m$ and $t = 0.15\ \mu m$). (d) Replotted (c) from 1.5 – 4.0 $\mu m$ wavelength range to reveal the coupling between the SPP (-1,0) mode and the third-order magnetic resonance. Dashed black curves represent SPPs in the Al hexagonal lattice following equation (5). Dashed white line in (d) represents the linear relation between the third-order magnetic resonance and the Al disk diameter. Solid curves in (d) denotes hybridized modes between the SPP (-1,0) and the third-order magnetic resonance.

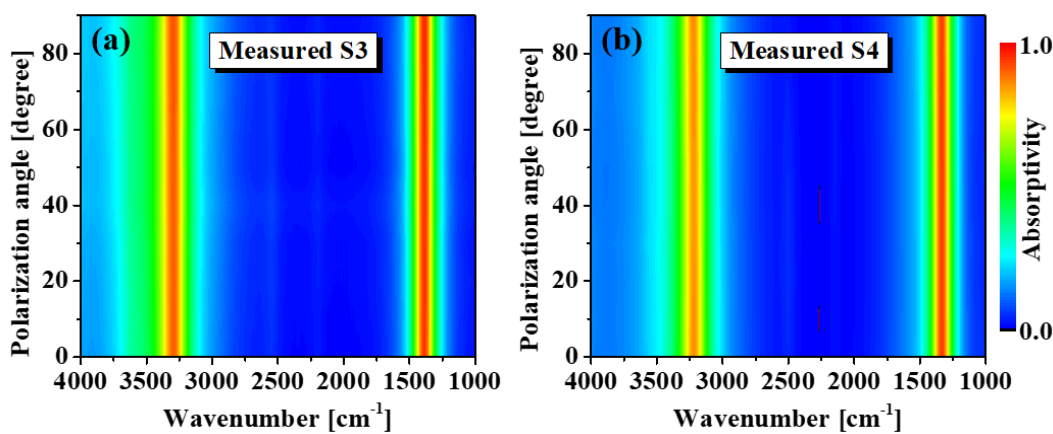

**Figure S3.** The measured polarization independence of the absorptivity spectra of (a) S3 and (b) S4.

**Spectroscopic ellipsometry of PDMS film.** A 10:1 (volume) mixture of the elastomer base and curing agent from a PDMS silicone elastomer kit (SYLGARD® 184, Sigma-Aldrich) was diluted in *n*-heptane with different volume ratios. The diluted mixtures in *n*-heptane solutions with different ratios were then spin-coated onto 1×1 Si (100) wafers at 6000 rpm for 60 seconds. After spin-coating, the mixture films were kept in ambient for 2 hours to allow the mixture films smoothly form on the Si substrates. Subsequently, the films were cured at 100 °C for 45 minutes to uniformly form PDMS elastomer, and then naturally cooled to room temperature. In the UV – NIR range, a Tauc-Lorentz model was used. The thicknesses of the PDMS films formed at different PDMS:*n*-heptane volume ratio were also obtained with ellipsometry (the inset in Figure S4) which is almost proportional to the PDMS:*n*-heptane volume ratio.

To get the optical properties of the PDMS film from the UV to MIR region, a sub-300 nm PDMS film coated on a Si substrate was measured in the range of 50000 – 400 cm$^{-1}$ (0.2 – 25 $\mu$m) using a spectroscopic ellipsometry. Beside using a Tauc-Lorentz model in the UV – NIR region, a Brendel multi-oscillator model with 11 oscillators was employed to simulate the molecular vibrations of the PDMS film in the MIR region. As seen in Figure S4a and Figure

S4b, the models used in the fitting are in good agreement with the measured amplitude change (Ψ) and the phase change (Δ).

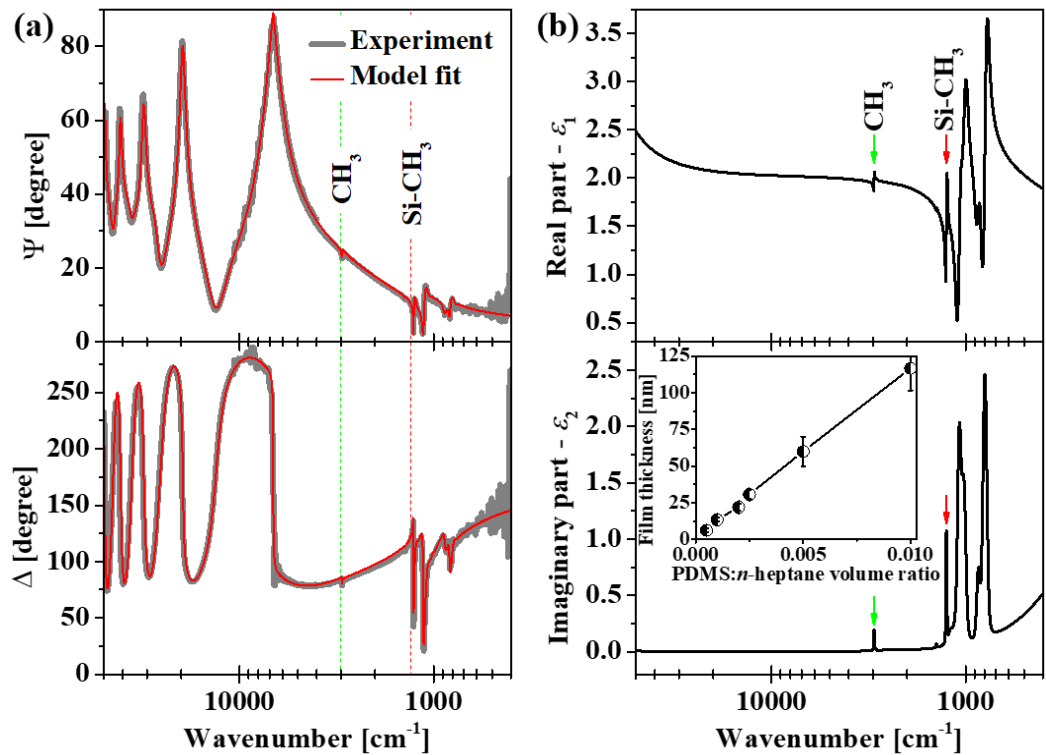

**Figure S4.** Spectroscopic ellipsometry of the PDMS thin films. (a) The relative amplitude change (Ψ) and the phase change (Δ) of the PDMS films measured with two ellipsometric angles (gray curves). Red curves show the fitting results. (b) Retrieved permittivity of the PDMS. The inset in (b) shows the thickness dependence of the PDMS thin film on the PDMS:$n$-heptane volume ratio.

**SEIRA enhancement factor of DPA.** The SEIRA enhancement factor of DPA-S5 was calculated as: $EF = \frac{I_{SEIRA}}{I_{bulk}} \frac{N_{bulk}}{N_{SEIRA}}$, where $I_{SEIRA}$ and $I_{bulk}$ are intensities of the molecular vibration mode from a 10-nm PDMS layer coated on the DPA-S5 (for SEIRA) and from a 450-nm PDMS film coated on an Al film, respectively. $N_{bulk}$ and $N_{SEIRA}$ are the number of PDMS molecules contributed to vibrational signals of a 450-nm PDMS film coated on the Al film and the SEIRA signal on DPA-S5, respectively. $N_{bulk}$ can be expressed as: $N_{bulk} = At_{PDMS}n$, where

$A$ is the IR beam area, $t_{PDMS}$ is the bulk PDMS thickness ($t_{PDMS}$ = 450 nm), and $n$ is the molecular number density. $N_{SEIRA}$ can be estimated by the number of molecules from the PDMS film with 10-nm-thick toroidal volumes around Al disks with respect to the hot-spot volume of each Al disk antenna: $N_{SEIRA} = \left\{\pi\left[(r+t_0)^2 - r^2\right](t+t_0)\right\} \left[\dfrac{\dfrac{\pi\sqrt{3}}{6}A}{\pi\left(\dfrac{p}{2}\right)^2}\right] n$, where $r$ is the Al disk diameter of DPA-S5 ($r$ = 1275 nm), $t$ is the Al disk thickness ($t$ = 100 nm), $t_0$ is the PDMS thickness coated on DPA-S5 ($t_0$ = 10 nm), $\dfrac{\pi\sqrt{3}}{6}$ is the filling factor of the 2D closed-package hexagonal lattice, $p$ is the periodicity of DPA-S5 ($p$ = 3000 nm). Here the first factor – $\left\{\pi\left[(r+t_0)^2 - r^2\right](t+t_0)\right\}$ is the toroidal volume of PDMS molecule on each Al disk, and the second factor – $\left[\dfrac{\dfrac{\pi\sqrt{3}}{6}A}{\pi\left(\dfrac{p}{2}\right)^2}\right]$ is the number of Al disks in the IR beam area. Thus, the ratio $\dfrac{N_{bulk}}{N_{SEIRA}}$ is given by:

$$\dfrac{N_{bulk}}{N_{SEIRA}} = \dfrac{A t_{PDMS} n}{\left\{\pi\left[(r+t_0)^2 - r^2\right](t+t_0)\right\}\left[\dfrac{\dfrac{\pi\sqrt{3}}{6}A}{\pi\left(\dfrac{p}{2}\right)^2}\right]n} = \dfrac{6 t_{PDMS}\left(\dfrac{p}{2}\right)^2}{\pi\sqrt{3}\left[(r+t_0)^2 - r^2\right](t+t_0)} = 396$$

. Finally, the enhancement factors for each resonance mode were calculated as high as 571 at M1 and 642 at M2.

**Further analysis of curing dynamics of PDMS by SEIRA.** Here we assume the conformational changes at the C-H stretching and the CH$_3$ deformation of PDMS elastomer were proportional to the absorbance reduction of the crosslinker's S-H stretching during the hydrosilylation reaction (Figure S5a). The conversions – $\alpha$ can be therefore predicted by the following equation: $\alpha_t = \dfrac{\log_{10}(R_t/R_0)}{\log_{10}(R_S/R_0)}$, where $R_S/R_0$ is the saturated SEIRA reflectance. Figure S5b plots the degrees of the conversions at the conformational changes of both C-H and Si-CH$_3$ vibrations as functions of the reaction times (Figure S5b). In addition, the kinetic of the reaction can be expressed by: $\dfrac{d\alpha}{dt} = K(1-\alpha)^n$, where $t$ is the reaction time, $K$ is the rate constant, $A$ is the preexponential factor. By fitting the degrees of conversions using the above reaction rate equation, we obtained the rate constant – $K$ of about 0.13 with the order – $n$ of the kinetic of about 0.45.

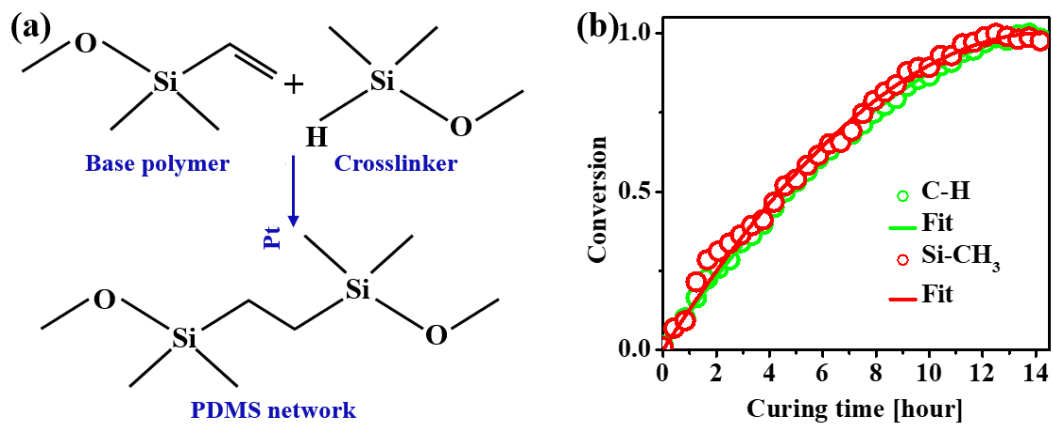

**Figure S5. (a)** Hydrosilylation reaction of PDMS. **(b)** Degrees of conversions versus curing times.

**Angle-resolved SEIRA spectroscopy.** Angle-resolved measurements on DPA-S5 and its SEIRA spectroscopy with a 10-nm PDMS layer was also carried out (Figure S6a – Figure S6c). The measured angle-dependent absorptivity agrees well with the simulation result shown in Figure S2a. The angle-dependent absorptivity of DPA-S5 targeting at the magnetic resonance – M2 are re-plotted in Figure S6b and their SEIRA spectra with a 10-nm PDMS are presented in Figure S6c.

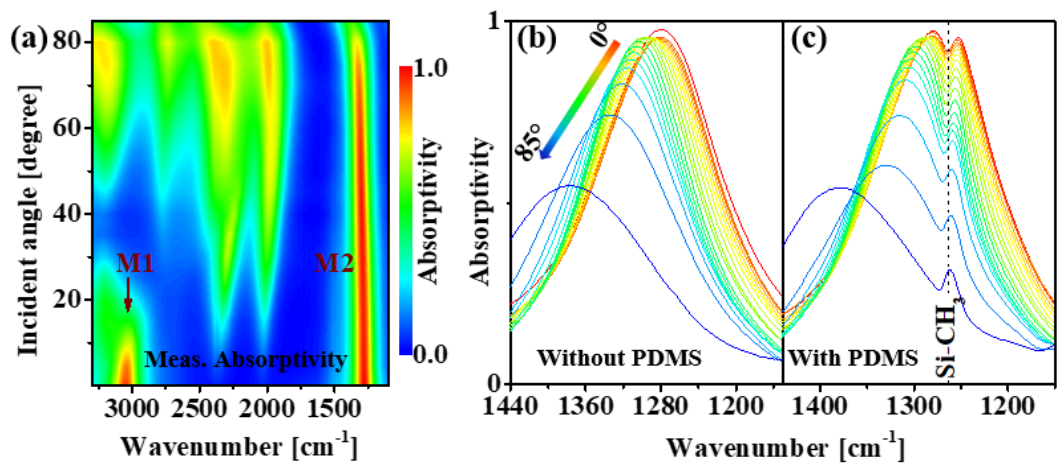

**Figure S6.** (a) Measured spectral map of the angle-dependent absorptivity of DPA-S5. M2 is almost unchanged up to 80° of incidence while M1 strongly depends on the incidence angle up to 15°. (b) Replotted angle-resolved absorptivity spectra at M2 and (c) Angle-resolved SEIRA spectra at M2 with 10-nm PDMS film.

**Detailed information of samples S6 and S7.** To fully investigate the coupling between Si-CH$_3$ vibration and DPAs, two additional DPAs named sample S6 and sample S7 were designed and fabricated (Figure S7). Both S6 and S7 have the same periodicity (4.4 $\mu$m) and the insulator thickness (0.2 $\mu$m) but were designed at different diameters to have different magnetic resonances at 1260 cm$^{-1}$ ($d$ = 2.97 $\mu$m, S6) and at 1130 cm$^{-1}$ ($d$ = 3.28 $\mu$m, S7). We also observed the SPP coupled magnetic (third order) resonances of S6 and S7 at around 2500 cm$^{-1}$ with absorptivities of 0.67 (S6) and 0.83 (S7).

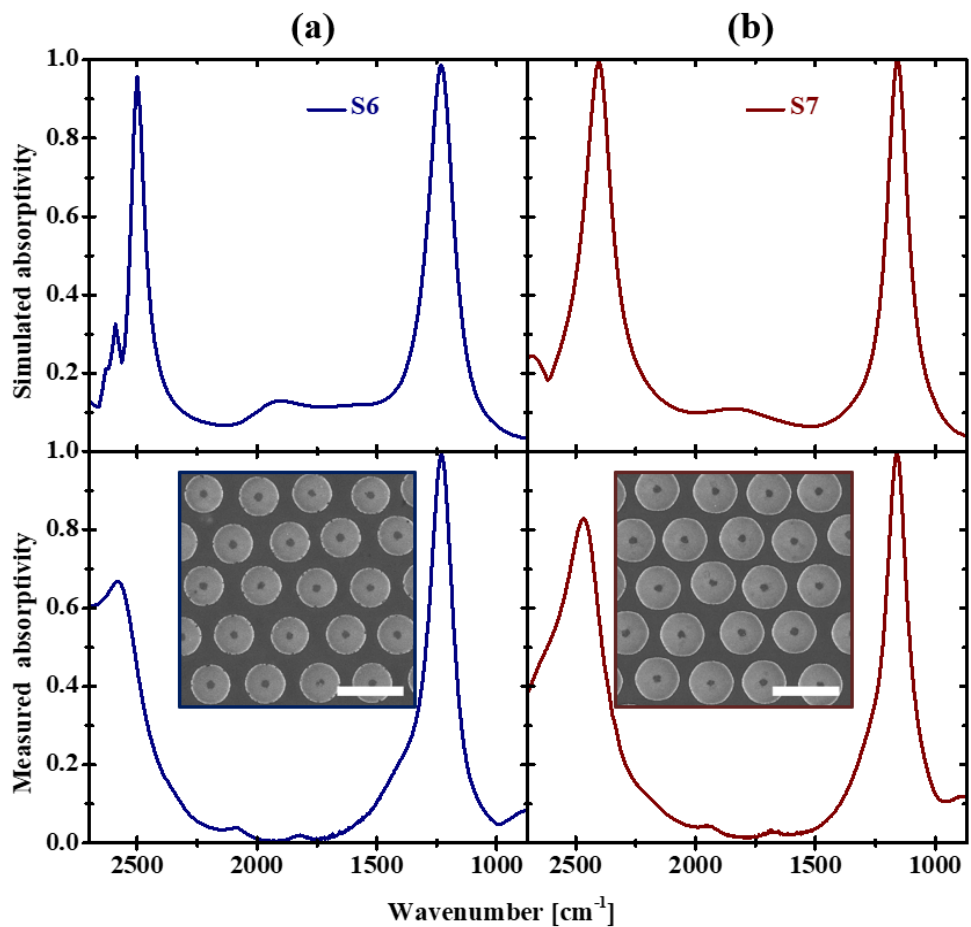

**Figure S7.** Simulated and measured absorptivities of two samples S6 ($d$ = 2.97 $\mu$m) and S7 ($d$ = 3.28 $\mu$m) with the same periodicity (4.4 $\mu$m) and insulator thickness (0.2 $\mu$m). The scale bars used in the SEM images represent 5 $\mu$m.

**Further SEIRA spectra of C-H stretching and its hybridization with M1 resonance.**
Together with SEIRA study of a 10-nm PDMS film coated on all DPAs (S1 – S7) at the fundamental magnetic resonance – M2, SEIRA spectra and the coupling between C-H stretching vibration with the SPP coupled magnetic (third order) resonance – M1 were also analyzed (Figure S8).

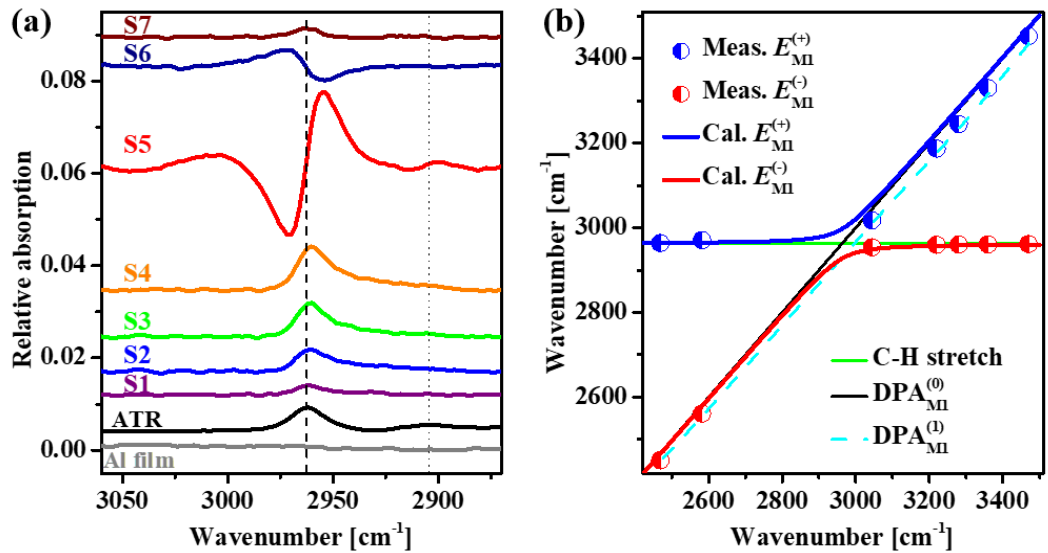

**Figure S8.** (a) Relative absorption spectra of asymmetric C-H stretching in $CH_3$ taken from 10-nm PDMS films coated on DPAs, and on Al films measured in both standard reflectance and ATR geometries. (b) Hybridized modes (half-filled circles) resulted from the coupling between the C-H stretching in $CH_3$ and DPAs' resonances plotted against the bare DPAs resonances (black line with slope = 1) and C-H stretching in $CH_3$ at 2962 cm$^{-1}$ (green line). The dashed cyan line denotes the DPAs' resonances with a redshift caused by the presence of the PDMS film. Blue and red curves denote the calculated hybridized modes.

**Table S1.** Complex permittivity of PDMS film retrieved from spectroscopic ellipsometry

| Wavenumber [cm$^{-1}$] | Real part ($\varepsilon_1$) | Imaginary part ($\varepsilon_2$) |
|---|---|---|
| 50000 | 2.48471 | 0.00831 |
| 47079.60502 | 2.41198 | 0.00598 |
| 44481.53018 | 2.3561 | 0.00433 |
| 42155.20865 | 2.312 | 0.00313 |
| 40060.12063 | 2.27643 | 0.00224 |
| 38163.4205 | 2.24724 | 0.00158 |
| 36438.20616 | 2.22291 | 0.00108 |
| 34862.22553 | 2.20239 | 7.00E-04 |
| 33416.91796 | 2.1849 | 4.30E-04 |
| 32086.67715 | 2.16984 | 2.40E-04 |
| 30858.28969 | 2.15677 | 1.10E-04 |
| 29720.48802 | 2.14535 | 3.00E-05 |
| 28663.60746 | 2.13532 | 0 |
| 27679.31323 | 2.12648 | 0 |
| 26760.37508 | 2.11863 | 0 |
| 25900.49212 | 2.11162 | 0 |
| 25094.15011 | 2.10533 | 0 |
| 24336.49882 | 2.09966 | 1.00E-05 |
| 23623.25668 | 2.09453 | 1.00E-05 |
| 22950.6314 | 2.08987 | 1.00E-05 |
| 22315.24904 | 2.08562 | 1.00E-05 |
| 21714.09927 | 2.08173 | 1.00E-05 |
| 21144.48903 | 2.07817 | 1.00E-05 |
| 20603.99923 | 2.07489 | 1.00E-05 |
| 20090.45244 | 2.07187 | 1.00E-05 |
| 19601.88263 | 2.06907 | 1.00E-05 |
| 19136.51161 | 2.06648 | 1.00E-05 |
| 18692.72502 | 2.06407 | 1.00E-05 |
| 18269.05491 | 2.06183 | 1.00E-05 |
| 17864.16443 | 2.05974 | 1.00E-05 |
| 17476.83171 | 2.05778 | 1.00E-05 |
| 17105.93862 | 2.05595 | 2.00E-05 |
| 16750.46085 | 2.05423 | 2.00E-05 |
| 16409.45664 | 2.05262 | 2.00E-05 |
| 16082.05942 | 2.0511 | 2.00E-05 |
| 15767.47117 | 2.04966 | 2.00E-05 |
| 15464.9544 | 2.04831 | 2.00E-05 |
| 15173.82735 | 2.04703 | 2.00E-05 |
| 14893.45845 | 2.04582 | 2.00E-05 |
| 14623.26264 | 2.04467 | 3.00E-05 |
| 14362.69587 | 2.04357 | 3.00E-05 |
| 14111.25224 | 2.04253 | 3.00E-05 |
| 13868.46122 | 2.04154 | 3.00E-05 |
| 13633.88357 | 2.0406 | 3.00E-05 |
| 13407.10926 | 2.0397 | 3.00E-05 |
| 13187.75563 | 2.03883 | 3.00E-05 |
| 12975.46415 | 2.03801 | 4.00E-05 |
| 12769.899 | 2.03722 | 4.00E-05 |
| 12570.7458 | 2.03646 | 4.00E-05 |
| 12377.70901 | 2.03573 | 4.00E-05 |
| 12190.51096 | 2.03503 | 4.00E-05 |
| 12008.891 | 2.03435 | 5.00E-05 |
| 11832.60331 | 2.0337 | 5.00E-05 |
| 11661.41646 | 2.03307 | 5.00E-05 |
| 11495.1121 | 2.03247 | 5.00E-05 |
| 11333.48453 | 2.03188 | 5.00E-05 |
| 11176.33908 | 2.03132 | 6.00E-05 |
| 11023.49174 | 2.03077 | 6.00E-05 |
| 10874.76879 | 2.03024 | 6.00E-05 |
| 10730.0054 | 2.02972 | 6.00E-05 |
| 10589.04541 | 2.02922 | 7.00E-05 |
| 10451.74108 | 2.02873 | 7.00E-05 |
| 10317.95192 | 2.02826 | 7.00E-05 |
| 10187.54455 | 2.0278 | 7.00E-05 |
| 10060.39254 | 2.02735 | 8.00E-05 |
| 9936.3754 | 2.02691 | 8.00E-05 |
| 9815.37862 | 2.02648 | 8.00E-05 |
| 9697.29308 | 2.02607 | 9.00E-05 |
| 9582.01516 | 2.02566 | 9.00E-05 |
| 9469.4458 | 2.02526 | 9.00E-05 |
| 9359.49056 | 2.02486 | 1.00E-04 |
| 9252.05961 | 2.02448 | 1.00E-04 |
| 9147.06693 | 2.0241 | 1.00E-04 |
| 9044.43034 | 2.02374 | 1.10E-04 |
| 8944.07158 | 2.02337 | 1.10E-04 |
| 8845.91559 | 2.02302 | 1.10E-04 |
| 8749.89054 | 2.02267 | 1.20E-04 |
| 8655.92794 | 2.02232 | 1.20E-04 |
| 8563.96197 | 2.02198 | 1.30E-04 |
| 8473.9296 | 2.02165 | 1.30E-04 |
| 8385.77061 | 2.02132 | 1.30E-04 |
| 8299.42706 | 2.02099 | 1.40E-04 |
| 8214.84346 | 2.02067 | 1.40E-04 |
| 8131.96646 | 2.02035 | 1.50E-04 |
| 8050.74507 | 2.02004 | 1.50E-04 |
| 7971.1301 | 2.01973 | 1.60E-04 |
| 7893.07429 | 2.01942 | 1.60E-04 |
| 7816.53241 | 2.01912 | 1.70E-04 |
| 7741.46079 | 2.01882 | 1.70E-04 |
| 7667.81739 | 2.01852 | 1.70E-04 |
| 7595.56197 | 2.01822 | 1.80E-04 |
| 7524.65559 | 2.01793 | 1.90E-04 |
| 7455.06076 | 2.01764 | 1.90E-04 |
| 7386.74155 | 2.01735 | 2.00E-04 |
| 7319.66314 | 2.01706 | 2.00E-04 |
| 7253.79203 | 2.01678 | 2.10E-04 |
| 7189.09587 | 2.01649 | 2.10E-04 |
| 7125.5436 | 2.01621 | 2.20E-04 |
| 7063.1051 | 2.01593 | 2.20E-04 |
| 7001.7513 | 2.01565 | 2.30E-04 |
| 6941.45427 | 2.01537 | 2.40E-04 |
| 6882.18689 | 2.0151 | 2.40E-04 |
| 6823.92297 | 2.01482 | 2.50E-04 |
| 6766.63733 | 2.01454 | 2.60E-04 |
| 6710.30548 | 2.01427 | 2.60E-04 |
| 6654.90377 | 2.014 | 2.70E-04 |
| 6600.40942 | 2.01372 | 2.80E-04 |
| 6546.8003 | 2.01345 | 2.80E-04 |
| 6494.05499 | 2.01318 | 2.90E-04 |
| 6442.15275 | 2.01291 | 3.00E-04 |
| 6391.0736 | 2.01264 | 3.10E-04 |
| 6340.79809 | 2.01237 | 3.10E-04 |
| 6291.30735 | 2.01209 | 3.20E-04 |
| 6242.58323 | 2.01182 | 3.30E-04 |
| 6194.60801 | 2.01155 | 3.40E-04 |
| 6147.36452 | 2.01128 | 3.40E-04 |
| 6100.83623 | 2.01101 | 3.50E-04 |
| 6055.00698 | 2.01074 | 3.60E-04 |
| 6009.86109 | 2.01047 | 3.70E-04 |
| 5965.38346 | 2.0102 | 3.80E-04 |
| 5921.55934 | 2.00992 | 3.90E-04 |
| 5878.37439 | 2.00965 | 4.00E-04 |
| 5835.81479 | 2.00938 | 4.10E-04 |
| 5793.86703 | 2.00911 | 4.20E-04 |
| 5752.518 | 2.00883 | 4.20E-04 |
| 5711.75495 | 2.00856 | 4.30E-04 |
| 5671.56557 | 2.00828 | 4.40E-04 |
| 5631.9378 | 2.008 | 4.50E-04 |
| 5592.85993 | 2.00773 | 4.60E-04 |
| 5554.32064 | 2.00745 | 4.80E-04 |
| 5516.30886 | 2.00717 | 4.90E-04 |
| 5478.81378 | 2.00689 | 5.00E-04 |
| 5441.82501 | 2.00661 | 5.10E-04 |
| 5405.33233 | 2.00633 | 5.20E-04 |
| 5369.3258 | 2.00605 | 5.30E-04 |
| 5333.79582 | 2.00576 | 5.40E-04 |
| 5298.73297 | 2.00548 | 5.60E-04 |
| 5264.1281 | 2.00519 | 5.70E-04 |
| 5229.97226 | 2.0049 | 5.80E-04 |
| 5196.25683 | 2.00461 | 5.90E-04 |
| 5162.97331 | 2.00432 | 6.10E-04 |
| 5130.11343 | 2.00403 | 6.20E-04 |
| 5097.6692 | 2.00374 | 6.30E-04 |
| 5065.63277 | 2.00344 | 6.50E-04 |
| 5033.99647 | 2.00315 | 6.60E-04 |
| 5002.75289 | 2.00285 | 6.80E-04 |
| 4971.89475 | 2.00255 | 6.90E-04 |
| 4941.41493 | 2.00225 | 7.10E-04 |
| 4911.30657 | 2.00194 | 7.20E-04 |
| 4881.56289 | 2.00164 | 7.40E-04 |
| 4852.17728 | 2.00133 | 7.60E-04 |
| 4823.14336 | 2.00102 | 7.80E-04 |
| 4794.45484 | 2.00071 | 7.90E-04 |
| 4766.10558 | 2.0004 | 8.10E-04 |
| 4738.08959 | 2.00008 | 8.30E-04 |
| 4710.40106 | 1.99977 | 8.50E-04 |
| 4683.03426 | 1.99945 | 8.70E-04 |
| 4655.9836 | 1.99913 | 8.90E-04 |
| 4629.24367 | 1.9988 | 9.10E-04 |
| 4602.80912 | 1.99848 | 9.40E-04 |
| 4576.67475 | 1.99815 | 9.60E-04 |
| 4550.83549 | 1.99782 | 9.80E-04 |
| 4525.28637 | 1.99749 | 0.00101 |
| 4500.0225 | 1.99715 | 0.00103 |
| 4475.03917 | 1.99681 | 0.00106 |
| 4450.33172 | 1.99647 | 0.00109 |
| 4425.89559 | 1.99613 | 0.00112 |
| 4401.72634 | 1.99579 | 0.00115 |
| 4377.81963 | 1.99544 | 0.00118 |
| 4354.17121 | 1.99509 | 0.00121 |
| 4330.77689 | 1.99473 | 0.00125 |
| 4307.63263 | 1.99438 | 0.00128 |
| 4284.73443 | 1.99402 | 0.00132 |
| 4262.07837 | 1.99366 | 0.00136 |
| 4239.66066 | 1.99329 | 0.0014 |
| 4217.47754 | 1.99292 | 0.00144 |
| 4195.52533 | 1.99255 | 0.00149 |
| 4173.80047 | 1.99218 | 0.00153 |
| 4152.29945 | 1.9918 | 0.00158 |
| 4131.0188 | 1.99142 | 0.00163 |
| 4109.95518 | 1.99104 | 0.00169 |
| 4089.10527 | 1.99066 | 0.00174 |
| 4068.46584 | 1.99027 | 0.0018 |
| 4048.03369 | 1.98988 | 0.00186 |
| 4027.80576 | 1.98949 | 0.00193 |
| 4007.77899 | 1.9891 | 0.002 |
| 3987.95036 | 1.9887 | 0.00206 |
| 3968.31699 | 1.9883 | 0.00214 |
| 3948.87599 | 1.9879 | 0.00221 |
| 3929.62453 | 1.9875 | 0.00229 |
| 3910.55988 | 1.98709 | 0.00237 |
| 3891.67932 | 1.98668 | 0.00246 |
| 3872.98019 | 1.98627 | 0.00255 |
| 3854.45991 | 1.98586 | 0.00264 |
| 3836.11591 | 1.98545 | 0.00273 |
| 3817.94569 | 1.98504 | 0.00283 |
| 3799.94677 | 1.98462 | 0.00293 |
| 3782.11678 | 1.98421 | 0.00303 |
| 3764.45332 | 1.98379 | 0.00313 |
| 3746.95407 | 1.98337 | 0.00324 |
| 3729.61678 | 1.98295 | 0.00335 |
| 3712.43918 | 1.98253 | 0.00346 |
| 3695.41908 | 1.98211 | 0.00358 |
| 3678.55434 | 1.98169 | 0.00369 |
| 3661.84283 | 1.98126 | 0.00381 |
| 3645.28246 | 1.98083 | 0.00393 |
| 3628.87122 | 1.98041 | 0.00405 |
| 3612.60708 | 1.97998 | 0.00417 |
| 3596.48807 | 1.97954 | 0.00429 |
| 3580.51227 | 1.97911 | 0.00442 |
| 3564.67777 | 1.97867 | 0.00454 |
| 3548.98271 | 1.97823 | 0.00467 |
| 3533.42525 | 1.97778 | 0.00479 |
| 3518.00359 | 1.97733 | 0.00492 |
| 3502.71597 | 1.97688 | 0.00505 |
| 3487.56062 | 1.97642 | 0.00518 |
| 3472.53587 | 1.97596 | 0.0053 |
| 3457.64002 | 1.97549 | 0.00544 |
| 3442.87141 | 1.97501 | 0.00557 |
| 3428.22843 | 1.97453 | 0.0057 |
| 3413.70949 | 1.97404 | 0.00583 |
| 3399.31299 | 1.97354 | 0.00597 |
| 3385.03742 | 1.97304 | 0.0061 |
| 3370.88125 | 1.97252 | 0.00624 |
| 3356.84299 | 1.972 | 0.00638 |
| 3342.92117 | 1.97146 | 0.00652 |
| 3329.11435 | 1.97091 | 0.00667 |
| 3315.42111 | 1.97035 | 0.00681 |
| 3301.84004 | 1.96978 | 0.00696 |
| 3288.3698 | 1.96918 | 0.00711 |
| 3275.00901 | 1.96857 | 0.00726 |
| 3261.75635 | 1.96794 | 0.00741 |
| 3248.61052 | 1.96729 | 0.00757 |
| 3235.57023 | 1.96662 | 0.00773 |
| 3222.6342 | 1.96591 | 0.00789 |
| 3209.80121 | 1.96517 | 0.00806 |
| 3197.07002 | 1.9644 | 0.00823 |
| 3184.43941 | 1.96357 | 0.00841 |
| 3171.90822 | 1.9627 | 0.00859 |
| 3159.47527 | 1.96177 | 0.00877 |
| 3147.1394 | 1.96077 | 0.00897 |
| 3134.89947 | 1.95967 | 0.00917 |
| 3122.7544 | 1.95848 | 0.00939 |
| 3110.70306 | 1.95715 | 0.00962 |
| 3098.74438 | 1.95565 | 0.00988 |
| 3086.8773 | 1.95395 | 0.01017 |
| 3075.10076 | 1.95197 | 0.0105 |
| 3063.41373 | 1.94962 | 0.01091 |
| 3051.81521 | 1.94677 | 0.01142 |
| 3040.30418 | 1.94318 | 0.0121 |
| 3028.87966 | 1.9385 | 0.01311 |
| 3017.54068 | 1.93203 | 0.01473 |
| 3006.28628 | 1.92243 | 0.01778 |
| 2995.11552 | 1.90668 | 0.02489 |
| 2984.02746 | 1.87907 | 0.0483 |
| 2973.02121 | 1.86885 | 0.13412 |
| 2962.09585 | 1.99915 | 0.18915 |
| 2951.25048 | 2.06275 | 0.08665 |
| 2940.48425 | 2.03268 | 0.03682 |
| 2929.79628 | 2.0101 | 0.02558 |
| 2919.18572 | 1.99713 | 0.02533 |
| 2908.65175 | 1.99404 | 0.02899 |
| 2898.19352 | 1.99692 | 0.02613 |
| 2887.81023 | 1.99488 | 0.01965 |
| 2877.50108 | 1.99046 | 0.01617 |
| 2867.26527 | 1.98665 | 0.01467 |
| 2857.10202 | 1.98365 | 0.01394 |
| 2847.01057 | 1.98123 | 0.01356 |
| 2836.99015 | 1.97921 | 0.01334 |
| 2827.04003 | 1.97748 | 0.01322 |
| 2817.15944 | 1.97598 | 0.01315 |
| 2807.3477 | 1.97464 | 0.01313 |
| 2797.60406 | 1.97344 | 0.01314 |
| 2787.92781 | 1.97234 | 0.01316 |
| 2778.31828 | 1.97133 | 0.0132 |
| 2768.77476 | 1.97039 | 0.01324 |
| 2759.29658 | 1.96951 | 0.0133 |
| 2749.88307 | 1.96869 | 0.01336 |
| 2740.53358 | 1.9679 | 0.01343 |
| 2731.24743 | 1.96716 | 0.0135 |
| 2722.02402 | 1.96644 | 0.01357 |
| 2712.86269 | 1.96576 | 0.01365 |
| 2703.76282 | 1.9651 | 0.01373 |
| 2694.72379 | 1.96446 | 0.01381 |
| 2685.745 | 1.96384 | 0.01389 |
| 2676.82584 | 1.96324 | 0.01397 |
| 2667.96572 | 1.96265 | 0.01404 |
| 2659.16407 | 1.96208 | 0.01412 |
| 2650.4203 | 1.96152 | 0.0142 |
| 2641.73384 | 1.96097 | 0.01428 |
| 2633.10414 | 1.96043 | 0.01435 |
| 2624.53063 | 1.9599 | 0.01443 |
| 2616.01277 | 1.95938 | 0.0145 |
| 2607.55002 | 1.95886 | 0.01457 |
| 2599.14186 | 1.95835 | 0.01464 |
| 2590.78774 | 1.95784 | 0.0147 |
| 2582.48714 | 1.95734 | 0.01477 |
| 2574.23958 | 1.95683 | 0.01483 |
| 2566.04452 | 1.95634 | 0.0149 |
| 2557.90147 | 1.95584 | 0.01496 |
| 2549.80994 | 1.95535 | 0.01501 |
| 2541.76945 | 1.95486 | 0.01507 |
| 2533.7795 | 1.95437 | 0.01513 |
| 2525.83963 | 1.95387 | 0.01518 |
| 2517.94937 | 1.95338 | 0.01523 |
| 2510.10824 | 1.95289 | 0.01529 |
| 2502.3158 | 1.9524 | 0.01534 |
| 2494.57159 | 1.95191 | 0.01539 |
| 2486.87517 | 1.95142 | 0.01544 |
| 2479.22609 | 1.95093 | 0.01549 |
| 2471.62393 | 1.95044 | 0.01554 |
| 2464.06824 | 1.94994 | 0.01559 |
| 2456.5586 | 1.94945 | 0.01564 |
| 2449.09461 | 1.94895 | 0.01569 |
| 2441.67583 | 1.94845 | 0.01573 |
| 2434.30186 | 1.94795 | 0.01578 |
| 2426.97229 | 1.94745 | 0.01583 |
| 2419.68674 | 1.94695 | 0.01588 |
| 2412.44479 | 1.94645 | 0.01593 |
| 2405.24606 | 1.94595 | 0.01597 |
| 2398.09017 | 1.94544 | 0.01602 |
| 2390.97672 | 1.94493 | 0.01607 |
| 2383.90536 | 1.94443 | 0.01612 |
| 2376.8757 | 1.94392 | 0.01616 |
| 2369.88738 | 1.9434 | 0.01621 |
| 2362.94002 | 1.94289 | 0.01626 |
| 2356.03328 | 1.94238 | 0.0163 |
| 2349.16681 | 1.94186 | 0.01635 |
| 2342.34023 | 1.94134 | 0.0164 |
| 2335.55322 | 1.94082 | 0.01644 |
| 2328.80543 | 1.9403 | 0.01649 |
| 2322.09651 | 1.93977 | 0.01653 |
| 2315.42614 | 1.93925 | 0.01658 |
| 2308.79398 | 1.93872 | 0.01662 |
| 2302.1997 | 1.93819 | 0.01667 |
| 2295.64299 | 1.93766 | 0.01671 |
| 2289.12352 | 1.93713 | 0.01675 |
| 2282.64096 | 1.93659 | 0.0168 |
| 2276.19503 | 1.93605 | 0.01684 |
| 2269.7854 | 1.93551 | 0.01688 |
| 2263.41176 | 1.93496 | 0.01692 |
| 2257.07382 | 1.93442 | 0.01696 |
| 2250.77128 | 1.93387 | 0.017 |
| 2244.50383 | 1.93332 | 0.01704 |
| 2238.27119 | 1.93276 | 0.01708 |
| 2232.07307 | 1.93221 | 0.01711 |
| 2225.90918 | 1.93165 | 0.01715 |
| 2219.77924 | 1.93108 | 0.01719 |
| 2213.68297 | 1.93052 | 0.01722 |
| 2207.6201 | 1.92995 | 0.01726 |
| 2201.59034 | 1.92938 | 0.01729 |
| 2195.59344 | 1.9288 | 0.01733 |
| 2189.62911 | 1.92822 | 0.01736 |
| 2183.6971 | 1.92764 | 0.0174 |
| 2177.79714 | 1.92706 | 0.01743 |
| 2171.92899 | 1.92647 | 0.01746 |
| 2166.09237 | 1.92587 | 0.01749 |
| 2160.28703 | 1.92528 | 0.01753 |
| 2154.51273 | 1.92468 | 0.01756 |
| 2148.76922 | 1.92408 | 0.01759 |
| 2143.05624 | 1.92347 | 0.01762 |
| 2137.37356 | 1.92286 | 0.01765 |
| 2131.72095 | 1.92224 | 0.01768 |
| 2126.09815 | 1.92162 | 0.01771 |
| 2120.50493 | 1.921 | 0.01774 |
| 2114.94107 | 1.92038 | 0.01777 |
| 2109.40633 | 1.91975 | 0.0178 |
| 2103.90048 | 1.91911 | 0.01783 |
| 2098.4233 | 1.91847 | 0.01786 |
| 2092.97456 | 1.91783 | 0.01788 |
| 2087.55404 | 1.91718 | 0.01791 |
| 2082.16153 | 1.91653 | 0.01794 |
| 2076.79681 | 1.91588 | 0.01797 |
| 2071.45966 | 1.91522 | 0.018 |
| 2066.14987 | 1.91456 | 0.01803 |
| 2060.86724 | 1.91389 | 0.01806 |
| 2055.61154 | 1.91322 | 0.01809 |
| 2050.38259 | 1.91254 | 0.01812 |
| 2045.18017 | 1.91186 | 0.01815 |
| 2040.00408 | 1.91118 | 0.01818 |
| 2034.85413 | 1.91049 | 0.01821 |
| 2029.73011 | 1.90979 | 0.01825 |
| 2024.63184 | 1.9091 | 0.01828 |
| 2019.55911 | 1.90839 | 0.01831 |
| 2014.51174 | 1.90769 | 0.01834 |
| 2009.48954 | 1.90698 | 0.01838 |
| 2004.49231 | 1.90626 | 0.01841 |
| 1999.51988 | 1.90554 | 0.01844 |
| 1994.57205 | 1.90481 | 0.01848 |
| 1989.64865 | 1.90409 | 0.01851 |
| 1984.7495 | 1.90335 | 0.01855 |
| 1979.87441 | 1.90261 | 0.01858 |
| 1975.02322 | 1.90187 | 0.01862 |
| 1970.19574 | 1.90112 | 0.01866 |
| 1965.3918 | 1.90037 | 0.01869 |
| 1960.61123 | 1.89962 | 0.01873 |
| 1955.85386 | 1.89885 | 0.01877 |
| 1951.11953 | 1.89809 | 0.01881 |
| 1946.40806 | 1.89732 | 0.01885 |
| 1941.71928 | 1.89654 | 0.01889 |
| 1937.05304 | 1.89576 | 0.01893 |
| 1932.40918 | 1.89498 | 0.01897 |
| 1927.78753 | 1.89419 | 0.01902 |
| 1923.18793 | 1.8934 | 0.01906 |
| 1918.61023 | 1.8926 | 0.0191 |
| 1914.05427 | 1.89179 | 0.01915 |
| 1909.5199 | 1.89098 | 0.01919 |
| 1905.00696 | 1.89017 | 0.01924 |
| 1900.5153 | 1.88935 | 0.01929 |
| 1896.04477 | 1.88853 | 0.01933 |
| 1891.59522 | 1.8877 | 0.01938 |
| 1887.16651 | 1.88687 | 0.01943 |
| 1882.75849 | 1.88603 | 0.01948 |
| 1878.37101 | 1.88519 | 0.01953 |
| 1874.00394 | 1.88434 | 0.01958 |
| 1869.65712 | 1.88349 | 0.01963 |
| 1865.33042 | 1.88263 | 0.01968 |
| 1861.0237 | 1.88176 | 0.01974 |
| 1856.73682 | 1.88089 | 0.01979 |
| 1852.46965 | 1.88002 | 0.01984 |
| 1848.22205 | 1.87914 | 0.0199 |
| 1843.99387 | 1.87826 | 0.01995 |
| 1839.78501 | 1.87736 | 0.02001 |
| 1835.59531 | 1.87647 | 0.02006 |
| 1831.42464 | 1.87557 | 0.02012 |
| 1827.27289 | 1.87466 | 0.02018 |
| 1823.13992 | 1.87375 | 0.02024 |
| 1819.02561 | 1.87283 | 0.0203 |
| 1814.92982 | 1.87191 | 0.02036 |
| 1810.85243 | 1.87098 | 0.02042 |
| 1806.79333 | 1.87004 | 0.02048 |
| 1802.75238 | 1.8691 | 0.02054 |
| 1798.72946 | 1.86816 | 0.0206 |
| 1794.72446 | 1.8672 | 0.02067 |
| 1790.73726 | 1.86625 | 0.02073 |
| 1786.76773 | 1.86528 | 0.02079 |
| 1782.81576 | 1.86431 | 0.02086 |
| 1778.88123 | 1.86333 | 0.02093 |
| 1774.96404 | 1.86235 | 0.02099 |
| 1771.06406 | 1.86136 | 0.02106 |
| 1767.18118 | 1.86037 | 0.02113 |
| 1763.31528 | 1.85936 | 0.0212 |
| 1759.46626 | 1.85836 | 0.02127 |
| 1755.63401 | 1.85734 | 0.02134 |
| 1751.81842 | 1.85632 | 0.02141 |
| 1748.01938 | 1.85529 | 0.02148 |
| 1744.23678 | 1.85426 | 0.02155 |
| 1740.47051 | 1.85322 | 0.02162 |
| 1736.72047 | 1.85217 | 0.0217 |
| 1732.98656 | 1.85111 | 0.02177 |
| 1729.26867 | 1.85005 | 0.02185 |
| 1725.5667 | 1.84898 | 0.02192 |
| 1721.88055 | 1.84791 | 0.022 |
| 1718.2101 | 1.84682 | 0.02207 |
| 1714.55528 | 1.84573 | 0.02215 |
| 1710.91597 | 1.84463 | 0.02223 |
| 1707.29208 | 1.84353 | 0.02231 |
| 1703.6835 | 1.84241 | 0.02239 |
| 1700.09015 | 1.84129 | 0.02247 |
| 1696.51192 | 1.84016 | 0.02255 |
| 1692.94873 | 1.83903 | 0.02263 |
| 1689.40047 | 1.83788 | 0.02272 |
| 1685.86705 | 1.83673 | 0.0228 |
| 1682.34839 | 1.83557 | 0.02289 |
| 1678.84438 | 1.8344 | 0.02297 |
| 1675.35493 | 1.83322 | 0.02306 |
| 1671.87996 | 1.83204 | 0.02315 |
| 1668.41938 | 1.83084 | 0.02323 |
| 1664.9731 | 1.82964 | 0.02332 |
| 1661.54102 | 1.82843 | 0.02341 |
| 1658.12306 | 1.82721 | 0.0235 |
| 1654.71914 | 1.82598 | 0.02359 |
| 1651.32916 | 1.82474 | 0.02369 |
| 1647.95304 | 1.82349 | 0.02378 |
| 1644.5907 | 1.82224 | 0.02387 |
| 1641.24206 | 1.82097 | 0.02397 |
| 1637.90702 | 1.8197 | 0.02406 |
| 1634.58551 | 1.81841 | 0.02416 |
| 1631.27744 | 1.81712 | 0.02426 |
| 1627.98273 | 1.81581 | 0.02436 |
| 1624.70131 | 1.8145 | 0.02446 |
| 1621.43309 | 1.81317 | 0.02456 |
| 1618.17799 | 1.81183 | 0.02466 |
| 1614.93594 | 1.81049 | 0.02476 |
| 1611.70685 | 1.80913 | 0.02487 |
| 1608.49064 | 1.80776 | 0.02497 |
| 1605.28725 | 1.80639 | 0.02508 |
| 1602.09659 | 1.805 | 0.02519 |
| 1598.91859 | 1.80359 | 0.02529 |
| 1595.75317 | 1.80218 | 0.0254 |
| 1592.60026 | 1.80076 | 0.02551 |
| 1589.45979 | 1.79932 | 0.02563 |
| 1586.33168 | 1.79787 | 0.02574 |
| 1583.21585 | 1.79641 | 0.02585 |
| 1580.11224 | 1.79494 | 0.02597 |
| 1577.02078 | 1.79346 | 0.02609 |
| 1573.94139 | 1.79196 | 0.0262 |
| 1570.874 | 1.79045 | 0.02632 |
| 1567.81855 | 1.78892 | 0.02644 |
| 1564.77495 | 1.78739 | 0.02657 |
| 1561.74315 | 1.78584 | 0.02669 |
| 1558.72308 | 1.78427 | 0.02681 |
| 1555.71466 | 1.78269 | 0.02694 |
| 1552.71784 | 1.7811 | 0.02707 |
| 1549.73254 | 1.77949 | 0.0272 |
| 1546.75869 | 1.77787 | 0.02733 |
| 1543.79624 | 1.77623 | 0.02746 |
| 1540.84511 | 1.77457 | 0.02759 |
| 1537.90525 | 1.7729 | 0.02773 |
| 1534.97658 | 1.77122 | 0.02787 |
| 1532.05904 | 1.76951 | 0.028 |
| 1529.15258 | 1.76779 | 0.02814 |
| 1526.25712 | 1.76605 | 0.02829 |
| 1523.37261 | 1.7643 | 0.02843 |
| 1520.49897 | 1.76253 | 0.02858 |
| 1517.63616 | 1.76073 | 0.02872 |
| 1514.78411 | 1.75892 | 0.02887 |
| 1511.94276 | 1.75709 | 0.02903 |
| 1509.11205 | 1.75524 | 0.02918 |
| 1506.29191 | 1.75337 | 0.02933 |
| 1503.4823 | 1.75148 | 0.02949 |
| 1500.68315 | 1.74956 | 0.02965 |
| 1497.89441 | 1.74763 | 0.02981 |
| 1495.116 | 1.74567 | 0.02998 |
| 1492.34789 | 1.74368 | 0.03015 |
| 1489.59001 | 1.74167 | 0.03032 |
| 1486.8423 | 1.73963 | 0.03049 |
| 1484.10471 | 1.73757 | 0.03066 |
| 1481.37718 | 1.73548 | 0.03084 |
| 1478.65966 | 1.73336 | 0.03102 |
| 1475.9521 | 1.7312 | 0.0312 |
| 1473.25442 | 1.72902 | 0.03139 |
| 1470.5666 | 1.7268 | 0.03158 |
| 1467.88856 | 1.72454 | 0.03177 |
| 1465.22026 | 1.72224 | 0.03196 |
| 1462.56164 | 1.71989 | 0.03216 |
| 1459.91265 | 1.7175 | 0.03237 |
| 1457.27324 | 1.71506 | 0.03257 |
| 1454.64336 | 1.71255 | 0.03279 |
| 1452.02295 | 1.70999 | 0.03301 |
| 1449.41197 | 1.70734 | 0.03323 |
| 1446.81036 | 1.7046 | 0.03347 |
| 1444.21807 | 1.70176 | 0.03371 |
| 1441.63506 | 1.69878 | 0.03398 |
| 1439.06126 | 1.69563 | 0.03427 |
| 1436.49664 | 1.69225 | 0.0346 |
| 1433.94115 | 1.68856 | 0.03503 |
| 1431.39473 | 1.68442 | 0.03565 |
| 1428.85734 | 1.67964 | 0.03671 |
| 1426.32893 | 1.67408 | 0.03881 |
| 1423.80946 | 1.66824 | 0.04303 |
| 1421.29886 | 1.66401 | 0.04969 |
| 1418.79711 | 1.66227 | 0.05743 |
| 1416.30415 | 1.66425 | 0.06598 |
| 1413.81993 | 1.67019 | 0.07146 |
| 1411.34441 | 1.67812 | 0.07316 |
| 1408.87755 | 1.68483 | 0.06993 |
| 1406.41929 | 1.68909 | 0.06313 |
| 1403.9696 | 1.68863 | 0.05551 |
| 1401.52843 | 1.68552 | 0.04891 |
| 1399.09573 | 1.68002 | 0.04391 |
| 1396.67146 | 1.67382 | 0.04136 |
| 1394.25558 | 1.66812 | 0.04037 |
| 1391.84804 | 1.66306 | 0.0401 |
| 1389.44881 | 1.65845 | 0.04014 |
| 1387.05783 | 1.65413 | 0.04033 |
| 1384.67506 | 1.64999 | 0.0406 |
| 1382.30047 | 1.64597 | 0.04091 |
| 1379.93401 | 1.64201 | 0.04125 |
| 1377.57563 | 1.63809 | 0.04161 |
| 1375.22531 | 1.63417 | 0.04199 |
| 1372.88299 | 1.63025 | 0.04239 |
| 1370.54863 | 1.6263 | 0.0428 |
| 1368.2222 | 1.62233 | 0.04323 |
| 1365.90366 | 1.61831 | 0.04367 |
| 1363.59296 | 1.61423 | 0.04413 |
| 1361.29006 | 1.6101 | 0.0446 |
| 1358.99493 | 1.60591 | 0.04509 |
| 1356.70752 | 1.60164 | 0.04559 |
| 1354.42781 | 1.5973 | 0.04611 |
| 1352.15574 | 1.59287 | 0.04664 |
| 1349.89128 | 1.58834 | 0.0472 |
| 1347.63439 | 1.58373 | 0.04777 |
| 1345.38504 | 1.579 | 0.04836 |
| 1343.14318 | 1.57417 | 0.04898 |
| 1340.90879 | 1.56921 | 0.04962 |
| 1338.68181 | 1.56413 | 0.05028 |
| 1336.46222 | 1.55891 | 0.05096 |
| 1334.24998 | 1.55354 | 0.05167 |
| 1332.04505 | 1.54801 | 0.05241 |
| 1329.84739 | 1.54231 | 0.05318 |
| 1327.65697 | 1.53643 | 0.05398 |
| 1325.47376 | 1.53035 | 0.05481 |
| 1323.29772 | 1.52406 | 0.05568 |
| 1321.12881 | 1.51753 | 0.05658 |
| 1318.967 | 1.51074 | 0.05753 |
| 1316.81225 | 1.50368 | 0.05852 |
| 1314.66453 | 1.49631 | 0.05956 |
| 1312.5238 | 1.48859 | 0.06064 |
| 1310.39004 | 1.4805 | 0.06179 |
| 1308.2632 | 1.47198 | 0.063 |
| 1306.14325 | 1.46298 | 0.06427 |
| 1304.03016 | 1.45344 | 0.06563 |
| 1301.9239 | 1.44329 | 0.06707 |
| 1299.82444 | 1.43242 | 0.06862 |
| 1297.73173 | 1.42073 | 0.07029 |
| 1295.64575 | 1.40807 | 0.07211 |
| 1293.56646 | 1.39426 | 0.07412 |
| 1291.49384 | 1.37907 | 0.07638 |
| 1289.42785 | 1.36219 | 0.07896 |
| 1287.36846 | 1.34322 | 0.08201 |
| 1285.31564 | 1.32163 | 0.08574 |
| 1283.26935 | 1.29668 | 0.0905 |
| 1281.22957 | 1.26736 | 0.09692 |
| 1279.19626 | 1.23228 | 0.10616 |
| 1277.1694 | 1.18964 | 0.12035 |
| 1275.14895 | 1.1375 | 0.14359 |
| 1273.13488 | 1.07512 | 0.18337 |
| 1271.12716 | 1.00725 | 0.25117 |
| 1269.12577 | 0.95032 | 0.35635 |
| 1267.13067 | 0.92616 | 0.49168 |
| 1265.14183 | 0.9406 | 0.64202 |
| 1263.15922 | 1.00402 | 0.80539 |
| 1261.18282 | 1.13767 | 0.94931 |
| 1259.2126 | 1.31261 | 1.03882 |
| 1257.24852 | 1.51695 | 1.07106 |
| 1255.29056 | 1.7124 | 1.02502 |
| 1253.33868 | 1.87883 | 0.92722 |
| 1251.39287 | 1.99328 | 0.77986 |
| 1249.45309 | 2.0416 | 0.62707 |
| 1247.51932 | 2.04813 | 0.48782 |
| 1245.59152 | 2.01775 | 0.36492 |
| 1243.66967 | 1.95949 | 0.27339 |
| 1241.75374 | 1.89408 | 0.21633 |
| 1239.8437 | 1.83469 | 0.18409 |
| 1237.93953 | 1.78472 | 0.16658 |
| 1236.04121 | 1.74337 | 0.15736 |
| 1234.14869 | 1.70889 | 0.15284 |
| 1232.26196 | 1.6797 | 0.15109 |
| 1230.38099 | 1.65457 | 0.15107 |
| 1228.50576 | 1.63261 | 0.15221 |
| 1226.63623 | 1.61315 | 0.15415 |
| 1224.77238 | 1.59574 | 0.1567 |
| 1222.91419 | 1.58003 | 0.15969 |
| 1221.06163 | 1.56578 | 0.16303 |
| 1219.21468 | 1.55278 | 0.16662 |
| 1217.3733 | 1.54088 | 0.17039 |
| 1215.53748 | 1.52996 | 0.17429 |
| 1213.70718 | 1.51991 | 0.17824 |
| 1211.88239 | 1.51068 | 0.18219 |
| 1210.06308 | 1.50218 | 0.18607 |
| 1208.24922 | 1.49433 | 0.18978 |
| 1206.44079 | 1.48704 | 0.19326 |
| 1204.63776 | 1.48021 | 0.19646 |
| 1202.84012 | 1.47374 | 0.19935 |
| 1201.04784 | 1.46753 | 0.20189 |
| 1199.26089 | 1.4615 | 0.20406 |
| 1197.47924 | 1.45552 | 0.20585 |
| 1195.70289 | 1.44951 | 0.20728 |
| 1193.93179 | 1.44338 | 0.20838 |
| 1192.16594 | 1.43706 | 0.20919 |
| 1190.4053 | 1.43048 | 0.20974 |
| 1188.64985 | 1.42358 | 0.21009 |
| 1186.89957 | 1.41631 | 0.21031 |
| 1185.15444 | 1.40864 | 0.21046 |
| 1183.41444 | 1.40058 | 0.21061 |
| 1181.67953 | 1.39213 | 0.21078 |
| 1179.94971 | 1.38328 | 0.21102 |
| 1178.22494 | 1.37403 | 0.21136 |
| 1176.50521 | 1.36438 | 0.21183 |
| 1174.79049 | 1.35434 | 0.21246 |
| 1173.08076 | 1.34391 | 0.21328 |
| 1171.376 | 1.3331 | 0.21431 |
| 1169.67619 | 1.32193 | 0.21556 |
| 1167.9813 | 1.31039 | 0.21703 |
| 1166.29132 | 1.29849 | 0.21872 |
| 1164.60622 | 1.28621 | 0.22064 |
| 1162.92599 | 1.27354 | 0.22278 |
| 1161.2506 | 1.26045 | 0.22517 |
| 1159.58002 | 1.24691 | 0.2278 |
| 1157.91425 | 1.2329 | 0.23071 |
| 1156.25325 | 1.21838 | 0.2339 |
| 1154.59702 | 1.20332 | 0.23741 |
| 1152.94552 | 1.18769 | 0.24127 |
| 1151.29874 | 1.17144 | 0.24553 |
| 1149.65666 | 1.15454 | 0.25022 |
| 1148.01925 | 1.13696 | 0.2554 |
| 1146.3865 | 1.11866 | 0.26114 |
| 1144.75839 | 1.09959 | 0.2675 |
| 1143.1349 | 1.07973 | 0.27457 |
| 1141.51601 | 1.05903 | 0.28243 |
| 1139.90169 | 1.03747 | 0.29119 |
| 1138.29194 | 1.01503 | 0.30095 |
| 1136.68672 | 0.99168 | 0.31186 |
| 1135.08603 | 0.96742 | 0.32406 |
| 1133.48983 | 0.94225 | 0.3377 |
| 1131.89812 | 0.91619 | 0.35296 |
| 1130.31088 | 0.88927 | 0.37004 |
| 1128.72808 | 0.86157 | 0.38915 |
| 1127.1497 | 0.83319 | 0.4105 |
| 1125.57574 | 0.80425 | 0.43433 |
| 1124.00616 | 0.77494 | 0.46088 |
| 1122.44096 | 0.74548 | 0.49038 |
| 1120.88011 | 0.71616 | 0.52304 |
| 1119.32359 | 0.6873 | 0.55906 |
| 1117.77139 | 0.65931 | 0.59858 |
| 1116.22349 | 0.63261 | 0.6417 |
| 1114.67987 | 0.60768 | 0.68844 |
| 1113.14051 | 0.58499 | 0.73874 |
| 1111.6054 | 0.56504 | 0.79247 |
| 1110.07452 | 0.54828 | 0.8494 |
| 1108.54785 | 0.53513 | 0.90926 |
| 1107.02537 | 0.52596 | 0.9717 |
| 1105.50707 | 0.52108 | 1.03637 |
| 1103.99293 | 0.52078 | 1.10288 |
| 1102.48293 | 0.52528 | 1.17085 |
| 1100.97706 | 0.53482 | 1.23988 |
| 1099.47529 | 0.54962 | 1.30954 |
| 1097.97761 | 0.56991 | 1.37934 |
| 1096.48401 | 0.59587 | 1.44874 |
| 1094.99447 | 0.62763 | 1.5171 |
| 1093.50897 | 0.66521 | 1.58369 |
| 1092.02749 | 0.7085 | 1.64777 |
| 1090.55002 | 0.75722 | 1.7086 |
| 1089.07655 | 0.81096 | 1.76549 |
| 1087.60705 | 0.86921 | 1.81788 |
| 1086.14151 | 0.93137 | 1.8653 |
| 1084.67991 | 0.99684 | 1.90738 |
| 1083.22225 | 1.06501 | 1.94384 |
| 1081.76849 | 1.13525 | 1.97444 |
| 1080.31864 | 1.20691 | 1.99901 |
| 1078.87266 | 1.27926 | 2.01741 |
| 1077.43055 | 1.35156 | 2.02965 |
| 1075.99229 | 1.42302 | 2.03585 |
| 1074.55787 | 1.4929 | 2.03631 |
| 1073.12726 | 1.56057 | 2.03142 |
| 1071.70046 | 1.62549 | 2.02166 |
| 1070.27745 | 1.68724 | 2.00756 |
| 1068.85821 | 1.74546 | 1.98965 |
| 1067.44273 | 1.79988 | 1.96845 |
| 1066.03099 | 1.85029 | 1.94454 |
| 1064.62299 | 1.89655 | 1.9185 |
| 1063.2187 | 1.93861 | 1.89092 |
| 1061.81811 | 1.97652 | 1.86241 |
| 1060.4212 | 2.01044 | 1.83348 |
| 1059.02796 | 2.04062 | 1.80461 |
| 1057.63838 | 2.06732 | 1.77614 |
| 1056.25244 | 2.09079 | 1.74835 |
| 1054.87013 | 2.11126 | 1.72146 |
| 1053.49144 | 2.12887 | 1.69568 |
| 1052.11634 | 2.14376 | 1.67125 |
| 1050.74483 | 2.15607 | 1.64849 |
| 1049.37688 | 2.16597 | 1.62779 |
| 1048.0125 | 2.17378 | 1.60957 |
| 1046.65166 | 2.17997 | 1.59427 |
| 1045.29435 | 2.18518 | 1.58221 |
| 1043.94055 | 2.19016 | 1.57357 |
| 1042.59026 | 2.19579 | 1.5683 |
| 1041.24345 | 2.2029 | 1.56607 |
| 1039.90012 | 2.21219 | 1.56628 |
| 1038.56025 | 2.22413 | 1.56815 |
| 1037.22383 | 2.23886 | 1.57084 |
| 1035.89085 | 2.25624 | 1.57357 |
| 1034.56128 | 2.27587 | 1.57574 |
| 1033.23513 | 2.29725 | 1.57705 |
| 1031.91237 | 2.31993 | 1.57751 |
| 1030.59299 | 2.34363 | 1.57731 |
| 1029.27698 | 2.36837 | 1.57677 |
| 1027.96433 | 2.39448 | 1.57609 |
| 1026.65502 | 2.42249 | 1.57518 |
| 1025.34905 | 2.45301 | 1.57355 |
| 1024.04639 | 2.48635 | 1.57025 |
| 1022.74704 | 2.52225 | 1.56411 |
| 1021.45098 | 2.55982 | 1.55416 |
| 1020.1582 | 2.59776 | 1.54004 |
| 1018.86869 | 2.63482 | 1.52211 |
| 1017.58244 | 2.67027 | 1.50123 |
| 1016.29943 | 2.704 | 1.47831 |
| 1015.01965 | 2.73635 | 1.45395 |
| 1013.74309 | 2.7678 | 1.42833 |
| 1012.46974 | 2.79874 | 1.40124 |
| 1011.19958 | 2.82927 | 1.37228 |
| 1009.9326 | 2.85921 | 1.34099 |
| 1008.6688 | 2.8881 | 1.30703 |
| 1007.40815 | 2.91532 | 1.27028 |
| 1006.15066 | 2.94019 | 1.23087 |
| 1004.89629 | 2.9621 | 1.18919 |
| 1003.64505 | 2.98062 | 1.14586 |
| 1002.39693 | 2.99555 | 1.10154 |
| 1001.1519 | 3.00693 | 1.05693 |
| 999.90996 | 3.01497 | 1.0126 |
| 998.6711 | 3.02004 | 0.96901 |
| 997.43531 | 3.02257 | 0.9264 |
| 996.20257 | 3.02294 | 0.88489 |
| 994.97287 | 3.0215 | 0.84445 |
| 993.74621 | 3.01849 | 0.80499 |
| 992.52257 | 3.01405 | 0.76637 |
| 991.30193 | 3.00822 | 0.72846 |
| 990.0843 | 3.00097 | 0.6912 |
| 988.86965 | 2.99223 | 0.65456 |
| 987.65798 | 2.98192 | 0.6186 |
| 986.44928 | 2.96996 | 0.58343 |
| 985.24353 | 2.95633 | 0.54921 |
| 984.04072 | 2.94104 | 0.51613 |
| 982.84085 | 2.92419 | 0.48438 |
| 981.6439 | 2.90589 | 0.45414 |
| 980.44986 | 2.88633 | 0.42555 |
| 979.25872 | 2.8657 | 0.39874 |
| 978.07048 | 2.84422 | 0.37376 |
| 976.88511 | 2.82211 | 0.35062 |
| 975.70261 | 2.79957 | 0.32932 |
| 974.52298 | 2.77681 | 0.30979 |
| 973.34619 | 2.75398 | 0.29196 |
| 972.17224 | 2.73124 | 0.27574 |
| 971.00112 | 2.70871 | 0.26102 |
| 969.83281 | 2.68648 | 0.24769 |
| 968.66732 | 2.66463 | 0.23562 |
| 967.50462 | 2.64321 | 0.22472 |
| 966.34471 | 2.62226 | 0.21487 |
| 965.18758 | 2.60181 | 0.20598 |
| 964.03322 | 2.58187 | 0.19795 |
| 962.88161 | 2.56245 | 0.19069 |
| 961.73275 | 2.54355 | 0.18413 |
| 960.58664 | 2.52515 | 0.17819 |
| 959.44324 | 2.50725 | 0.17281 |
| 958.30257 | 2.48983 | 0.16794 |
| 957.16461 | 2.47287 | 0.16352 |
| 956.02935 | 2.45635 | 0.15951 |
| 954.89677 | 2.44026 | 0.15586 |
| 953.76688 | 2.42457 | 0.15254 |
| 952.63966 | 2.40926 | 0.14952 |
| 951.51509 | 2.39432 | 0.14677 |
| 950.39318 | 2.37972 | 0.14426 |
| 949.27392 | 2.36544 | 0.14197 |
| 948.15728 | 2.35146 | 0.13988 |
| 947.04327 | 2.33778 | 0.13798 |
| 945.93188 | 2.32435 | 0.13624 |
| 944.82309 | 2.31118 | 0.13465 |
| 943.71689 | 2.29824 | 0.13321 |
| 942.61329 | 2.28552 | 0.13189 |
| 941.51226 | 2.27301 | 0.13069 |
| 940.4138 | 2.26068 | 0.1296 |
| 939.3179 | 2.24852 | 0.12862 |
| 938.22455 | 2.23652 | 0.12773 |
| 937.13375 | 2.22466 | 0.12692 |
| 936.04548 | 2.21294 | 0.12621 |
| 934.95973 | 2.20134 | 0.12557 |
| 933.8765 | 2.18985 | 0.12501 |
| 932.79577 | 2.17845 | 0.12453 |
| 931.71755 | 2.16714 | 0.12411 |
| 930.64181 | 2.15589 | 0.12376 |
| 929.56856 | 2.14471 | 0.12348 |
| 928.49778 | 2.13358 | 0.12326 |
| 927.42946 | 2.12248 | 0.1231 |
| 926.36359 | 2.11142 | 0.12301 |
| 925.30018 | 2.10036 | 0.12298 |
| 924.2392 | 2.08932 | 0.12302 |
| 923.18066 | 2.07826 | 0.12312 |
| 922.12453 | 2.06719 | 0.12329 |
| 921.07082 | 2.05608 | 0.12353 |
| 920.01951 | 2.04494 | 0.12384 |
| 918.97061 | 2.03374 | 0.12423 |
| 917.92409 | 2.02247 | 0.1247 |
| 916.87995 | 2.01112 | 0.12525 |
| 915.83818 | 1.99968 | 0.1259 |
| 914.79878 | 1.98814 | 0.12665 |
| 913.76174 | 1.97647 | 0.12751 |
| 912.72704 | 1.96467 | 0.12849 |
| 911.69469 | 1.95273 | 0.1296 |
| 910.66466 | 1.94062 | 0.13085 |
| 909.63696 | 1.92833 | 0.13227 |
| 908.61158 | 1.91584 | 0.13387 |
| 907.58851 | 1.90315 | 0.13567 |
| 906.56774 | 1.89024 | 0.13769 |
| 905.54926 | 1.87709 | 0.13997 |
| 904.53307 | 1.86368 | 0.14253 |
| 903.51916 | 1.85002 | 0.14541 |
| 902.50752 | 1.83609 | 0.14866 |
| 901.49814 | 1.82188 | 0.15231 |
| 900.49101 | 1.8074 | 0.15642 |
| 899.48614 | 1.79264 | 0.16104 |
| 898.4835 | 1.77763 | 0.16623 |
| 897.4831 | 1.76239 | 0.17207 |
| 896.48492 | 1.74694 | 0.17861 |
| 895.48896 | 1.73134 | 0.18593 |
| 894.49521 | 1.71564 | 0.1941 |
| 893.50366 | 1.69994 | 0.20316 |
| 892.51431 | 1.68432 | 0.21319 |
| 891.52715 | 1.66889 | 0.2242 |
| 890.54217 | 1.65378 | 0.23623 |
| 889.55936 | 1.63912 | 0.24925 |
| 888.57872 | 1.62504 | 0.26323 |
| 887.60024 | 1.61166 | 0.2781 |
| 886.62391 | 1.59907 | 0.29377 |
| 885.64973 | 1.58734 | 0.31012 |
| 884.67768 | 1.57652 | 0.32705 |
| 883.70777 | 1.56661 | 0.34443 |
| 882.73998 | 1.55758 | 0.3622 |
| 881.77431 | 1.54938 | 0.38029 |
| 880.81075 | 1.54198 | 0.39868 |
| 879.84929 | 1.53535 | 0.4174 |
| 878.88993 | 1.52952 | 0.43649 |
| 877.93266 | 1.52454 | 0.45601 |
| 876.97748 | 1.52053 | 0.47597 |
| 876.02437 | 1.51766 | 0.49638 |
| 875.07332 | 1.51611 | 0.51715 |
| 874.12435 | 1.51608 | 0.53812 |
| 873.17742 | 1.51769 | 0.55904 |
| 872.23255 | 1.52103 | 0.57962 |
| 871.28972 | 1.52605 | 0.5995 |
| 870.34892 | 1.5326 | 0.61839 |
| 869.41016 | 1.54044 | 0.63604 |
| 868.47342 | 1.54926 | 0.65232 |
| 867.53869 | 1.55877 | 0.6672 |
| 866.60597 | 1.56871 | 0.68078 |
| 865.67526 | 1.57894 | 0.69321 |
| 864.74655 | 1.58943 | 0.70466 |
| 863.81982 | 1.60027 | 0.71523 |
| 862.89508 | 1.61162 | 0.72489 |
| 861.97232 | 1.62361 | 0.73345 |
| 861.05153 | 1.63625 | 0.74058 |
| 860.1327 | 1.64932 | 0.74592 |
| 859.21583 | 1.66239 | 0.74922 |
| 858.30092 | 1.67494 | 0.75047 |
| 857.38795 | 1.68651 | 0.74992 |
| 856.47692 | 1.69684 | 0.74795 |
| 855.56782 | 1.70589 | 0.74493 |
| 854.66066 | 1.71376 | 0.74112 |
| 853.75541 | 1.72061 | 0.73663 |
| 852.85208 | 1.72655 | 0.73143 |
| 851.95066 | 1.73158 | 0.72542 |
| 851.05115 | 1.73559 | 0.71849 |
| 850.15353 | 1.7384 | 0.71057 |
| 849.2578 | 1.73972 | 0.70169 |
| 848.36396 | 1.73929 | 0.69198 |
| 847.472 | 1.73685 | 0.68165 |
| 846.58191 | 1.73223 | 0.67102 |
| 845.69369 | 1.72532 | 0.66043 |
| 844.80733 | 1.71613 | 0.65024 |
| 843.92283 | 1.70475 | 0.64075 |
| 843.04017 | 1.6913 | 0.63223 |
| 842.15937 | 1.67597 | 0.6249 |
| 841.2804 | 1.65891 | 0.6189 |
| 840.40326 | 1.64028 | 0.61434 |
| 839.52795 | 1.62022 | 0.6113 |
| 838.65446 | 1.59881 | 0.60984 |
| 837.78279 | 1.57613 | 0.61004 |
| 836.91293 | 1.55224 | 0.61199 |
| 836.04487 | 1.52719 | 0.61581 |
| 835.17861 | 1.50104 | 0.62166 |
| 834.31414 | 1.47389 | 0.6297 |
| 833.45146 | 1.44585 | 0.64012 |
| 832.59057 | 1.4171 | 0.65311 |
| 831.73145 | 1.38784 | 0.66884 |
| 830.8741 | 1.35833 | 0.68747 |
| 830.01852 | 1.32887 | 0.70912 |
| 829.1647 | 1.29977 | 0.73387 |
| 828.31263 | 1.27138 | 0.76173 |
| 827.46231 | 1.24402 | 0.79267 |
| 826.61374 | 1.21804 | 0.8266 |
| 825.7669 | 1.19374 | 0.86338 |
| 824.9218 | 1.17139 | 0.90282 |
| 824.07843 | 1.15122 | 0.94471 |
| 823.23677 | 1.13341 | 0.98881 |
| 822.39684 | 1.11809 | 1.03488 |
| 821.55862 | 1.10534 | 1.08269 |
| 820.7221 | 1.0952 | 1.13204 |
| 819.88729 | 1.08767 | 1.18276 |
| 819.05417 | 1.08276 | 1.23471 |
| 818.22275 | 1.08048 | 1.28783 |
| 817.39301 | 1.08087 | 1.34205 |
| 816.56495 | 1.084 | 1.39734 |
| 815.73857 | 1.09 | 1.45369 |
| 814.91386 | 1.09904 | 1.51104 |
| 814.09082 | 1.11135 | 1.56931 |
| 813.26943 | 1.12719 | 1.62835 |
| 812.4497 | 1.14684 | 1.68792 |
| 811.63163 | 1.17054 | 1.7477 |
| 810.8152 | 1.19851 | 1.80727 |
| 810.0004 | 1.23086 | 1.86617 |
| 809.18725 | 1.26762 | 1.92386 |
| 808.37573 | 1.30867 | 1.97981 |
| 807.56583 | 1.3538 | 2.03351 |
| 806.75755 | 1.40266 | 2.08454 |
| 805.95089 | 1.45486 | 2.13258 |
| 805.14584 | 1.50996 | 2.17743 |
| 804.3424 | 1.56754 | 2.21901 |
| 803.54056 | 1.62724 | 2.25737 |
| 802.74032 | 1.6888 | 2.2926 |
| 801.94167 | 1.75208 | 2.32481 |
| 801.14461 | 1.81704 | 2.35411 |
| 800.34913 | 1.88374 | 2.38047 |
| 799.55522 | 1.95225 | 2.40377 |
| 798.7629 | 2.02262 | 2.42374 |
| 797.97214 | 2.09477 | 2.44001 |
| 797.18294 | 2.16845 | 2.45213 |
| 796.39531 | 2.24324 | 2.45975 |
| 795.60922 | 2.31853 | 2.46263 |
| 794.82469 | 2.39369 | 2.46072 |
| 794.04171 | 2.46809 | 2.45423 |
| 793.26026 | 2.54124 | 2.4435 |
| 792.48036 | 2.61285 | 2.42897 |
| 791.70198 | 2.68279 | 2.41103 |
| 790.92513 | 2.75107 | 2.39002 |
| 790.14981 | 2.81778 | 2.36613 |
| 789.376 | 2.88299 | 2.33943 |
| 788.60371 | 2.94674 | 2.30989 |
| 787.83293 | 3.00895 | 2.27743 |
| 787.06365 | 3.06946 | 2.24195 |
| 786.29587 | 3.12802 | 2.20339 |
| 785.52959 | 3.1843 | 2.16177 |
| 784.76481 | 3.23794 | 2.11719 |
| 784.00151 | 3.28858 | 2.06987 |
| 783.23969 | 3.3359 | 2.0201 |
| 782.47935 | 3.37967 | 1.96829 |
| 781.72049 | 3.41971 | 1.91484 |
| 780.9631 | 3.45598 | 1.86022 |
| 780.20717 | 3.4885 | 1.80486 |
| 779.45271 | 3.51741 | 1.74915 |
| 778.6997 | 3.54287 | 1.69343 |
| 777.94815 | 3.56512 | 1.63796 |
| 777.19804 | 3.5844 | 1.58294 |
| 776.44939 | 3.60093 | 1.5285 |
| 775.70217 | 3.61494 | 1.4747 |
| 774.95639 | 3.6266 | 1.42157 |
| 774.21204 | 3.63605 | 1.36911 |
| 773.46912 | 3.6434 | 1.31732 |
| 772.72762 | 3.64871 | 1.26618 |
| 771.98755 | 3.652 | 1.2157 |
| 771.24889 | 3.65331 | 1.16591 |
| 770.51164 | 3.65261 | 1.11684 |
| 769.77581 | 3.64993 | 1.06858 |
| 769.04137 | 3.64526 | 1.02122 |
| 768.30834 | 3.63863 | 0.97487 |
| 767.5767 | 3.63008 | 0.92967 |
| 766.84645 | 3.61969 | 0.88574 |
| 766.1176 | 3.60755 | 0.8432 |
| 765.39012 | 3.59378 | 0.80219 |
| 764.66403 | 3.57851 | 0.76279 |
| 763.93931 | 3.56189 | 0.72511 |
| 763.21597 | 3.54407 | 0.68918 |
| 762.49399 | 3.52524 | 0.65507 |
| 761.77338 | 3.50555 | 0.62279 |
| 761.05413 | 3.48517 | 0.59233 |
| 760.33624 | 3.46424 | 0.56368 |
| 759.6197 | 3.44293 | 0.5368 |
| 758.90451 | 3.42136 | 0.51163 |
| 758.19066 | 3.39965 | 0.48811 |
| 757.47815 | 3.37792 | 0.46618 |
| 756.76699 | 3.35625 | 0.44575 |
| 756.05716 | 3.33474 | 0.42676 |
| 755.34865 | 3.31344 | 0.4091 |
| 754.64148 | 3.29243 | 0.39271 |
| 753.93563 | 3.27174 | 0.3775 |
| 753.23109 | 3.25142 | 0.3634 |
| 752.52788 | 3.2315 | 0.35032 |
| 751.82597 | 3.212 | 0.33819 |
| 751.12537 | 3.19293 | 0.32695 |
| 750.42608 | 3.17431 | 0.31652 |
| 749.72809 | 3.15615 | 0.30686 |
| 749.03139 | 3.13845 | 0.29789 |
| 748.33599 | 3.1212 | 0.28957 |
| 747.64188 | 3.10441 | 0.28184 |
| 746.94905 | 3.08806 | 0.27467 |
| 746.25751 | 3.07216 | 0.26801 |
| 745.56725 | 3.05669 | 0.26181 |
| 744.87826 | 3.04165 | 0.25605 |
| 744.19055 | 3.02702 | 0.25069 |
| 743.5041 | 3.01279 | 0.24569 |
| 742.81892 | 2.99895 | 0.24104 |
| 742.135 | 2.98549 | 0.2367 |
| 741.45234 | 2.9724 | 0.23266 |
| 740.77094 | 2.95966 | 0.22888 |
| 740.09078 | 2.94726 | 0.22535 |
| 739.41187 | 2.9352 | 0.22206 |
| 738.73421 | 2.92346 | 0.21898 |
| 738.05779 | 2.91203 | 0.21609 |
| 737.38261 | 2.9009 | 0.2134 |
| 736.70866 | 2.89005 | 0.21087 |
| 736.03594 | 2.87948 | 0.2085 |
| 735.36444 | 2.86918 | 0.20628 |
| 734.69418 | 2.85914 | 0.2042 |
| 734.02513 | 2.84935 | 0.20225 |
| 733.3573 | 2.8398 | 0.20042 |
| 732.69069 | 2.83048 | 0.19869 |
| 732.02528 | 2.82138 | 0.19708 |
| 731.36109 | 2.8125 | 0.19556 |
| 730.69809 | 2.80382 | 0.19413 |
| 730.0363 | 2.79535 | 0.19279 |
| 729.37571 | 2.78707 | 0.19153 |
| 728.71631 | 2.77898 | 0.19034 |
| 728.0581 | 2.77106 | 0.18923 |
| 727.40108 | 2.76332 | 0.18818 |
| 726.74524 | 2.75575 | 0.18719 |
| 726.09059 | 2.74834 | 0.18626 |
| 725.43711 | 2.74109 | 0.18539 |
| 724.78481 | 2.73399 | 0.18457 |
| 724.13368 | 2.72704 | 0.1838 |
| 723.48372 | 2.72022 | 0.18307 |
| 722.83493 | 2.71355 | 0.18239 |
| 722.1873 | 2.70701 | 0.18175 |
| 721.54083 | 2.70059 | 0.18115 |
| 720.89551 | 2.69431 | 0.18059 |
| 720.25135 | 2.68814 | 0.18006 |
| 719.60834 | 2.68209 | 0.17956 |
| 718.96647 | 2.67615 | 0.1791 |
| 718.32575 | 2.67032 | 0.17866 |
| 717.68617 | 2.6646 | 0.17825 |
| 717.04773 | 2.65898 | 0.17787 |
| 716.41042 | 2.65347 | 0.17752 |
| 715.77425 | 2.64805 | 0.17719 |
| 715.1392 | 2.64272 | 0.17688 |
| 714.50528 | 2.63749 | 0.1766 |
| 713.87248 | 2.63234 | 0.17633 |
| 713.2408 | 2.62729 | 0.17609 |
| 712.61024 | 2.62231 | 0.17586 |
| 711.9808 | 2.61742 | 0.17566 |
| 711.35246 | 2.61261 | 0.17547 |
| 710.72523 | 2.60787 | 0.17529 |
| 710.09911 | 2.60322 | 0.17514 |
| 709.47409 | 2.59863 | 0.175 |
| 708.85017 | 2.59412 | 0.17487 |
| 708.22734 | 2.58967 | 0.17476 |
| 707.60561 | 2.58529 | 0.17466 |
| 706.98497 | 2.58098 | 0.17457 |
| 706.36542 | 2.57674 | 0.1745 |
| 705.74695 | 2.57255 | 0.17443 |
| 705.12956 | 2.56843 | 0.17438 |
| 704.51325 | 2.56437 | 0.17434 |
| 703.89802 | 2.56036 | 0.17431 |
| 703.28387 | 2.55641 | 0.17429 |
| 702.67078 | 2.55252 | 0.17428 |
| 702.05876 | 2.54868 | 0.17428 |
| 701.44781 | 2.54489 | 0.17429 |
| 700.83792 | 2.54116 | 0.1743 |
| 700.22909 | 2.53747 | 0.17433 |
| 699.62132 | 2.53384 | 0.17436 |
| 699.0146 | 2.53025 | 0.1744 |
| 698.40893 | 2.52671 | 0.17445 |
| 697.80431 | 2.52322 | 0.1745 |
| 697.20073 | 2.51977 | 0.17457 |
| 696.5982 | 2.51636 | 0.17464 |
| 695.99671 | 2.513 | 0.17471 |
| 695.39626 | 2.50967 | 0.17479 |
| 694.79684 | 2.50639 | 0.17488 |
| 694.19846 | 2.50315 | 0.17497 |
| 693.6011 | 2.49995 | 0.17507 |
| 693.00478 | 2.49679 | 0.17518 |
| 692.40947 | 2.49367 | 0.17529 |
| 691.81519 | 2.49058 | 0.1754 |
| 691.22193 | 2.48753 | 0.17552 |
| 690.62969 | 2.48451 | 0.17565 |
| 690.03845 | 2.48153 | 0.17578 |
| 689.44823 | 2.47858 | 0.17591 |
| 688.85902 | 2.47566 | 0.17605 |
| 688.27082 | 2.47278 | 0.17619 |
| 687.68362 | 2.46993 | 0.17634 |
| 687.09742 | 2.46711 | 0.17649 |
| 686.51222 | 2.46432 | 0.17665 |
| 685.92801 | 2.46157 | 0.17681 |
| 685.3448 | 2.45884 | 0.17697 |
| 684.76258 | 2.45614 | 0.17714 |
| 684.18134 | 2.45347 | 0.17731 |
| 683.6011 | 2.45082 | 0.17748 |
| 683.02183 | 2.44821 | 0.17766 |
| 682.44355 | 2.44562 | 0.17784 |
| 681.86625 | 2.44306 | 0.17802 |
| 681.28992 | 2.44052 | 0.17821 |
| 680.71456 | 2.43801 | 0.1784 |
| 680.14018 | 2.43552 | 0.17859 |
| 679.56676 | 2.43306 | 0.17878 |
| 678.99431 | 2.43062 | 0.17898 |
| 678.42283 | 2.42821 | 0.17918 |
| 677.8523 | 2.42582 | 0.17938 |
| 677.28274 | 2.42345 | 0.17959 |
| 676.71413 | 2.42111 | 0.1798 |
| 676.14647 | 2.41878 | 0.18001 |
| 675.57977 | 2.41648 | 0.18022 |
| 675.01401 | 2.4142 | 0.18044 |
| 674.44921 | 2.41194 | 0.18066 |
| 673.88534 | 2.40971 | 0.18088 |
| 673.32242 | 2.40749 | 0.1811 |
| 672.76044 | 2.40529 | 0.18132 |
| 672.19939 | 2.40311 | 0.18155 |
| 671.63928 | 2.40095 | 0.18178 |
| 671.08011 | 2.39881 | 0.18201 |
| 670.52186 | 2.39669 | 0.18224 |
| 669.96454 | 2.39459 | 0.18248 |
| 669.40815 | 2.3925 | 0.18272 |
| 668.85268 | 2.39044 | 0.18295 |
| 668.29813 | 2.38839 | 0.1832 |
| 667.7445 | 2.38635 | 0.18344 |
| 667.19179 | 2.38434 | 0.18368 |
| 666.63999 | 2.38234 | 0.18393 |
| 666.0891 | 2.38036 | 0.18417 |
| 665.53912 | 2.37839 | 0.18442 |
| 664.99005 | 2.37644 | 0.18467 |
| 664.44189 | 2.37451 | 0.18493 |
| 663.89463 | 2.37259 | 0.18518 |
| 663.34827 | 2.37069 | 0.18544 |
| 662.8028 | 2.3688 | 0.18569 |
| 662.25824 | 2.36693 | 0.18595 |
| 661.71457 | 2.36507 | 0.18621 |
| 661.17179 | 2.36323 | 0.18647 |
| 660.6299 | 2.3614 | 0.18674 |
| 660.08889 | 2.35958 | 0.187 |
| 659.54877 | 2.35778 | 0.18727 |
| 659.00954 | 2.35599 | 0.18753 |
| 658.47119 | 2.35421 | 0.1878 |
| 657.93371 | 2.35245 | 0.18807 |
| 657.39712 | 2.3507 | 0.18834 |
| 656.86139 | 2.34897 | 0.18861 |
| 656.32654 | 2.34724 | 0.18889 |
| 655.79256 | 2.34553 | 0.18916 |
| 655.25945 | 2.34384 | 0.18944 |
| 654.7272 | 2.34215 | 0.18971 |
| 654.19582 | 2.34048 | 0.18999 |
| 653.6653 | 2.33881 | 0.19027 |
| 653.13564 | 2.33716 | 0.19055 |
| 652.60684 | 2.33552 | 0.19083 |
| 652.07889 | 2.3339 | 0.19112 |
| 651.55179 | 2.33228 | 0.1914 |
| 651.02555 | 2.33068 | 0.19168 |
| 650.50016 | 2.32908 | 0.19197 |
| 649.97561 | 2.3275 | 0.19226 |
| 649.45191 | 2.32593 | 0.19254 |
| 648.92906 | 2.32436 | 0.19283 |
| 648.40704 | 2.32281 | 0.19312 |
| 647.88586 | 2.32127 | 0.19341 |
| 647.36552 | 2.31974 | 0.19371 |
| 646.84602 | 2.31822 | 0.194 |
| 646.32735 | 2.31671 | 0.19429 |
| 645.80951 | 2.31521 | 0.19459 |
| 645.2925 | 2.31371 | 0.19488 |
| 644.77631 | 2.31223 | 0.19518 |
| 644.26095 | 2.31076 | 0.19548 |
| 643.74642 | 2.30929 | 0.19577 |
| 643.2327 | 2.30784 | 0.19607 |
| 642.71981 | 2.30639 | 0.19637 |
| 642.20773 | 2.30496 | 0.19667 |
| 641.69647 | 2.30353 | 0.19698 |
| 641.18602 | 2.30211 | 0.19728 |
| 640.67638 | 2.3007 | 0.19758 |
| 640.16755 | 2.2993 | 0.19788 |
| 639.65953 | 2.29791 | 0.19819 |
| 639.15232 | 2.29652 | 0.1985 |
| 638.6459 | 2.29514 | 0.1988 |
| 638.14029 | 2.29378 | 0.19911 |
| 637.63549 | 2.29242 | 0.19942 |
| 637.13147 | 2.29106 | 0.19973 |
| 636.62826 | 2.28972 | 0.20004 |
| 636.12584 | 2.28838 | 0.20035 |
| 635.62421 | 2.28705 | 0.20066 |
| 635.12337 | 2.28573 | 0.20097 |
| 634.62332 | 2.28442 | 0.20128 |
| 634.12406 | 2.28311 | 0.20159 |
| 633.62558 | 2.28181 | 0.20191 |
| 633.12789 | 2.28052 | 0.20222 |
| 632.63097 | 2.27924 | 0.20254 |
| 632.13484 | 2.27796 | 0.20285 |
| 631.63948 | 2.27669 | 0.20317 |
| 631.1449 | 2.27543 | 0.20349 |
| 630.65109 | 2.27417 | 0.2038 |
| 630.15806 | 2.27292 | 0.20412 |
| 629.6658 | 2.27168 | 0.20444 |
| 629.1743 | 2.27044 | 0.20476 |
| 628.68357 | 2.26921 | 0.20508 |
| 628.19361 | 2.26799 | 0.2054 |
| 627.7044 | 2.26677 | 0.20573 |
| 627.21596 | 2.26556 | 0.20605 |
| 626.72828 | 2.26436 | 0.20637 |
| 626.24136 | 2.26316 | 0.2067 |
| 625.75519 | 2.26197 | 0.20702 |
| 625.26978 | 2.26078 | 0.20734 |
| 624.78512 | 2.2596 | 0.20767 |
| 624.30121 | 2.25843 | 0.208 |
| 623.81805 | 2.25726 | 0.20832 |
| 623.33564 | 2.2561 | 0.20865 |
| 622.85397 | 2.25495 | 0.20898 |
| 622.37305 | 2.2538 | 0.20931 |
| 621.89287 | 2.25265 | 0.20963 |
| 621.41343 | 2.25152 | 0.20996 |
| 620.93473 | 2.25038 | 0.21029 |
| 620.45676 | 2.24926 | 0.21062 |
| 619.97953 | 2.24814 | 0.21096 |
| 619.50303 | 2.24702 | 0.21129 |
| 619.02727 | 2.24591 | 0.21162 |
| 618.55224 | 2.2448 | 0.21195 |
| 618.07793 | 2.2437 | 0.21229 |
| 617.60435 | 2.24261 | 0.21262 |
| 617.1315 | 2.24152 | 0.21295 |
| 616.65937 | 2.24043 | 0.21329 |
| 616.18796 | 2.23936 | 0.21362 |
| 615.71727 | 2.23828 | 0.21396 |
| 615.2473 | 2.23721 | 0.2143 |
| 614.77805 | 2.23615 | 0.21463 |
| 614.30951 | 2.23509 | 0.21497 |
| 613.84168 | 2.23403 | 0.21531 |
| 613.37457 | 2.23298 | 0.21565 |
| 612.90817 | 2.23194 | 0.21599 |
| 612.44248 | 2.2309 | 0.21633 |
| 611.97749 | 2.22986 | 0.21667 |
| 611.51321 | 2.22883 | 0.21701 |
| 611.04964 | 2.2278 | 0.21735 |
| 610.58676 | 2.22678 | 0.21769 |
| 610.12459 | 2.22576 | 0.21803 |
| 609.66311 | 2.22475 | 0.21837 |
| 609.20234 | 2.22374 | 0.21872 |
| 608.74226 | 2.22274 | 0.21906 |
| 608.28287 | 2.22174 | 0.2194 |
| 607.82418 | 2.22074 | 0.21975 |
| 607.36618 | 2.21975 | 0.22009 |
| 606.90886 | 2.21877 | 0.22044 |
| 606.45224 | 2.21778 | 0.22078 |
| 605.9963 | 2.2168 | 0.22113 |
| 605.54105 | 2.21583 | 0.22147 |
| 605.08648 | 2.21486 | 0.22182 |
| 604.63259 | 2.21389 | 0.22217 |
| 604.17939 | 2.21293 | 0.22252 |
| 603.72686 | 2.21197 | 0.22286 |
| 603.27501 | 2.21102 | 0.22321 |
| 602.82383 | 2.21007 | 0.22356 |
| 602.37333 | 2.20912 | 0.22391 |
| 601.92351 | 2.20818 | 0.22426 |
| 601.47435 | 2.20724 | 0.22461 |
| 601.02586 | 2.20631 | 0.22496 |
| 600.57805 | 2.20537 | 0.22531 |
| 600.13089 | 2.20445 | 0.22566 |
| 599.68441 | 2.20352 | 0.22602 |
| 599.23859 | 2.2026 | 0.22637 |
| 598.79343 | 2.20169 | 0.22672 |
| 598.34893 | 2.20077 | 0.22708 |
| 597.90509 | 2.19986 | 0.22743 |
| 597.46191 | 2.19896 | 0.22778 |
| 597.01938 | 2.19805 | 0.22814 |
| 596.57751 | 2.19716 | 0.22849 |
| 596.1363 | 2.19626 | 0.22885 |
| 595.69573 | 2.19537 | 0.2292 |
| 595.25582 | 2.19448 | 0.22956 |
| 594.81656 | 2.19359 | 0.22992 |
| 594.37794 | 2.19271 | 0.23027 |
| 593.93997 | 2.19183 | 0.23063 |
| 593.50265 | 2.19096 | 0.23099 |
| 593.06596 | 2.19009 | 0.23135 |
| 592.62992 | 2.18922 | 0.2317 |
| 592.19453 | 2.18835 | 0.23206 |
| 591.75977 | 2.18749 | 0.23242 |
| 591.32565 | 2.18663 | 0.23278 |
| 590.89216 | 2.18577 | 0.23314 |
| 590.45931 | 2.18492 | 0.2335 |
| 590.0271 | 2.18407 | 0.23386 |
| 589.59551 | 2.18323 | 0.23422 |
| 589.16456 | 2.18238 | 0.23458 |
| 588.73424 | 2.18154 | 0.23495 |
| 588.30454 | 2.1807 | 0.23531 |
| 587.87547 | 2.17987 | 0.23567 |
| 587.44703 | 2.17904 | 0.23603 |
| 587.01921 | 2.17821 | 0.2364 |
| 586.59201 | 2.17738 | 0.23676 |
| 586.16544 | 2.17656 | 0.23712 |
| 585.73949 | 2.17574 | 0.23749 |
| 585.31415 | 2.17492 | 0.23785 |
| 584.88943 | 2.17411 | 0.23822 |
| 584.46533 | 2.1733 | 0.23858 |
| 584.04184 | 2.17249 | 0.23895 |
| 583.61896 | 2.17168 | 0.23931 |
| 583.1967 | 2.17088 | 0.23968 |
| 582.77505 | 2.17008 | 0.24005 |
| 582.354 | 2.16928 | 0.24042 |
| 581.93357 | 2.16848 | 0.24078 |
| 581.51374 | 2.16769 | 0.24115 |
| 581.09452 | 2.1669 | 0.24152 |
| 580.6759 | 2.16611 | 0.24189 |
| 580.25788 | 2.16533 | 0.24226 |
| 579.84046 | 2.16455 | 0.24262 |
| 579.42365 | 2.16377 | 0.24299 |
| 579.00743 | 2.16299 | 0.24336 |
| 578.59181 | 2.16222 | 0.24373 |
| 578.17679 | 2.16144 | 0.2441 |
| 577.76236 | 2.16067 | 0.24448 |
| 577.34853 | 2.15991 | 0.24485 |
| 576.93529 | 2.15914 | 0.24522 |
| 576.52264 | 2.15838 | 0.24559 |
| 576.11058 | 2.15762 | 0.24596 |
| 575.6991 | 2.15686 | 0.24633 |
| 575.28822 | 2.15611 | 0.24671 |
| 574.87792 | 2.15536 | 0.24708 |
| 574.46821 | 2.15461 | 0.24745 |
| 574.05908 | 2.15386 | 0.24783 |
| 573.65053 | 2.15311 | 0.2482 |
| 573.24256 | 2.15237 | 0.24858 |
| 572.83518 | 2.15163 | 0.24895 |
| 572.42837 | 2.15089 | 0.24933 |
| 572.02214 | 2.15015 | 0.2497 |
| 571.61648 | 2.14942 | 0.25008 |
| 571.2114 | 2.14869 | 0.25045 |
| 570.8069 | 2.14796 | 0.25083 |
| 570.40296 | 2.14723 | 0.25121 |
| 569.9996 | 2.14651 | 0.25158 |
| 569.59681 | 2.14578 | 0.25196 |
| 569.19459 | 2.14506 | 0.25234 |
| 568.79293 | 2.14434 | 0.25272 |
| 568.39184 | 2.14363 | 0.25309 |
| 567.99132 | 2.14291 | 0.25347 |
| 567.59136 | 2.1422 | 0.25385 |
| 567.19196 | 2.14149 | 0.25423 |
| 566.79312 | 2.14078 | 0.25461 |
| 566.39485 | 2.14008 | 0.25499 |
| 565.99713 | 2.13937 | 0.25537 |
| 565.59998 | 2.13867 | 0.25575 |
| 565.20338 | 2.13797 | 0.25613 |
| 564.80733 | 2.13727 | 0.25651 |
| 564.41184 | 2.13658 | 0.25689 |
| 564.01691 | 2.13588 | 0.25727 |
| 563.62252 | 2.13519 | 0.25765 |
| 563.22869 | 2.1345 | 0.25804 |
| 562.83541 | 2.13381 | 0.25842 |
| 562.44267 | 2.13313 | 0.2588 |
| 562.05049 | 2.13244 | 0.25918 |
| 561.65885 | 2.13176 | 0.25957 |
| 561.26775 | 2.13108 | 0.25995 |
| 560.8772 | 2.1304 | 0.26034 |
| 560.48719 | 2.12973 | 0.26072 |
| 560.09773 | 2.12905 | 0.2611 |
| 559.70881 | 2.12838 | 0.26149 |
| 559.32042 | 2.12771 | 0.26187 |
| 558.93258 | 2.12704 | 0.26226 |
| 558.54527 | 2.12637 | 0.26264 |
| 558.1585 | 2.12571 | 0.26303 |
| 557.77226 | 2.12504 | 0.26342 |
| 557.38656 | 2.12438 | 0.2638 |
| 557.00139 | 2.12372 | 0.26419 |
| 556.61675 | 2.12306 | 0.26458 |
| 556.23264 | 2.12241 | 0.26496 |
| 555.84907 | 2.12175 | 0.26535 |
| 555.46602 | 2.1211 | 0.26574 |
| 555.0835 | 2.12045 | 0.26613 |
| 554.7015 | 2.1198 | 0.26651 |
| 554.32004 | 2.11915 | 0.2669 |
| 553.93909 | 2.1185 | 0.26729 |
| 553.55867 | 2.11786 | 0.26768 |
| 553.17877 | 2.11722 | 0.26807 |
| 552.79939 | 2.11658 | 0.26846 |
| 552.42054 | 2.11594 | 0.26885 |
| 552.0422 | 2.1153 | 0.26924 |
| 551.66438 | 2.11466 | 0.26963 |
| 551.28707 | 2.11403 | 0.27002 |
| 550.91028 | 2.1134 | 0.27041 |
| 550.53401 | 2.11276 | 0.2708 |
| 550.15825 | 2.11213 | 0.2712 |
| 549.783 | 2.11151 | 0.27159 |
| 549.40827 | 2.11088 | 0.27198 |
| 549.03404 | 2.11025 | 0.27237 |
| 548.66032 | 2.10963 | 0.27276 |
| 548.28712 | 2.10901 | 0.27316 |
| 547.91442 | 2.10839 | 0.27355 |
| 547.54222 | 2.10777 | 0.27394 |
| 547.17053 | 2.10715 | 0.27434 |
| 546.79935 | 2.10654 | 0.27473 |
| 546.42867 | 2.10592 | 0.27513 |
| 546.05849 | 2.10531 | 0.27552 |
| 545.68881 | 2.1047 | 0.27592 |
| 545.31964 | 2.10409 | 0.27631 |
| 544.95096 | 2.10348 | 0.27671 |
| 544.58278 | 2.10288 | 0.2771 |
| 544.21509 | 2.10227 | 0.2775 |
| 543.84791 | 2.10167 | 0.27789 |
| 543.48122 | 2.10106 | 0.27829 |
| 543.11502 | 2.10046 | 0.27869 |
| 542.74931 | 2.09986 | 0.27908 |
| 542.3841 | 2.09927 | 0.27948 |
| 542.01938 | 2.09867 | 0.27988 |
| 541.65515 | 2.09807 | 0.28027 |
| 541.29141 | 2.09748 | 0.28067 |
| 540.92816 | 2.09689 | 0.28107 |
| 540.56539 | 2.0963 | 0.28147 |
| 540.20311 | 2.09571 | 0.28187 |
| 539.84132 | 2.09512 | 0.28227 |
| 539.48001 | 2.09453 | 0.28267 |
| 539.11918 | 2.09395 | 0.28306 |
| 538.75883 | 2.09336 | 0.28346 |
| 538.39897 | 2.09278 | 0.28386 |
| 538.03959 | 2.0922 | 0.28426 |
| 537.68068 | 2.09162 | 0.28466 |
| 537.32226 | 2.09104 | 0.28506 |
| 536.96431 | 2.09046 | 0.28547 |
| 536.60684 | 2.08989 | 0.28587 |
| 536.24985 | 2.08931 | 0.28627 |
| 535.89333 | 2.08874 | 0.28667 |
| 535.53728 | 2.08817 | 0.28707 |
| 535.1817 | 2.0876 | 0.28747 |
| 534.8266 | 2.08703 | 0.28788 |
| 534.47197 | 2.08646 | 0.28828 |
| 534.11781 | 2.08589 | 0.28868 |
| 533.76412 | 2.08532 | 0.28908 |
| 533.41089 | 2.08476 | 0.28949 |
| 533.05814 | 2.0842 | 0.28989 |
| 532.70585 | 2.08363 | 0.29029 |
| 532.35402 | 2.08307 | 0.2907 |
| 532.00266 | 2.08251 | 0.2911 |
| 531.65176 | 2.08196 | 0.29151 |
| 531.30133 | 2.0814 | 0.29191 |
| 530.95136 | 2.08084 | 0.29232 |
| 530.60184 | 2.08029 | 0.29272 |
| 530.25279 | 2.07973 | 0.29313 |
| 529.9042 | 2.07918 | 0.29353 |
| 529.55606 | 2.07863 | 0.29394 |
| 529.20838 | 2.07808 | 0.29434 |
| 528.86116 | 2.07753 | 0.29475 |
| 528.5144 | 2.07698 | 0.29516 |
| 528.16808 | 2.07644 | 0.29556 |
| 527.82222 | 2.07589 | 0.29597 |
| 527.47682 | 2.07535 | 0.29638 |
| 527.13186 | 2.0748 | 0.29679 |
| 526.78736 | 2.07426 | 0.29719 |
| 526.44331 | 2.07372 | 0.2976 |
| 526.0997 | 2.07318 | 0.29801 |
| 525.75655 | 2.07264 | 0.29842 |
| 525.41384 | 2.07211 | 0.29883 |
| 525.07158 | 2.07157 | 0.29924 |
| 524.72976 | 2.07104 | 0.29964 |
| 524.38839 | 2.0705 | 0.30005 |
| 524.04746 | 2.06997 | 0.30046 |
| 523.70698 | 2.06944 | 0.30087 |
| 523.36693 | 2.06891 | 0.30128 |
| 523.02733 | 2.06838 | 0.30169 |
| 522.68817 | 2.06785 | 0.3021 |
| 522.34945 | 2.06732 | 0.30251 |
| 522.01117 | 2.06679 | 0.30293 |
| 521.67332 | 2.06627 | 0.30334 |
| 521.33591 | 2.06574 | 0.30375 |
| 520.99894 | 2.06522 | 0.30416 |
| 520.66241 | 2.0647 | 0.30457 |
| 520.3263 | 2.06418 | 0.30498 |
| 519.99064 | 2.06366 | 0.3054 |
| 519.6554 | 2.06314 | 0.30581 |
| 519.3206 | 2.06262 | 0.30622 |
| 518.98622 | 2.0621 | 0.30663 |
| 518.65228 | 2.06159 | 0.30705 |
| 518.31877 | 2.06107 | 0.30746 |
| 517.98569 | 2.06056 | 0.30787 |
| 517.65303 | 2.06004 | 0.30829 |
| 517.3208 | 2.05953 | 0.3087 |
| 516.989 | 2.05902 | 0.30911 |
| 516.65762 | 2.05851 | 0.30953 |
| 516.32667 | 2.058 | 0.30994 |
| 515.99614 | 2.05749 | 0.31036 |
| 515.66603 | 2.05699 | 0.31077 |
| 515.33635 | 2.05648 | 0.31119 |
| 515.00709 | 2.05598 | 0.3116 |
| 514.67824 | 2.05547 | 0.31202 |
| 514.34982 | 2.05497 | 0.31244 |
| 514.02182 | 2.05447 | 0.31285 |
| 513.69423 | 2.05397 | 0.31327 |
| 513.36706 | 2.05346 | 0.31369 |
| 513.04031 | 2.05297 | 0.3141 |
| 512.71397 | 2.05247 | 0.31452 |
| 512.38805 | 2.05197 | 0.31494 |
| 512.06254 | 2.05147 | 0.31535 |
| 511.73745 | 2.05098 | 0.31577 |
| 511.41277 | 2.05048 | 0.31619 |
| 511.0885 | 2.04999 | 0.31661 |
| 510.76464 | 2.0495 | 0.31702 |
| 510.44119 | 2.049 | 0.31744 |
| 510.11815 | 2.04851 | 0.31786 |
| 509.79552 | 2.04802 | 0.31828 |
| 509.4733 | 2.04753 | 0.3187 |
| 509.15148 | 2.04705 | 0.31912 |
| 508.83007 | 2.04656 | 0.31954 |
| 508.50907 | 2.04607 | 0.31996 |
| 508.18847 | 2.04559 | 0.32038 |
| 507.86827 | 2.0451 | 0.3208 |
| 507.54848 | 2.04462 | 0.32122 |
| 507.22909 | 2.04414 | 0.32164 |
| 506.91011 | 2.04365 | 0.32206 |
| 506.59152 | 2.04317 | 0.32248 |
| 506.27333 | 2.04269 | 0.3229 |
| 505.95554 | 2.04221 | 0.32332 |
| 505.63816 | 2.04173 | 0.32374 |
| 505.32117 | 2.04126 | 0.32417 |
| 505.00457 | 2.04078 | 0.32459 |
| 504.68838 | 2.0403 | 0.32501 |
| 504.37257 | 2.03983 | 0.32543 |
| 504.05717 | 2.03935 | 0.32585 |
| 503.74216 | 2.03888 | 0.32628 |
| 503.42754 | 2.03841 | 0.3267 |
| 503.11331 | 2.03794 | 0.32712 |
| 502.79948 | 2.03746 | 0.32755 |
| 502.48604 | 2.03699 | 0.32797 |
| 502.17298 | 2.03652 | 0.32839 |
| 501.86032 | 2.03606 | 0.32882 |
| 501.54805 | 2.03559 | 0.32924 |
| 501.23617 | 2.03512 | 0.32967 |
| 500.92467 | 2.03466 | 0.33009 |
| 500.61356 | 2.03419 | 0.33052 |
| 500.30283 | 2.03373 | 0.33094 |
| 499.9925 | 2.03326 | 0.33137 |
| 499.68254 | 2.0328 | 0.33179 |
| 499.37297 | 2.03234 | 0.33222 |
| 499.06379 | 2.03188 | 0.33264 |
| 498.75498 | 2.03142 | 0.33307 |
| 498.44656 | 2.03096 | 0.33349 |
| 498.13852 | 2.0305 | 0.33392 |
| 497.83086 | 2.03004 | 0.33435 |
| 497.52358 | 2.02958 | 0.33477 |
| 497.21668 | 2.02912 | 0.3352 |
| 496.91016 | 2.02867 | 0.33563 |
| 496.60401 | 2.02821 | 0.33605 |
| 496.29825 | 2.02776 | 0.33648 |
| 495.99285 | 2.02731 | 0.33691 |
| 495.68784 | 2.02685 | 0.33734 |
| 495.3832 | 2.0264 | 0.33777 |
| 495.07893 | 2.02595 | 0.33819 |
| 494.77504 | 2.0255 | 0.33862 |
| 494.47152 | 2.02505 | 0.33905 |
| 494.16837 | 2.0246 | 0.33948 |
| 493.86559 | 2.02415 | 0.33991 |
| 493.56319 | 2.0237 | 0.34034 |
| 493.26115 | 2.02326 | 0.34077 |
| 492.95948 | 2.02281 | 0.3412 |
| 492.65819 | 2.02237 | 0.34163 |
| 492.35726 | 2.02192 | 0.34206 |
| 492.05669 | 2.02148 | 0.34249 |
| 491.7565 | 2.02104 | 0.34292 |
| 491.45667 | 2.02059 | 0.34335 |
| 491.1572 | 2.02015 | 0.34378 |
| 490.85811 | 2.01971 | 0.34421 |
| 490.55937 | 2.01927 | 0.34464 |
| 490.261 | 2.01883 | 0.34507 |
| 489.96299 | 2.01839 | 0.3455 |
| 489.66534 | 2.01795 | 0.34593 |
| 489.36806 | 2.01752 | 0.34636 |
| 489.07113 | 2.01708 | 0.3468 |
| 488.77457 | 2.01664 | 0.34723 |
| 488.47836 | 2.01621 | 0.34766 |
| 488.18252 | 2.01577 | 0.34809 |
| 487.88703 | 2.01534 | 0.34853 |
| 487.5919 | 2.01491 | 0.34896 |
| 487.29712 | 2.01447 | 0.34939 |
| 487.0027 | 2.01404 | 0.34982 |
| 486.70864 | 2.01361 | 0.35026 |
| 486.41493 | 2.01318 | 0.35069 |
| 486.12158 | 2.01275 | 0.35112 |
| 485.82858 | 2.01232 | 0.35156 |
| 485.53594 | 2.01189 | 0.35199 |
| 485.24364 | 2.01146 | 0.35243 |
| 484.9517 | 2.01104 | 0.35286 |
| 484.66011 | 2.01061 | 0.3533 |
| 484.36887 | 2.01018 | 0.35373 |
| 484.07798 | 2.00976 | 0.35417 |
| 483.78743 | 2.00934 | 0.3546 |
| 483.49724 | 2.00891 | 0.35504 |
| 483.20739 | 2.00849 | 0.35547 |
| 482.9179 | 2.00807 | 0.35591 |
| 482.62875 | 2.00764 | 0.35634 |
| 482.33994 | 2.00722 | 0.35678 |
| 482.05148 | 2.0068 | 0.35722 |
| 481.76336 | 2.00638 | 0.35765 |
| 481.47559 | 2.00596 | 0.35809 |
| 481.18817 | 2.00554 | 0.35853 |
| 480.90108 | 2.00513 | 0.35896 |
| 480.61434 | 2.00471 | 0.3594 |
| 480.32794 | 2.00429 | 0.35984 |
| 480.04188 | 2.00388 | 0.36027 |
| 479.75616 | 2.00346 | 0.36071 |
| 479.47078 | 2.00304 | 0.36115 |
| 479.18574 | 2.00263 | 0.36159 |
| 478.90104 | 2.00222 | 0.36203 |
| 478.61668 | 2.0018 | 0.36246 |
| 478.33266 | 2.00139 | 0.3629 |
| 478.04897 | 2.00098 | 0.36334 |
| 477.76562 | 2.00057 | 0.36378 |
| 477.4826 | 2.00016 | 0.36422 |
| 477.19992 | 1.99975 | 0.36466 |
| 476.91757 | 1.99934 | 0.3651 |
| 476.63556 | 1.99893 | 0.36554 |
| 476.35388 | 1.99852 | 0.36598 |
| 476.07253 | 1.99811 | 0.36642 |
| 475.79152 | 1.99771 | 0.36686 |
| 475.51084 | 1.9973 | 0.3673 |
| 475.23048 | 1.9969 | 0.36774 |
| 474.95046 | 1.99649 | 0.36818 |
| 474.67077 | 1.99609 | 0.36862 |
| 474.39141 | 1.99568 | 0.36906 |
| 474.11237 | 1.99528 | 0.3695 |
| 473.83367 | 1.99488 | 0.36994 |
| 473.55529 | 1.99447 | 0.37038 |
| 473.27724 | 1.99407 | 0.37082 |
| 472.99951 | 1.99367 | 0.37126 |
| 472.72211 | 1.99327 | 0.37171 |
| 472.44504 | 1.99287 | 0.37215 |
| 472.16829 | 1.99247 | 0.37259 |
| 471.89186 | 1.99207 | 0.37303 |
| 471.61576 | 1.99167 | 0.37347 |
| 471.33998 | 1.99128 | 0.37392 |
| 471.06453 | 1.99088 | 0.37436 |
| 470.78939 | 1.99048 | 0.3748 |
| 470.51458 | 1.99009 | 0.37525 |
| 470.24008 | 1.98969 | 0.37569 |
| 469.96591 | 1.9893 | 0.37613 |
| 469.69206 | 1.9889 | 0.37658 |
| 469.41852 | 1.98851 | 0.37702 |
| 469.14531 | 1.98812 | 0.37746 |
| 468.87241 | 1.98772 | 0.37791 |
| 468.59983 | 1.98733 | 0.37835 |
| 468.32756 | 1.98694 | 0.3788 |
| 468.05561 | 1.98655 | 0.37924 |
| 467.78398 | 1.98616 | 0.37969 |
| 467.51266 | 1.98577 | 0.38013 |
| 467.24166 | 1.98538 | 0.38058 |
| 466.97097 | 1.98499 | 0.38102 |
| 466.7006 | 1.9846 | 0.38147 |
| 466.43053 | 1.98422 | 0.38191 |
| 466.16078 | 1.98383 | 0.38236 |
| 465.89135 | 1.98344 | 0.3828 |
| 465.62222 | 1.98306 | 0.38325 |
| 465.3534 | 1.98267 | 0.3837 |
| 465.0849 | 1.98229 | 0.38414 |
| 464.8167 | 1.9819 | 0.38459 |
| 464.54881 | 1.98152 | 0.38504 |
| 464.28124 | 1.98113 | 0.38548 |
| 464.01396 | 1.98075 | 0.38593 |
| 463.747 | 1.98037 | 0.38638 |
| 463.48035 | 1.97999 | 0.38682 |
| 463.214 | 1.97961 | 0.38727 |
| 462.94795 | 1.97922 | 0.38772 |
| 462.68221 | 1.97884 | 0.38817 |
| 462.41678 | 1.97846 | 0.38861 |
| 462.15165 | 1.97809 | 0.38906 |
| 461.88683 | 1.97771 | 0.38951 |
| 461.62231 | 1.97733 | 0.38996 |
| 461.35809 | 1.97695 | 0.39041 |
| 461.09417 | 1.97657 | 0.39086 |
| 460.83056 | 1.9762 | 0.3913 |
| 460.56724 | 1.97582 | 0.39175 |
| 460.30423 | 1.97545 | 0.3922 |
| 460.04152 | 1.97507 | 0.39265 |
| 459.7791 | 1.9747 | 0.3931 |
| 459.51699 | 1.97432 | 0.39355 |
| 459.25518 | 1.97395 | 0.394 |
| 458.99366 | 1.97358 | 0.39445 |
| 458.73244 | 1.9732 | 0.3949 |
| 458.47152 | 1.97283 | 0.39535 |
| 458.21089 | 1.97246 | 0.3958 |
| 457.95056 | 1.97209 | 0.39625 |
| 457.69053 | 1.97172 | 0.3967 |
| 457.43079 | 1.97135 | 0.39715 |
| 457.17135 | 1.97098 | 0.3976 |
| 456.9122 | 1.97061 | 0.39805 |
| 456.65334 | 1.97024 | 0.39851 |
| 456.39478 | 1.96987 | 0.39896 |
| 456.13651 | 1.9695 | 0.39941 |
| 455.87853 | 1.96914 | 0.39986 |
| 455.62084 | 1.96877 | 0.40031 |
| 455.36345 | 1.9684 | 0.40076 |
| 455.10634 | 1.96804 | 0.40122 |
| 454.84953 | 1.96767 | 0.40167 |
| 454.593 | 1.96731 | 0.40212 |
| 454.33677 | 1.96694 | 0.40257 |
| 454.08082 | 1.96658 | 0.40303 |
| 453.82516 | 1.96622 | 0.40348 |
| 453.56979 | 1.96585 | 0.40393 |
| 453.31471 | 1.96549 | 0.40438 |
| 453.05991 | 1.96513 | 0.40484 |
| 452.8054 | 1.96477 | 0.40529 |
| 452.55118 | 1.96441 | 0.40574 |
| 452.29724 | 1.96405 | 0.4062 |
| 452.04358 | 1.96369 | 0.40665 |
| 451.79021 | 1.96333 | 0.40711 |
| 451.53712 | 1.96297 | 0.40756 |
| 451.28432 | 1.96261 | 0.40801 |
| 451.0318 | 1.96225 | 0.40847 |
| 450.77956 | 1.96189 | 0.40892 |
| 450.52761 | 1.96153 | 0.40938 |
| 450.27593 | 1.96118 | 0.40983 |
| 450.02454 | 1.96082 | 0.41029 |
| 449.77343 | 1.96046 | 0.41074 |
| 449.52259 | 1.96011 | 0.4112 |
| 449.27204 | 1.95975 | 0.41165 |
| 449.02177 | 1.9594 | 0.41211 |
| 448.77177 | 1.95905 | 0.41256 |
| 448.52205 | 1.95869 | 0.41302 |
| 448.27261 | 1.95834 | 0.41348 |
| 448.02345 | 1.95799 | 0.41393 |
| 447.77457 | 1.95763 | 0.41439 |
| 447.52596 | 1.95728 | 0.41484 |
| 447.27763 | 1.95693 | 0.4153 |
| 447.02957 | 1.95658 | 0.41576 |
| 446.78179 | 1.95623 | 0.41621 |
| 446.53428 | 1.95588 | 0.41667 |
| 446.28704 | 1.95553 | 0.41713 |
| 446.04008 | 1.95518 | 0.41759 |
| 445.7934 | 1.95483 | 0.41804 |
| 445.54698 | 1.95448 | 0.4185 |
| 445.30084 | 1.95413 | 0.41896 |
| 445.05497 | 1.95379 | 0.41942 |
| 444.80937 | 1.95344 | 0.41987 |
| 444.56404 | 1.95309 | 0.42033 |
| 444.31898 | 1.95274 | 0.42079 |
| 444.0742 | 1.9524 | 0.42125 |
| 443.82968 | 1.95205 | 0.42171 |
| 443.58543 | 1.95171 | 0.42217 |
| 443.34145 | 1.95136 | 0.42262 |
| 443.09774 | 1.95102 | 0.42308 |
| 442.85429 | 1.95068 | 0.42354 |
| 442.61112 | 1.95033 | 0.424 |
| 442.36821 | 1.94999 | 0.42446 |
| 442.12556 | 1.94965 | 0.42492 |
| 441.88319 | 1.9493 | 0.42538 |
| 441.64108 | 1.94896 | 0.42584 |
| 441.39923 | 1.94862 | 0.4263 |
| 441.15765 | 1.94828 | 0.42676 |
| 440.91633 | 1.94794 | 0.42722 |
| 440.67528 | 1.9476 | 0.42768 |
| 440.43449 | 1.94726 | 0.42814 |
| 440.19396 | 1.94692 | 0.4286 |
| 439.95369 | 1.94658 | 0.42906 |
| 439.71369 | 1.94624 | 0.42952 |
| 439.47395 | 1.94591 | 0.42998 |
| 439.23447 | 1.94557 | 0.43044 |
| 438.99525 | 1.94523 | 0.4309 |
| 438.75629 | 1.94489 | 0.43136 |
| 438.5176 | 1.94456 | 0.43183 |
| 438.27916 | 1.94422 | 0.43229 |
| 438.04098 | 1.94389 | 0.43275 |
| 437.80306 | 1.94355 | 0.43321 |
| 437.56539 | 1.94322 | 0.43367 |
| 437.32799 | 1.94288 | 0.43413 |
| 437.09084 | 1.94255 | 0.4346 |
| 436.85395 | 1.94222 | 0.43506 |
| 436.61732 | 1.94188 | 0.43552 |
| 436.38094 | 1.94155 | 0.43598 |
| 436.14482 | 1.94122 | 0.43645 |
| 435.90895 | 1.94088 | 0.43691 |
| 435.67334 | 1.94055 | 0.43737 |
| 435.43799 | 1.94022 | 0.43783 |
| 435.20288 | 1.93989 | 0.4383 |
| 434.96804 | 1.93956 | 0.43876 |
| 434.73344 | 1.93923 | 0.43922 |
| 434.4991 | 1.9389 | 0.43969 |
| 434.26501 | 1.93857 | 0.44015 |
| 434.03117 | 1.93824 | 0.44062 |
| 433.79758 | 1.93791 | 0.44108 |
| 433.56425 | 1.93759 | 0.44154 |
| 433.33117 | 1.93726 | 0.44201 |
| 433.09833 | 1.93693 | 0.44247 |
| 432.86575 | 1.9366 | 0.44294 |
| 432.63342 | 1.93628 | 0.4434 |
| 432.40133 | 1.93595 | 0.44387 |
| 432.1695 | 1.93563 | 0.44433 |
| 431.93791 | 1.9353 | 0.4448 |
| 431.70657 | 1.93498 | 0.44526 |
| 431.47548 | 1.93465 | 0.44573 |
| 431.24463 | 1.93433 | 0.44619 |
| 431.01404 | 1.934 | 0.44666 |
| 430.78369 | 1.93368 | 0.44712 |
| 430.55358 | 1.93336 | 0.44759 |
| 430.32372 | 1.93303 | 0.44805 |
| 430.09411 | 1.93271 | 0.44852 |
| 429.86474 | 1.93239 | 0.44898 |
| 429.63562 | 1.93207 | 0.44945 |
| 429.40674 | 1.93175 | 0.44992 |
| 429.1781 | 1.93142 | 0.45038 |
| 428.94971 | 1.9311 | 0.45085 |
| 428.72156 | 1.93078 | 0.45132 |
| 428.49365 | 1.93046 | 0.45178 |
| 428.26598 | 1.93014 | 0.45225 |
| 428.03856 | 1.92983 | 0.45272 |
| 427.81138 | 1.92951 | 0.45318 |
| 427.58444 | 1.92919 | 0.45365 |
| 427.35774 | 1.92887 | 0.45412 |
| 427.13128 | 1.92855 | 0.45459 |
| 426.90506 | 1.92824 | 0.45505 |
| 426.67907 | 1.92792 | 0.45552 |
| 426.45333 | 1.9276 | 0.45599 |
| 426.22783 | 1.92729 | 0.45646 |
| 426.00257 | 1.92697 | 0.45693 |
| 425.77754 | 1.92665 | 0.45739 |
| 425.55275 | 1.92634 | 0.45786 |
| 425.3282 | 1.92602 | 0.45833 |
| 425.10388 | 1.92571 | 0.4588 |
| 424.87981 | 1.9254 | 0.45927 |
| 424.65596 | 1.92508 | 0.45974 |
| 424.43236 | 1.92477 | 0.46021 |
| 424.20899 | 1.92446 | 0.46067 |
| 423.98585 | 1.92414 | 0.46114 |
| 423.76295 | 1.92383 | 0.46161 |
| 423.54028 | 1.92352 | 0.46208 |
| 423.31785 | 1.92321 | 0.46255 |
| 423.09565 | 1.9229 | 0.46302 |
| 422.87368 | 1.92259 | 0.46349 |
| 422.65195 | 1.92227 | 0.46396 |
| 422.43044 | 1.92196 | 0.46443 |
| 422.20917 | 1.92165 | 0.4649 |
| 421.98814 | 1.92135 | 0.46537 |
| 421.76733 | 1.92104 | 0.46584 |
| 421.54675 | 1.92073 | 0.46631 |
| 421.32641 | 1.92042 | 0.46678 |
| 421.10629 | 1.92011 | 0.46725 |
| 420.88641 | 1.9198 | 0.46772 |
| 420.66675 | 1.9195 | 0.46819 |
| 420.44733 | 1.91919 | 0.46866 |
| 420.22813 | 1.91888 | 0.46914 |
| 420.00916 | 1.91857 | 0.46961 |
| 419.79042 | 1.91827 | 0.47008 |
| 419.57191 | 1.91796 | 0.47055 |
| 419.35362 | 1.91766 | 0.47102 |
| 419.13556 | 1.91735 | 0.47149 |
| 418.91773 | 1.91705 | 0.47196 |
| 418.70012 | 1.91674 | 0.47244 |
| 418.48274 | 1.91644 | 0.47291 |
| 418.26559 | 1.91614 | 0.47338 |
| 418.04866 | 1.91583 | 0.47385 |
| 417.83196 | 1.91553 | 0.47432 |
| 417.61548 | 1.91523 | 0.4748 |
| 417.39922 | 1.91492 | 0.47527 |
| 417.18319 | 1.91462 | 0.47574 |
| 416.96738 | 1.91432 | 0.47621 |
| 416.7518 | 1.91402 | 0.47669 |
| 416.53643 | 1.91372 | 0.47716 |
| 416.32129 | 1.91342 | 0.47763 |
| 416.10638 | 1.91312 | 0.47811 |
| 415.89168 | 1.91282 | 0.47858 |
| 415.67721 | 1.91252 | 0.47905 |
| 415.46295 | 1.91222 | 0.47953 |
| 415.24892 | 1.91192 | 0.48 |
| 415.03511 | 1.91162 | 0.48047 |
| 414.82152 | 1.91132 | 0.48095 |
| 414.60814 | 1.91102 | 0.48142 |
| 414.39499 | 1.91072 | 0.4819 |
| 414.18206 | 1.91043 | 0.48237 |
| 413.96934 | 1.91013 | 0.48284 |
| 413.75685 | 1.90983 | 0.48332 |
| 413.54457 | 1.90954 | 0.48379 |
| 413.33251 | 1.90924 | 0.48427 |
| 413.12066 | 1.90894 | 0.48474 |
| 412.90904 | 1.90865 | 0.48522 |
| 412.69763 | 1.90835 | 0.48569 |
| 412.48643 | 1.90806 | 0.48617 |
| 412.27546 | 1.90776 | 0.48664 |
| 412.06469 | 1.90747 | 0.48712 |
| 411.85415 | 1.90717 | 0.48759 |
| 411.64382 | 1.90688 | 0.48807 |
| 411.4337 | 1.90659 | 0.48854 |
| 411.2238 | 1.90629 | 0.48902 |
| 411.01411 | 1.906 | 0.48949 |
| 410.80463 | 1.90571 | 0.48997 |
| 410.59537 | 1.90542 | 0.49045 |
| 410.38633 | 1.90512 | 0.49092 |
| 410.17749 | 1.90483 | 0.4914 |
| 409.96887 | 1.90454 | 0.49187 |
| 409.76046 | 1.90425 | 0.49235 |
| 409.55226 | 1.90396 | 0.49283 |
| 409.34427 | 1.90367 | 0.4933 |
| 409.13649 | 1.90338 | 0.49378 |
| 408.92893 | 1.90309 | 0.49426 |
| 408.72157 | 1.9028 | 0.49473 |
| 408.51443 | 1.90251 | 0.49521 |
| 408.3075 | 1.90222 | 0.49569 |
| 408.10077 | 1.90193 | 0.49617 |
| 407.89425 | 1.90165 | 0.49664 |
| 407.68795 | 1.90136 | 0.49712 |
| 407.48185 | 1.90107 | 0.4976 |
| 407.27596 | 1.90078 | 0.49807 |
| 407.07028 | 1.9005 | 0.49855 |
| 406.8648 | 1.90021 | 0.49903 |
| 406.65953 | 1.89992 | 0.49951 |
| 406.45447 | 1.89964 | 0.49999 |
| 406.24962 | 1.89935 | 0.50046 |
| 406.04497 | 1.89907 | 0.50094 |
| 405.84053 | 1.89878 | 0.50142 |
| 405.6363  | 1.8985  | 0.5019  |
| 405.43226 | 1.89821 | 0.50238 |
| 405.22844 | 1.89793 | 0.50286 |
| 405.02482 | 1.89764 | 0.50333 |
| 404.8214  | 1.89736 | 0.50381 |
| 404.61819 | 1.89708 | 0.50429 |
| 404.41519 | 1.89679 | 0.50477 |
| 404.21238 | 1.89651 | 0.50525 |
| 404.00978 | 1.89623 | 0.50573 |
| 403.80738 | 1.89595 | 0.50621 |
| 403.60519 | 1.89566 | 0.50669 |
| 403.4032  | 1.89538 | 0.50717 |
| 403.20141 | 1.8951  | 0.50765 |
| 402.99982 | 1.89482 | 0.50813 |
| 402.79843 | 1.89454 | 0.50861 |
| 402.59725 | 1.89426 | 0.50909 |
| 402.39626 | 1.89398 | 0.50957 |
| 402.19548 | 1.8937  | 0.51005 |
| 401.99489 | 1.89342 | 0.51053 |
| 401.79451 | 1.89314 | 0.51101 |
| 401.59432 | 1.89286 | 0.51149 |
| 401.39434 | 1.89258 | 0.51197 |
| 401.19455 | 1.8923  | 0.51245 |
| 400.99497 | 1.89202 | 0.51293 |
| 400.79558 | 1.89175 | 0.51341 |
| 400.59639 | 1.89147 | 0.51389 |
| 400.39739 | 1.89119 | 0.51437 |
| 400.1986  | 1.89091 | 0.51485 |
| 400       | 1.89064 | 0.51533 |